\documentclass[10pt,journal]{IEEEtran}
\usepackage{amsthm}
\usepackage{graphicx} %
\usepackage{mathptmx} %
\usepackage{times} %
\usepackage{amssymb}  %
\usepackage{pstricks,pst-plot,psfrag}
\usepackage{float}
\usepackage{multirow}
\usepackage[draft]{hyperref}
\usepackage{booktabs}
\usepackage{cite}
\usepackage{lettrine}
\usepackage{import}
\usepackage{tikz}
\usepackage{epigraph}
\def\ppbb{path picture bounding box}

\usetikzlibrary{positioning, calc, decorations.pathmorphing,shapes.multipart, shadows, patterns}

\tikzset{>=stealth} %

\tikzstyle{block} = [draw, rectangle, minimum height=2.2em, minimum width=6.5em]
\tikzstyle{coord} = [coordinate]
\tikzstyle{sum} = [draw,circle]
\tikzstyle{input} = [coordinate]
\tikzstyle{output} = [coordinate]
\tikzstyle{pinstyle} = [pin edge={<-,>=stealth,solid,line width = 0.4pt,black}]

\tikzstyle{tank} = %
	[transform shape,draw, path picture={ 
			\coordinate	(L) at ($ (\ppbb.south west)!.7!(\ppbb.north west) $);
			\coordinate	(R) at ($ (\ppbb.south east)!.7!(\ppbb.north east) $);
			\draw[decorate,decoration={snake, amplitude=1pt, segment length=6pt}] (L) -- (R);
	}]

\tikzstyle{circular node} = [circle, draw, font=\huge]

\tikzstyle{mechanical shaft} = [very thick]

\tikzset{
  adjust height/.style={minimum height=#1*\pgfkeysvalueof{/pgf/minimum width}},
  adjust width/.style={minimum width=#1*\pgfkeysvalueof{/pgf/minimum height}}
}

\tikzstyle{wheel} = [rectangle, draw, rounded corners, adjust height=0.5, path picture={
\draw[-] ($(\ppbb.south west)!0.8!(\ppbb.north west)$) -- ($(\ppbb.south east)!0.8!(\ppbb.north east)$);
\draw[-] ($(\ppbb.south west)!0.2!(\ppbb.north west)$) -- ($(\ppbb.south east)!0.2!(\ppbb.north east)$);
}]

\tikzstyle{wheelRotated} = [rectangle, draw, rounded corners, adjust width=0.5,path picture={
\draw[-] ($(\ppbb.north east)!0.8!(\ppbb.north west)$) -- ($(\ppbb.south east)!0.8!(\ppbb.south west)$);
\draw[-] ($(\ppbb.north east)!0.2!(\ppbb.north west)$) -- ($(\ppbb.south east)!0.2!(\ppbb.south west)$);
}]

\tikzstyle{inverter} = %
[transform shape,draw, path picture ={ %
		\path let \p1=(\ppbb.north west), \p2=(\ppbb.south west), \n1={1.73*.25*veclen(\x2-\x1,\y2-\y1)} in \pgfextra{\xdef\var{\n1}};
		\draw ($(\ppbb.west)!.05!(\ppbb.east)$) -- ($(\ppbb.west)!.95!(\ppbb.east)$);
		\path ([shift=(0:-.05cm)]\ppbb.center) -- +(0:\var) coordinate (A)  +(120:\var) coordinate(B) -- +(240:\var) coordinate(C);
		\draw[fill=white] (A)--(B)--(C)--cycle;
		\node[circle, draw, minimum size=.05cm, anchor=west, inner sep=0pt, fill=white] (circ) at (A){};
}]

\tikzstyle{blockDyn} = [draw, rectangle, minimum height=2.5em, minimum width=3.5em, align=center, inner sep=10pt, thick, fill=white, copy shadow={draw=black,fill=black,opacity=1,shadow xshift=0.5ex,shadow yshift=-0.5ex}]

\tikzstyle{blockAlg} = [draw, rectangle, minimum height=1.5em, minimum width=2.5em, align=center, inner sep=10pt, thick]

\tikzstyle{sum} = [draw,circle]

\tikzset{add reference/.style={insert path={%
			coordinate [pos=0,xshift=-0.5\pgflinewidth,yshift=-0.5\pgflinewidth] (#1 south west) 
			coordinate [pos=1,xshift=0.5\pgflinewidth,yshift=0.5\pgflinewidth]   (#1 north east)
			coordinate [pos=.5] (#1 center)                        
			(#1 south west |- #1 north east)     coordinate (#1 north west)
			(#1 center     |- #1 north east)     coordinate (#1 north)
			(#1 center     |- #1 south west)     coordinate (#1 south)
			(#1 south west -| #1 north east)     coordinate (#1 south east)
			(#1 center     -| #1 south west)     coordinate (#1 west)
			(#1 center     -| #1 north east)     coordinate (#1 east)   
		}}}

\tikzstyle{simbolo} = [draw, path picture={ %
	\node[circle, fill, minimum size=6pt, below, inner sep=0] (A) at (\ppbb.north){};
	\draw[densely dotted] (A.south) -- ($(0,0 |- \ppbb.south)+(0, 10pt)$) coordinate (temp);
	\draw[thick] (temp) -- +(-.1, 0) -- +(.1, 0);
	\node at ($(0,0 -| \ppbb.south)+(0, 5pt)$)(testo) {$<100\%?$};
	\node[circle, fill, minimum size=3pt, inner sep=0, left] (B) at (\ppbb.15){};
	\node[circle, fill, minimum size=3pt, inner sep=0, right] (C) at (\ppbb.165){};
	\draw[thick] (A) -- (C);
	\node (testo) at (testo-|\ppbb.west){};
}]
\input{pictures/engineGearbox}
\usepackage{luatex85}
\usepackage{booktabs}
\usepackage{cancel}	
\usepackage[acronym]{glossaries}
\usepackage{float}
\usepackage{amsmath}
\usepackage{amsfonts}
\usepackage{bm}
\usepackage{cleveref}
\crefname{figure}{fig.}{figs.}

\usepackage{siunitx}
\usepackage{pgfplots}
\usepackage{algorithm}
\usepackage{algorithmic}
\usetikzlibrary{positioning}
\usetikzlibrary{arrows.meta}

\newcommand{\ds}{\textnormal{d}s}

\newcommand{\dds}{\frac{\textnormal{d}}{\textnormal{d}s}}
\definecolor{lightblue}{rgb}{0.60784,0.76078,0.90196}
\definecolor{darkblue}{rgb}{0.26667,0.44706,0.76863}
\definecolor{lightgreen}{rgb}{0.66275,0.81569,0.55686}
\definecolor{darkgreen}{rgb}{0.43922,0.67843,0.27843}
\definecolor{orange}{rgb}{0.92941,0.49020,0.19216}
\definecolor{yellow}{rgb}{1.00000,0.75294,0.00000}
\definecolor{grey}{rgb}{0.64706,0.64706,0.64706}
\definecolor{purple}{rgb}{0.51373,0.23529,0.04706}

\newtheorem{problem}{Problem}

\newacronym{acr:aecms}{A-ECMS}{adaptive ECMS}
\newacronym{acr:dp}{DP}{dynamic programming}
\newacronym{acr:ecms}{ECMS}{Equivalent Consumption Minimization Strategy}
\newacronym{acr:eltms}{ELTMS}{Equivalent Lap Time Minimization Strategies}
\newacronym{acr:ecu}{ECU}{electronic control unit}
\newacronym{acr:ers}{ERS}{Energy Recovery System}
\newacronym{acr:F1}{F1}{Formula 1}
\newacronym{acr:FIA}{FIA}{Fédération Internationale de l'Automobile}
\newacronym{acr:fcc}{FCC}{feedforward cylinder controller}
\newacronym{acr:ICE}{ICE}{internal combustion engine}
\newacronym{acr:mguh}{MGU-H}{motor generator unit - heat}
\newacronym{acr:mguk}{MGU-K}{motor generator unit - kinetic}
\newacronym{acr:mpc}{MPC}{model predictive control}
\newacronym{acr:milp}{MILP}{mixed-integer linear programs}
\newacronym{acr:nmpc}{NMPC}{nonlinear MPC}
\newacronym{acr:nlp}{NLP}{nonlinear program}
\newacronym{acr:ocp}{OCP}{optimal control problem}
\newacronym{acr:pmp}{PMP}{Pontryagin's minimum principle}
\newacronym{acr:pi}{PI}{proportional–integral}
\newacronym{acr:wg}{WG}{waste-gate}

\renewcommand\vec{\mathbf}
\usepackage{tkz-euclide}

\usepgfplotslibrary{external} 
\newif\ifmargincomments %
\margincommentstrue
\ifmargincomments

\else

\fi

\begin{document}
\title{Low-level Online Control of the Formula 1 Power Unit with Feedforward Cylinder Deactivation}
\author{\IEEEauthorblockN{
        Marc-Philippe Neumann\IEEEauthorrefmark{1}, 
        Giona Fieni\IEEEauthorrefmark{1},
        Camillo Balerna\IEEEauthorrefmark{1},
        Pol Duhr\IEEEauthorrefmark{1},
        Alberto Cerofolini\IEEEauthorrefmark{2},  
        Christopher H. Onder\IEEEauthorrefmark{1}\\
    }
    \IEEEauthorblockA{
        \IEEEauthorrefmark{1} Institute of Dynamic Systems and Control, ETH Zürich, Sonneggstrasse 3, 8092 Zürich, Switzerland.\\
        \IEEEauthorrefmark{2} Power Unit Performance Group, Ferrari S.p.A., 41053 Maranello, Italy}
}

\IEEEpubid{\begin{minipage}{\textwidth}\ \centering \\[12pt]
		DOI \href{https://doi.org/10.1109/TVT.2023.3246130}{10.1109/TVT.2023.3246130},
		\copyright~2023 IEEE.  Personal use of this material is permitted.  Permission from IEEE must be obtained for all other uses, in any current or future media, including reprinting/republishing this material for advertising or promotional purposes, creating new collective works, for resale or redistribution to servers or lists, or reuse of any copyrighted component of this work in other works.
\end{minipage}}

\maketitle
\begin{abstract}
Since 2014, the Fédération Internationale de l'Automobile has prescribed a parallel hybrid powertrain for the Formula 1 race cars.
The complex low-level interactions between the thermal and the electrical part represent a non-trivial and challenging system to be controlled online. 
We present a novel controller architecture composed of a supervisory controller for the energy management, a feedforward cylinder deactivation controller, and a track region-dependent low-level nonlinear model predictive controller to optimize the engine actuators. 
Except for the nonlinear model predictive controller, the proposed controller subsystems are computationally inexpensive and are real time capable. 
The framework is tested and validated in a simulation environment for several realistic scenarios disturbed by driver actions or grip conditions on the track.
In particular, we analyze how the control architecture deals with an unexpected gearshift trajectory during an acceleration phase.
Further, we demonstrate how an increased maximum velocity trajectory impacts the online low-level controller.
Our results show a suboptimality over an entire lap with respect to the benchmark solution of 49 ms and 64 ms, respectively, which we deem acceptable. 
Compared to the same control architecture with full knowledge of the disturbances, the suboptimality amounted to only 2 ms and 17 ms.
For all case studies we show that the cylinder deactivation capability decreases the suboptimality by 7 to 8 ms.
\end{abstract}
\begin{IEEEkeywords}
Online Control, Nonlinear Model Predictive Control, Cylinder Deactivation, Equivalent Lap Time Minimization Strategies, Energy Management, Low-level Control, Formula 1, Hybrid Vehicles.
\end{IEEEkeywords}

\section{Introduction}\label{sec:introduction}
\begin{figure}
	\centering
	\includegraphics[width=\columnwidth]{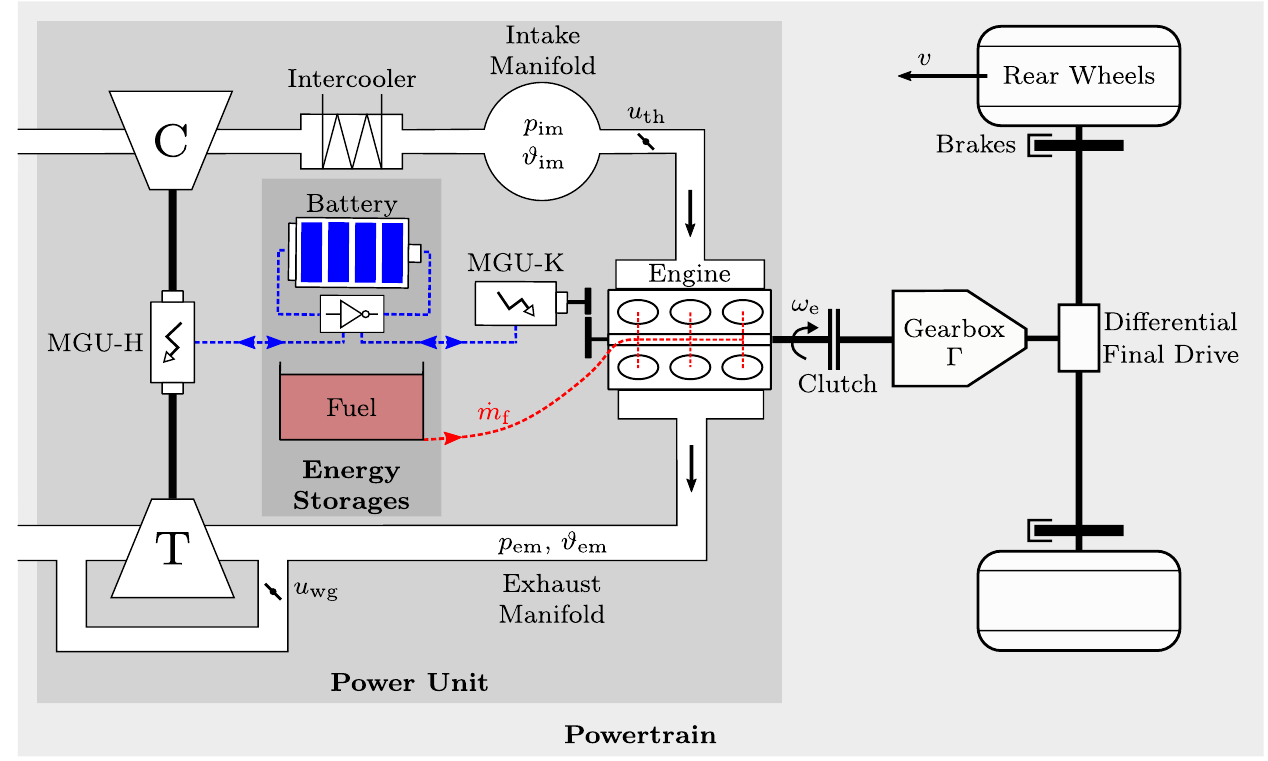}
	\caption{Schematic of the F1 powertrain. In the darker gray-shaded area there is the turbocharged V6 internal combustion engine with the two electric motors (MGU-K and MGU-H). Highlighted in lighter gray the coupling to the vehicles' wheels is shown, featuring the clutch, the gearbox, the differential drive, and the friction brakes.}
	\label{fig:Powertrain}\vspace{-0.4cm}
\end{figure}

\IEEEPARstart{F}{ormula} 1 (F1) cars belong to the fastest circuit racing vehicles in the world, competing in the most prestigious racing championship.
Since its foundation in the early 1950s, the \gls{acr:FIA} released multiple rule updates concerning safety, racing regulations, as well as technical requirements~\cite{FIA2021Sporting,FIA2021Technical}.
One revolutionary technical update was introduced in the 2014 season, featuring the parallel electric hybrid powertrain architecture shown in Fig. \ref{fig:Powertrain}.
Since then, a downsized \SI{1.6}{L} V6 \gls{acr:ICE} with individual cylinder deactivation capability is boosted by a turbocharger, which is further electrified by means of an electric motor-generator unit (MGU-H, H for heat).
This electric machine is exploited to compensate for suboptimal transient phenomena such as the turbo lag, and also recuperates energy from the exhaust gases thanks to an oversized turbine producing more power than used by the compressor.
A second motor, called MGU-K (K for kinetic), is mounted on the crankshaft of the engine, providing extra torque when accelerating and recuperating energy in braking phases.
It is restricted by the \gls{acr:FIA} with a power constraint of $\pm \SI{120}{kW}$.
Henceforth denoted as power unit, this propulsive package is connected to the wheels through an eight-speed sequential gearbox and a limited slip differential. 
Together with the friction brakes actuated by the driver and controlled via a brake-by-wire system, this setup represents the complete powertrain of the considered F1 racing car.
To incentivize the efficient operation of this power unit, the regulations allow the use of an energy management controller.
To understand its degrees of freedom, we need to distinguish between two regions that depend on the driving behavior of the driver: the power-limited region (PL) and the grip-limited region (GL). \IEEEpubidadjcol
In the power-limited region the velocity of the car is only limited by the maximum power output of the power unit: This usually occurs on the straights where the driver is fully pressing the throttle pedal and the maximum velocity profile is not reached.
Here, the \gls{acr:ecu} is allowed to overwrite the driver's request and provide a different amount of traction power depending on the selected strategy.
In the grip-limited region, however, the car is limited by the grip of the tires: This usually occurs in the corners where the driver is not requesting full power. 
Since the regulations do not allow active traction control, the power requested by the driver needs to be fulfilled and only the split between \gls{acr:ICE}, MGU-K and friction brakes can be chosen.
Additionally, refueling during pit stops is prohibited, the battery capacity is limited to \SI{4}{MJ}, and electrical boosting and recuperating is restricted to \SI{4}{MJ} and \SI{2}{MJ} per lap, respectively.

The lap time optimization of this complex system is of paramount importance to ensure that the driver has a car that makes optimal use of its propulsive capabilities within a predefined regulatory framework. 
The setup of this hybrid power unit introduces highly complex interactions between the thermal system and the electric motors, e.g., we do not only have to consider trivial scenarios such as pure electric boost or recuperation, but also need to account for nonlinear turbocharger transients in the energy balance.
Coupled to the numerous \gls{acr:FIA} rules that constrain energy availability, this problem statement calls for model-based optimization of the powertrain's operation.
In this research paper, we present a two-level online control architecture combining a computationally inexpensive supervisory energy management with a model-based low-level power unit controller.
Furthermore, we include our controller setup in a simulation framework to assess the online solution and compare it to offline optimizations.

\subsection{Literature Review}\label{subsec:literaturereview}
Owing to the power unit topology of modern F1 cars, the relevant methodologies investigated are found in the hybrid electric vehicle research field.
Therefore, we identify three different streams of research related to our topic.

The first one deals with the offline optimization of the energy management of hybrid electric vehicles.
Given the steadily increasing restrictions on CO$_2$ emissions, many non-causal fuel optimal control strategies using \gls{acr:dp}~\cite{hooker1988optimal,perez2009optimal,heppeler2014fuel}, convex optimization~\cite{sciarretta2004optimal,nuesch2014convex}, and \gls{acr:pmp}~\cite{sciarretta2007control,kim2011optimal,sciarretta2015optimal} were investigated.
Further, models with integer variables such as engine on/off and gearshifts were considered and optimized by means of iterative linear programming~\cite{robuschi2018minimum}, via \gls{acr:milp}~\cite{BALERNA2020115248} and \gls{acr:pmp}~\cite{ritzmann2019}.
However, given the racing application, our main interest lies in lap time optimizations.
For F1 racing vehicles the focus has been put on the optimal energy recovery system control~\cite{limebeer2014optimal}, whilst modeling the internal combustion engine from a high-level perspective.
Additionally, a three-dimensional track was identified from GPS data, and subsequently used for joint optimization of lap time and driven line~\cite{perantoni2015optimal,limebeer2015optimal,perantoni2014optimal}.
In our research group, time-optimal control strategies have been investigated, leveraging convex approximations and relaxations~\cite{ebbesen2018time}. 
Finally, to the best of the authors' knowledge, in~\cite{Balerna} the joint optimization of a detailed low-level model of the F1 powertrain and the energy management from a lap time perspective was presented for the first time.

In the second stream we consider the online control of hybrid electric powertrains.
In~\cite{serrao2011comparative} the authors compare the non-causal \gls{acr:dp} and \gls{acr:pmp} methods to the real time feasible \gls{acr:ecms}~\cite{paganelli2002equivalent}.
A comparison between rule-based control, \gls{acr:aecms} and $\mathcal{H}_\infty$ control has been pursued in~\cite{Pisu2007}, concluding that the \gls{acr:aecms} strategy is the best performing one.
These strategies have been augmented to consider additional constraints such as pollutant emissions~\cite{nuesch2014equivalent} or the battery state-of-health~\cite{ebbesen2012battery}.
Moreover, other tuning approaches such as the fuzzy-tuned \gls{acr:ecms} were proposed as real-time solution~\cite{zhao2015real}.
Various model-based approaches were investigated to tackle the optimal energy management: In~\cite{Borhan2009,borhan2012mpc} the authors linearized the model at each sample time and employed \gls{acr:mpc}, while in \cite{zhao2017characterisation} a supervisory controller was merged with an $\mathcal{H}_\infty$ for electrified turbocharger diesel engines.
Further, non-causal trajectories of the battery's state of charge generated with \gls{acr:dp} optimizations were tracked by means of an \gls{acr:mpc} in~\cite{Zhou2021}.
In~\cite{lot2013lap} the author also considered the optimal trajectory optimization.
In our research group, the \gls{acr:ecms} was adapted for the goal of lap time optimization by defining the optimal control policy through \gls{acr:pmp}~\cite{salazar2017time}, resulting in the \gls{acr:eltms}~\cite{salazar2018equivalent,salazar2018minimum}.
The online control of a linearized model of an F1 car was tackled also by means of a two-level \gls{acr:mpc} scheme~\cite{salazar2017real}.
Most recently, research on lap time optimal control gained traction in the field of pure electric race cars, particularly focusing on the transmission design and control~\cite{Borsboom2020,Borsboom2021}, as well as thermal limitations~\cite{Locatello2021}.

Finally, in the third research stream we looked at state-of-the-art cylinder deactivation online control.
In \cite{Michelini2003} the authors examine the engine efficiency improvement due to cylinder deactivation introducing various engine modes that employ specific control actions. 
However, in literature the investigated operation modes mostly include binary decision variables for the state of the \gls{acr:ICE}: either on or off \cite{elbert2014engine,6476747}.
In \cite{Corno2019} the authors incremented the modes with a ``half" running engine and propose a probabilistic approach using Markov's chains to determine the engine operation mode.
Further, additional variations such as three, four, and six running cylinders were analyzed \cite{Fujiwara2008}.
In \cite{Sujan} the authors discuss the case of fully individual cylinder deactivation. 
They used engine speed vs. torque maps to determine the optimal amount of cylinders.
To include the possibility of shutting off the engine in an optimization framework and avoid mixed-integer programs, partial outer convexification \cite{Josevski2016} and iterative schemes \cite{robuschi2018minimum} have been investigated.
Finally, in racing applications without refueling possibility, the crucial influence of the fuel quantity at the race start on the vehicle's weight and thus on the lap time was shown in~\cite{Bekker2009,Heilmeier2018,RacingLimited15/02/2022}.

\subsection{Research Statement}\label{subsec:statement}
To the best of the authors' knowledge, there exists a gap in the low-level online control of hybrid electric racing vehicles.
In particular, low-level actuators of the power unit are not considered in the state-of-the-art lap time optimal energy management. 
The literature focuses either on fuel optimality which is irrelevant in racing applications, or online controllers that rely on high-level models.
The latter allow for computationally efficient implementations.
However, the underlying assumption is that a requested power is always achievable.
This entirely neglects the actual actuation that allows for its delivery.
Therefore, in this paper we focus on a model-based online control architecture that aims for lap time optimal disturbance rejection by considering actual power unit actuators and accounting for fast changing system dynamics.
Such an architecture allows us to react in a lap time optimal fashion to energy budget deviations resulting from various disturbances by controlling low-level actuators, such as the air-to-fuel ratio or the spark advance efficiency.
Due to confidentiality reasons, a comparison with the online controller currently employed in our specific application field is not possible.
We therefore focus on the analysis of the suboptimality under given disturbances compared to a benchmark solution, commenting on the origins of the lost lap time and on the strength of our model-based approach.
Finally, we want to provide a tool to test online controllers and infer potential heuristics that can be applied on a race weekend.
Due to the immense cost of testing on a physical system, we therefore design a simulation environment.

\subsection{Contribution}\label{subsec:contribution}
The scientific value of this paper is threefold:
The main contribution is the implementable online control architecture.
Its novelty is the combination of a supervisory high-level controller \gls{acr:eltms} taking care of slow changing dynamics, a detailed low-level \gls{acr:nmpc} accounting for all relevant system states, and an estimator providing predictive data on the driver actions.
Second, we devise a feedforward cylinder deactivation strategy relying on look-up tables that are derived from behavioral patterns of various lap time optimizations.
This novel architecture avoids computationally expensive mixed-integer programming and allows the deactivation of single cylinders during transients rather than just an on/off behavior.
Finally, we augment the online controller with a simulation environment. 
The resulting framework provides the possibility to run the controller without expensive test bench sessions and assess its suboptimality under given disturbances with respect to a benchmark. 
Furthermore, control heuristics for the race can be inferred and other controllers can be compared.
\section{Framework Overview}\label{sec:overview}
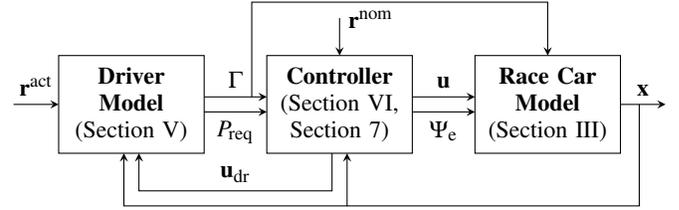
\begin{figure}
	\centering
	\tikzstyle{block} = [draw, rectangle, minimum height=2em, minimum width=3em]
\tikzstyle{input} = [coordinate]
\tikzstyle{output} = [coordinate]

\begin{tikzpicture}[scale=1, every node/.append style={outer sep=0pt}, >=stealth, font=\small] %

\def\altezza{1.3cm}
\def\larghezza{1.7cm}

\node [block, minimum height=\altezza, minimum width=\larghezza, text centered, text width=\larghezza] at (0,0) (controller) {\textbf{Controller} (Section \ref{sec:controller}, Section \ref{fig:ControllerInside})};
\node [block, minimum height=\altezza, minimum width=\larghezza, text centered, text width=\larghezza] at ($(controller.east) + (1.8,0)$) (plant) {\textbf{Race Car Model} (Section \ref{sec:model})};
\node [block, minimum height=\altezza, minimum width=\larghezza, text centered, text width=\larghezza] at ($(controller.west) + (-1.8,0)$) (driver) {\textbf{Driver Model} (Section \ref{sec:driver})};

\node[coordinate] at ($(driver.south) + (-0.1,0)$) (driverSottoSinistra) {};
\node[coordinate] at ($(driver.south) + (-0.1,-0.7)$) (driverSottoSinistra1) {};
\node[coordinate] at ($(driver.south) + (0.1,0)$) (driverSottoDestra) {};
\node[coordinate] at ($(driver.south) + (0.1,-0.5)$) (driverSottoDestra1) {};
\node[coordinate] at ($(driver.east) + (0,0.1)$) (driverDestra1) {};
\node[coordinate] at ($(driver.east) + (0,-0.1)$) (driverDestra2) {};
\node[coordinate] at ($(driver.west)$) (driverSinistra) {};
\node[coordinate] at ($(driver.west) + (-0.6, 0)$) (driverSinistra1) {};

\node[coordinate] at ($(controller.west) + (0, 0.1)$) (controllerSinistra1) {};
\node[coordinate] at ($(controller.west) + (0,-0.1)$) (controllerSinistra2) {};
\node[coordinate] at ($(controller.west) + (-0.2, 0.1)$) (controllerSinistra3) {};
\node[coordinate] at ($(controllerSinistra3) + (0, 1cm+0.2*\altezza)$) (controllerSopraSinistra) {};
\node[coordinate] at ($(controller.south) + (0.1,0)$) (controllerSottoDestra) {};
\node[coordinate] at ($(controller.south) + (0.1,-0.7)$) (controllerSottoDestra1) {};
\node[coordinate] at ($(controller.south) + (-0.1,0)$) (controllerSottoSinistra) {};
\node[coordinate] at ($(controller.south) + (-0.1,-0.5)$) (controllerSottoSinistra1) {};
\node[coordinate] at ($(controller.north)$) (controllerSopra) {};
\node[coordinate] at ($(controller.north) + (0, 0.5)$) (controllerSopra1) {};
\node[coordinate] at ($(controller.east) + (0,0.1)$) (controllerDestra1) {};
\node[coordinate] at ($(controller.east) + (0,-0.1)$) (controllerDestra2) {};

\node[coordinate] at ($(plant.west) + (0,0.1)$) (plantSinistra1) {};
\node[coordinate] at ($(plant.west) + (0,-0.1)$) (plantSinistra2) {};
\node[coordinate] at ($(plant.east)$) (plantDestra) {};
\node[coordinate] at ($(plant.east) + (0.6,0)$) (plantDestra1) {};
\node[coordinate] at ($(plant.east) + (0.25,0)$) (plantDestra2) {};
\node[coordinate] at ($(plantDestra2) + (0,-\altezza/2-0.7cm)$) (plantSottoDestra) {};
\node[coordinate] at ($(plant.north)$) (plantSopra) {};
\node[coordinate] at ($(plant) + (0,1.1cm+0.2*\altezza)$) (plantSopra1) {};

\draw [->] (driverSinistra1) to node[above] {$\vec{r}^\mathrm{act}$} (driverSinistra);

\draw [->] (driverDestra1) to node[above] {$\Gamma$} (controllerSinistra1);
\draw [->] (driverDestra2) to node[below] {$P_\mathrm{req}$} (controllerSinistra2);
\draw [->] (controllerSinistra3) -- (controllerSopraSinistra) -- (plantSopra1) -- (plantSopra);

\draw [->] (controllerDestra1) to node[above] {$\vec{u}$} (plantSinistra1);
\draw [->] (controllerDestra2) to node[below] {${\Psi}_\mathrm{e}$} (plantSinistra2);
\draw [->] (controllerSopra1) to node[above right] {$\vec{r}^\mathrm{nom}$} (controllerSopra);
\draw [->] (controllerSottoSinistra) -- (controllerSottoSinistra1) -- node[above] {$\vec{u}_\mathrm{dr}$} (driverSottoDestra1) -- (driverSottoDestra);

\draw [->] (plantDestra) to node[above] {$\vec{x}$} (plantDestra1);
\draw [->] (plantDestra2) |- (controllerSottoDestra1) -- (controllerSottoDestra);
\draw [->] (controllerSottoDestra1) -| (driverSottoSinistra);

\end{tikzpicture}
	\caption{Overview of the framework, containing the controller, the driver model and the race car model. For readability reasons we dropped the space dependencies of the shown variables. We distinguish between actual and nominal reference $\vec{r}^\mathrm{act}(s)$ and $\vec{r}^\mathrm{nom}(s)$, respectively, as they differ in a disturbed scenario. The outputs of the driver model are the requested power $P_\mathrm{req}(s)$ and the engaged gear $\Gamma(s)$. The controller provides the control vector $\vec{u}(s)$ and the number of active cylinders $\Psi_{\mathrm{e}}(s)$ going into the race car model, as well as some control information $\vec{u}_\mathrm{dr}(s)$ needed in the driver model. Finally, $\vec{x}(s)$ represents the state vector.}
	\label{fig:Overview}\vspace{-0.35cm}
\end{figure}

To facilitate the understanding of this paper, in this section we introduce the structure of our work and highlight the interconnections between the elements.
Fig. \ref{fig:Overview} illustrates the three main components of our framework: the driver model, the controller, and the race car model. 
To test the performance of the controller in a simulation environment we need a driver model.
This component is responsible for providing driver inputs that are not subject to optimization and act as a disturbance to the controller.
Next, we dive into the main contribution outlined in Section \ref{subsec:contribution}: the control architecture.
The heart of this controller is an \gls{acr:nmpc}, which performs reference tracking by optimizing low-level powertrain actuators.
The reference trajectories are obtained by solving the offline optimization problem presented in \cite{Balerna} for the race car model.
Additionally, the controller includes an estimator that is designed to generate future driver actions relying on the same offline optimization results and feedback from the race car. 
To cope with disturbances from an energetic point of view, we also include a supervisory controller, which needs the optimal reference trajectories to recognize discrepancies and react accordingly.
To compute the suboptimality with respect to a non-causal benchmark, we recompute the optimal solution offline with the energy consumption obtained during online control and known disturbances.
Finally, the driver's gear choice and the control inputs are fed to the race car model.

\subsection{Paper Structure}\label{subsec:structure}
In Section \ref{sec:model} we introduce the low-level race car model that we aim to control.
In Section \ref{sec:offlineoptimization}, we illustrate the offline optimization of such a system that has been carried out previously in our research group.
Section \ref{sec:driver} addresses the driver model needed for the simulation of the online control loop. 
Then, in Section \ref{sec:controller}, we outline the estimator that provides the predictive information for the \gls{acr:nmpc}. 
Furthermore, we include the supervisory \gls{acr:eltms} controller to comply with slow dynamical disturbances in energy trajectories, and a feedforward cylinder deactivation controller to tackle the mixed-integer nature of such a system. 
We conclude the section with the definition of the optimal control problems solved by the \gls{acr:nmpc}. 
In Section \ref{sec:results} we showcase the performance and robustness of our framework by means of two case studies where we analyze multiple disturbances generated by an unforeseen driver behavior. 
Finally, in Section \ref{sec:conclusion} we draw the conclusions, comment on the relevant insights gained by means of our framework, and give an outlook on future research.
\section{Race Car Model}\label{sec:model}
As introduced in Section \ref{sec:introduction}, the system that we control is an F1 hybrid electric race car.
Its modeling relies on \cite{guzzella2004introduction} and \cite{Balerna}, to which the reader is referred for the detailed model validation.
The race car model is embedded in our framework as shown in Fig. \ref{fig:Overview}.
The state vector $\vec{x}$ of the race car model reads as
\begin{equation}
	\vec{x}(s) = \begin{bmatrix}
		v(s) & p_\mathrm{im}(s) & E_\mathrm{f}(s) & E_\mathrm{b}(s) & E_\mathrm{tc}(s)
	\end{bmatrix}^\top,
\end{equation}
where $v$ is the velocity of the car, $p_\mathrm{im}$ is the intake manifold pressure, $E_\mathrm{f}$ is the fuel energy, $E_\mathrm{b}$ is the battery energy and $E_\mathrm{tc}$ is the kinetic energy of the turbocharger.
The control input vector $\vec{u}$ is
\begin{equation}\label{eq:inputs}
	\begin{split}
		\vec{u}(s) = \big[
		u_\mathrm{th}(s) \quad &\dot{m}_\mathrm{f,cyl}(s) \quad P_\mathrm{k}(s) \quad P_\mathrm{h}(s)\quad \dots\\
		\dots \quad &u_\mathrm{wg}(s) \quad u_\mathrm{sa}(s) \quad P_\mathrm{brk}(s) \big]^\top,
	\end{split}
\end{equation} 
where $u_\mathrm{th}$ is the throttle position, $\dot{m}_\mathrm{f,cyl}$ is the cylinder fuel mass flow, $P_\mathrm{k}$ is the MGU-K power, $P_\mathrm{h}$ is the MGU-H power, $u_\mathrm{wg}$ is the waste-gate position, $u_\mathrm{sa}$ is the spark advance efficiency and $P_\mathrm{brk}$ is the brake power.
Additionally, also the engaged gear $\Gamma$ is an input to the plant, which is not controlled by the controller.
The nominal reference vector $\vec{r}^\mathrm{nom}$ will be introduced in Section \ref{sec:offlineoptimization}, whilst the actual reference vector $\vec{r}^\mathrm{act}$, the fed back controller information $\vec{u}_\mathrm{dr}$, and the requested power $P_\mathrm{req}$ will be defined in Section \ref{sec:driver}.
As we include track dependent parameters and optimize over the lap time, we write all equations as a function of the continuous path variable $s\in[0,\dots,S]$, where $S$ is the length of the track. 
This variable denotes the distance covered along the racing line on track.
\subsection{Internal Combustion Engine}\label{subsec:ice}
The heart of the power unit is the internal combustion engine.
The dynamics of the air path are essentially determined by the dynamics of the intake manifold pressure $p_\mathrm{im}$, namely
\begin{equation}\label{eq:pIM}
	\dds p_\mathrm{im}(s) =\frac{1}{v(s)}\cdot \frac{R_\mathrm{air}\cdot\vartheta_\mathrm{im}}{V_\mathrm{im}}\cdot\left(\dot{m}_\mathrm{c}(s)-\dot{m}_\beta(s)\right),
\end{equation}
where $R_\mathrm{air}$ is the specific gas constant of air, $\vartheta_\mathrm{im}$ is the assumed to be constant intake manifold temperature, $V_\mathrm{im}$ is the intake manifold volume, $\dot{m}_\mathrm{c}$ is the mass flow through the compressor and $\dot{m}_\beta$ is the air mass flow entering the cylinders. 
The division by $v$ stems from the fact that the spatial derivative is considered.
By approximating the engine as a volumetric pump, $\dot{m}_\beta$ can be modeled as
\begin{equation}\label{eq:mbeta}
	\dot{m}_{\beta}(s) = \frac{p_\mathrm{im}(s)\cdot V_\mathrm{d}}{R_\mathrm{air}\cdot\vartheta_\mathrm{im}}\cdot \frac{\omega_\mathrm{e}(s)}{4\pi} \cdot \lambda_\mathrm{vol}\big(\omega_\mathrm{e}(s)\big)\cdot u_\mathrm{th}(s),
\end{equation}
where $V_\mathrm{d}$ is the displacement volume, $\omega_\mathrm{e}$ is the engine speed, and $\lambda_\mathrm{vol}$ is the engine speed dependent volumetric efficiency.
By design, we can shut off single cylinders, i.e., no fuel injection and no ignition occurs.
The number of active cylinders is
\begin{equation}\label{eq:psidef}
	\Psi_{\mathrm{e}}(s)\in\{0,\dots,N_\mathrm{cyl}\}, 
\end{equation}
where $N_\mathrm{cyl}=6$ is the number of the \gls{acr:ICE}'s cylinders.
Only in those cylinders we inject the cylinder fuel mass flow $\dot{m}_\mathrm{f,cyl}$, that we assume to be equal for each active cylinder.
The normalized air-to-fuel ratio $\lambda_\mathrm{af}$, defined as
\begin{equation} \label{eq:lambda}
	\lambda_{\mathrm{af}}(s) = \frac{\dot{m}_\beta(s) / N_\mathrm{cyl}}{\dot{m}_\mathrm{f,cyl}(s)}\cdot\frac{1}{\sigma_0},
\end{equation}
with $\sigma_0$ as the stoichiometric constant, must lie in its bounds at all times
\begin{equation} \label{eq:lambdalimits}
	\lambda_\mathrm{af}^\mathrm{min} \leqslant \lambda_{\mathrm{af}}(s) \leqslant \lambda_\mathrm{af}^\mathrm{max},
\end{equation}
where $\lambda_\mathrm{af}^\mathrm{min}$ and $\lambda_\mathrm{af}^\mathrm{max}$ are operational bounds outside of which a proper combustion in a gasoline engine can no longer occur.
The \gls{acr:FIA} regulations constrain the control input $\dot{m}_\mathrm{f,cyl}$ with an engine speed dependent limit, i.e.,
\begin{equation}\label{eq:fialimit}
	0 \leqslant \dot{m}_\mathrm{f,cyl}(s) \leqslant \dot{m}_{\mathrm{f,cyl}}^\mathrm{max}\left(\omega_\mathrm{e}(s)\right) := \dot{m}_{\mathrm{f}}^\mathrm{max}\left(\omega_\mathrm{e}(s)\right)/N_\mathrm{cyl},
\end{equation}
where $\dot{m}_\mathrm{f}^\mathrm{max}\left(\omega_{\mathrm{e}}\right)$ is the speed dependent maximum fuel mass flow imposed:
\begin{equation}\label{eq:fiamax}
	\dot{m}_{\mathrm{f}}^\mathrm{max}\big(\omega_\mathrm{e}(s)\big) =
	\begin{cases}
		f_{\omega_\mathrm{e}}^\mathrm{FIA}\big(\omega_\mathrm{e}(s)\big) & \text{if } \omega_\mathrm{e}(s) \leqslant 10.5 \text{ krpm}, \\
		f^\mathrm{FIA}_{\mathrm{const}} & \text{if } \omega_\mathrm{e}(s) > 10.5 \text{ krpm},
	\end{cases}
\end{equation}
with $f_{\omega_\mathrm{e}}^\mathrm{FIA}$ being an affine function decreasing with engine speed, and $f^\mathrm{FIA}_{\mathrm{const}}$ a constant limit.
The resulting total fuel mass flow reads as   
\begin{equation}\label{eq:mf}
	\dot{m}_\mathrm{f}(s) = \dot{m}_\mathrm{f,cyl}(s)\cdot\Psi_{\mathrm{e}}(s).
\end{equation}
The amount of fuel that we inject gives us the fuel power $P_\mathrm{f}$ according to
\begin{equation}\label{eq:fuelpower}
	P_\mathrm{f}(s) = \dot{m}_\mathrm{f}(s)\cdot H_l,
\end{equation}
where $H_l$ is the fuel's lower heating value, which leads to the state equation of the fuel energy $E_\mathrm{f}$, i.e.,
\begin{equation}\label{eq:Ef}
	\dds E_\mathrm{f}(s) = \frac{1}{v(s)} \cdot P_\mathrm{f}(s).
\end{equation}
The engine power due to combustion is
\begin{equation}\label{eq:Pecomb}
	P_\mathrm{e,comb}(s) = \eta(\omega_{\mathrm{e}}(s),\lambda_\mathrm{af}(s)) \cdot u_\mathrm{sa}(s) \cdot P_\mathrm{f}(s),
\end{equation}
where $\eta$ is a lumped efficiency depending on various engine quantities, and $u_\mathrm{sa}$ is the efficiency due to the spark advance angle retardation.
Finally, we obtain the total engine power according to
\begin{equation}\label{eq:Pe}
	P_\mathrm{e}(s) = P_\mathrm{e,comb}(s) + P_\mathrm{e,fp}(s), 
\end{equation}
where $P_\mathrm{e,fp}$ contains the friction and pumping powers, derived in \cite{Balerna}.
\subsection{Turbocharger}\label{subsec:tc}
The turbocharger of the F1 vehicle consists of a turbine, a compressor and an electric motor (MGU-H) mounted on its shaft.
The state equation according to which the rotational kinetic energy $E_\mathrm{tc}$ evolves reads as
\begin{equation}\label{eq:Etc}
	\dds E_\mathrm{tc}(s) = \frac{1}{v(s)}\cdot\left(P_\mathrm{t}(s)-P_\mathrm{c}(s)+P_\mathrm{h}(s)\right),
\end{equation}
where $P_\mathrm{t}$ is the turbine power, $P_\mathrm{c}$ is the compressor power, and $P_\mathrm{h}$ is the MGU-H control input power.
The turbine and compressor powers and mass flows result from experimental turbocharger maps denoted by $\mathcal{M}$, i.e.,
\begin{equation}\label{eq:TCmaps}
	\begin{split}
	\dot{m}_\mathrm{c}(s) &= \mathcal{M}_{ \dot{m}_\mathrm{c}}\big(E_\mathrm{tc}(s),p_\mathrm{im}(s)\big), \\
	P_\mathrm{c}(s) &= \mathcal{M}_{ P_\mathrm{c}}\big(E_\mathrm{tc}(s),p_\mathrm{im}(s)\big), \\
	\dot{m}_\mathrm{t}(s) &= \mathcal{M}_{ \dot{m}_\mathrm{t}}\big(E_\mathrm{tc}(s),p_\mathrm{em}(s),\vartheta_\mathrm{em}(s)\big), \\
	P_\mathrm{t}(s) &= \mathcal{M}_{ P_\mathrm{t}}\big(E_\mathrm{tc}(s),p_\mathrm{em}(s),\vartheta_\mathrm{em}(s)\big),
	\end{split}
\end{equation}
where $\dot{m}_\mathrm{t}$ is the mass flow through the turbine, and $p_\mathrm{em}$ and $\vartheta_\mathrm{em}$ are the exhaust manifold pressure and temperature, respectively.
As shown in Fig. \ref{fig:Powertrain}, the turbine can be bypassed by the waste-gate, actuated by $u_\mathrm{wg}\in[0,1]$, with $u_\mathrm{wg}=0$ representing the closed position and $u_\mathrm{wg}=1$ the opened one. 
\subsection{\gls{acr:ers}}\label{subsec:ers}
The \gls{acr:ers} is composed of the MGU-K, the MGU-H and the battery.
The battery energy $E_\mathrm{b}$ evolves according to
\begin{equation}\label{eq:Eb}
	\dds E_\mathrm{b} = \frac{1}{v(s)}\cdot f_\mathrm{b}\left(P_\mathrm{k}(s), P_\mathrm{h}(s)\right),
\end{equation}
where $f_\mathrm{b}$ accounts for all mechanical-to-electrical and electrical-to-mechanical losses.
\subsection{Vehicle Dynamics}\label{subsec:vehicledynamics}
To increase the power of the \gls{acr:ICE} or recuperate kinetic energy, we use the MGU-K, which is mounted on the crankshaft of the engine.
In combination with the braking power $P_\mathrm{brk}$ generated by the friction brakes we obtain the final traction power
\begin{equation}\label{eq:tracpower}
	P_{\mathrm{trac}}(s) = f_\mathrm{trac}\left(P_\mathrm{e}(s),P_\mathrm{k}(s),P_\mathrm{brk}(s)\right),
\end{equation}
where $f_\mathrm{trac}$ considers all friction losses and the tires' slip, and $P_\mathrm{k}$ is the MGU-K power.
The longitudinal dynamics of the car evolve according to the state equation
\begin{equation}\label{eq:eqdiffvelocity}
	m_\mathrm{car}\dds v(s) = \frac{1}{v(s)}\left(P_\mathrm{trac} - P_\mathrm{res}(v(s),s)\right),
\end{equation}
where $m_\mathrm{car}$ is the mass of the car, which is assumed to be constant, and $P_\mathrm{res}$ is a lumped power considering all external powers opposing the motion (e.g., aerodynamic drag and rolling resistance).
To account for the lateral dynamics, we integrate the complex interactions between tire and road into a track dependent maximum velocity $v_\mathrm{max}$ profile shown in the upper plot of Fig. \ref{fig:v_max}~\cite{ebbesen2018time}.
The resulting constraint reads as 
\begin{figure}
	\centering
	\begin{tikzpicture}[trim axis left, trim axis right, >=stealth, font=\small]

\def\plotwidth{0.9\columnwidth}%
\def\plotheight{0.4\columnwidth}%
\def\yshift{-0.02cm}%

\begin{axis}[%
name=one,
width=\plotwidth,
height=\plotheight,
xmin=800,
xmax=2200,
xlabel style={font=\color{white!15!black}},
xticklabels={},
ymin=60,
ymax=415,
ylabel style={font=\color{white!15!black}},
ylabel style={at={(0.04,0.5)}},
ylabel={$v$ [kph]},
axis background/.style={fill=white},
xmajorgrids,
ymajorgrids,
legend style={at={(0.1,0.35)}, anchor=north west, legend cell align=left, align=left, draw=white!15!black}
]
\addplot [color=black, line width=1.0pt, forget plot]
  table[]{./pictures/Chapter2/picData/v_max-1.tsv};

\addplot [color=black, dotted, line width=1.0pt]
  table[]{./pictures/Chapter2/picData/v_max-2.tsv};
\addlegendentry{$v_\mathrm{max}$}

\addplot[area legend, draw=none, fill=white!65!black, fill opacity=0.4, forget plot]
table[] {./pictures/Chapter2/picData/v_max-3.tsv}--cycle;

\addplot[area legend, draw=none, fill=white!65!black, fill opacity=0.4, forget plot]
table[] {./pictures/Chapter2/picData/v_max-4.tsv}--cycle;

\addplot[area legend, draw=none, fill=white!65!black, fill opacity=0.4, forget plot]
table[] {./pictures/Chapter2/picData/v_max-5.tsv}--cycle;

\addplot[area legend, draw=none, fill=white!65!black, fill opacity=0.4, forget plot]
table[] {./pictures/Chapter2/picData/v_max-6.tsv}--cycle;

\addplot[area legend, draw=none, fill=white!65!black, fill opacity=0.4, forget plot]
table[] {./pictures/Chapter2/picData/v_max-7.tsv}--cycle;
\end{axis}

\begin{axis}[%
	name=two,
	width=\plotwidth,
	height=\plotheight,
	at=(one.below south west),
	yshift=\yshift,
	anchor=north west,
	xmin=800,
	xmax=2200,
	xlabel style={font=\color{white!15!black}},
	xlabel={Position [m]},
	xticklabels={, , 1000, , 1400, , 1800, , 2200},
	ymin=-1.77,
	ymax=0.06,
	ylabel style={font=\color{white!15!black}},
	ylabel style={at={(0.041,0.5)}},
	ylabel={$\lambda_\mathrm{kin}$ [-]},
	axis background/.style={fill=white},
	xmajorgrids,
	ymajorgrids,
	legend style={legend cell align=left, align=left, draw=white!15!black}
	]
	\addplot [color=black, line width=1.0pt]
	table[]{./pictures/Chapter2/picData/lambda_kin-1.tsv};
	
	\addplot[area legend, draw=none, fill=white!65!black, fill opacity=0.4, forget plot]
	table[] {./pictures/Chapter2/picData/lambda_kin-2.tsv}--cycle;
	
	\addplot[area legend, draw=none, fill=white!65!black, fill opacity=0.4, forget plot]
	table[] {./pictures/Chapter2/picData/lambda_kin-3.tsv}--cycle;
	
	\addplot[area legend, draw=none, fill=white!65!black, fill opacity=0.4, forget plot]
	table[] {./pictures/Chapter2/picData/lambda_kin-4.tsv}--cycle;
	
	\addplot[area legend, draw=none, fill=white!65!black, fill opacity=0.4, forget plot]
	table[] {./pictures/Chapter2/picData/lambda_kin-5.tsv}--cycle;
	
	\addplot[area legend, draw=none, fill=white!65!black, fill opacity=0.4, forget plot]
	table[] {./pictures/Chapter2/picData/lambda_kin-6.tsv}--cycle;
\end{axis}

\end{tikzpicture}%
	\caption{Nominal and maximal velocity trajectory, as well as the kinetic costate $\lambda_{\mathrm{kin}}$ over a portion of lap. The gray-patched intervals represent the grip-limited regions, where the maximum velocity constraint is active.}
	\label{fig:v_max}\vspace{-0.3cm}
\end{figure}
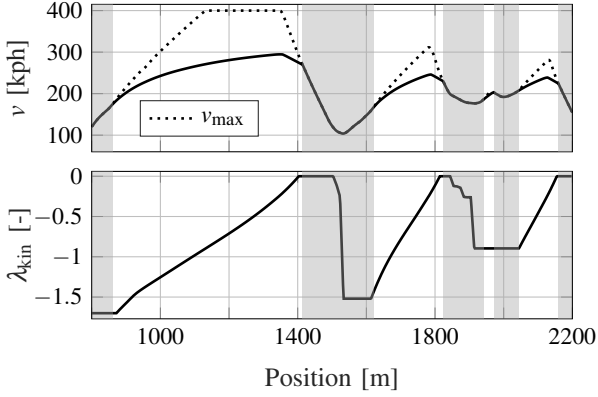
\begin{equation}\label{eq:vmaxconstraint}
	v(s) \leqslant v_\mathrm{max}(s), \quad \forall s.
\end{equation}
The maximum velocity profile is obtained by considering the throttle pedal position of the telemetry data acquired during a representative lap.
At those indices where the value is below \SI{100}{\percent}, the driver is in a corner and we assume that the vehicle is at the limit of the tire's grip, implying that the telemetry velocity represents the maximum velocity achievable.
Starting from those grip-limited regions, the vehicle dynamics are simulated backwards and forwards subject to tire adhesion limits only.
An alternative was presented in \cite{9769925}, where the tire adhesion is described through velocity dependent maximal forces.
However, to keep the simulation algorithms simple, we opted for the description by means of maximum velocity.
The link between the velocity of the car and the engine speed is
\begin{equation}\label{eq:vtoomega}
	\omega_\mathrm{e}(s) = \left(\frac{\Gamma(s)\cdot\Gamma_{\mathrm{d}}}{r_\mathrm{w}}\right)\cdot v(s) = \tilde{\Gamma}(s) \cdot v(s),
\end{equation}
where the overall transmission ratio $\tilde{\Gamma}$ includes the gear ratio input $\Gamma \in \{1,\dots,8\}$, the constant differential transmission ratio $\Gamma_\mathrm{d}$, and the wheel radius $r_\mathrm{w}$.
This conversion needs to account for the engine speed limits
\begin{equation} \label{eq:omegaenginelimits}
	\omega_\mathrm{e}^\mathrm{min} \leqslant \omega_{\mathrm{e}}(s) \leqslant \omega_\mathrm{e}^\mathrm{max},
\end{equation}
where $\omega_\mathrm{e}^\mathrm{min}$ and $\omega_\mathrm{e}^\mathrm{max}$ are chosen for mechanical and regulatory reasons \cite{FIA2021Technical}.
\section{Offline Optimization}\label{sec:offlineoptimization}
The objective of this study is to control the F1 powertrain in an optimal way to minimize the lap time.
To obtain the reference trajectories we solve the lap time \gls{acr:ocp} that was presented in Appendix 1 of \cite{Balerna}, which was formulated for the above mentioned plant.
In our paper, we limit the definition of the \gls{acr:ocp} to the constraints and model equations introduced in Section \ref{sec:model}.
\begin{problem}\label{prob:ocpBalerna}
	The lap-time-optimal low-level control strategy satisfying the fuel and battery targets is the solution of
	\begin{equation*}
		\min_{\vec{u}} \int_0^{S} \frac{\ds}{v(s)},
	\end{equation*}
	subject to the following constraints:
	\begin{alignat*}{2}
		& \text{ICE:} \quad && \eqref{eq:pIM} - \eqref{eq:Pe}, \eqref{eq:omegaenginelimits}\\
		& \text{Turbocharger:} \quad && \eqref{eq:Etc}, \eqref{eq:TCmaps}, \\
		& \text{ERS:} \quad && \eqref{eq:Eb}, \\
		& \text{Vehicle Dynamics:} \quad && \eqref{eq:tracpower}, \eqref{eq:eqdiffvelocity}, \eqref{eq:vmaxconstraint}, \eqref{eq:vtoomega}, \\
		& \text{Fuel Target:} \quad && E_\mathrm{f}(S) \leqslant E_\mathrm{f}(0) + \Delta E_\mathrm{f,target}, \\
		& \text{Battery Target:} \quad && E_\mathrm{b}(S) \geqslant E_\mathrm{b}(0) + \Delta E_\mathrm{b,target},
	\end{alignat*}
	where $\Delta E_\mathrm{f,target}$ and $\Delta E_\mathrm{b,target}$ are design parameters determined by the race strategy representing the desired fuel and battery targets, respectively.
\end{problem}
\Cref{prob:ocpBalerna} contains two integer variables, resulting in an undesired mixed-integer problem: the engaged gear and the number of active cylinders. 
To remove the first one, the model is convexified with respect to the gear selection using the outer convexification method \cite{Kirches2011,Sager2005}.
Next, the result is rounded applying a rounding strategy that does not violate the SOS-1 property according to the methodology proposed in \cite{sager2007solving}.
A detailed explanation of this adaptation can be found in Section 3.2 of \cite{Balerna}.
The number of active cylinders, however, is treated as continuous variable for the task of reference trajectory generation.
To compute the benchmark solution used to assess the suboptimality of the online controller, we round the continuous solution and reoptimize the \gls{acr:ocp} with given integers.
The \gls{acr:ocp} is implemented in \textsc{Matlab} using the symbolic framework CasADi \cite{Casadi} and discretized into a \gls{acr:nlp} using the multiple shooting method and Euler forward integration.
The optimization problem is then solved with the interior point optimizer IPOPT.
Computation times for a lap interval of \SI{1000}{m} are up to 20 minutes while for a complete lap they reach 4 hours.
As it will be shown in Section \ref{sec:controller}, various inputs and states resulting from the non-causal solution of \Cref{prob:ocpBalerna} with a step size of $\Delta s= \SI{2}{m}$ are needed in our controller architecture.
In particular, we need the gear trajectory for the estimator, while we use the states as reference trajectories for the high-level (\gls{acr:eltms}) and the low-level (\gls{acr:nmpc}) controllers.
Therefore, to comply with the schematic in Fig. \ref{fig:Overview} we define the nominal reference vector $\vec{r}^\mathrm{nom}$ as
\begin{equation}\label{eq:rnom}
	\begin{split}
		\vec{r}^\mathrm{nom}(s) = \big[
		v^\mathrm{nom}_\mathrm{max}(s) \quad \Gamma^\mathrm{nom}(s) \quad &p^\mathrm{nom}_\mathrm{im}(s) \quad E^\mathrm{nom}_\mathrm{f}(s) \quad\dots\\ \dots\quad &E^\mathrm{nom}_\mathrm{b}(s) \quad E^\mathrm{nom}_\mathrm{tc}(s) \big]^\top,
	\end{split}
\end{equation}
where all trajectories are not guaranteed to be globally optimal because of the nonlinear problem statement.
Finally, the solution of \Cref{prob:ocpBalerna} is recomputed with the energy consumption achieved by the online controller and known disturbances to obtain the benchmark solution shown in Section \ref{sec:results}.
\subsection{Notation}\label{subsec:notation}
\begin{figure}
	\centering
	\input{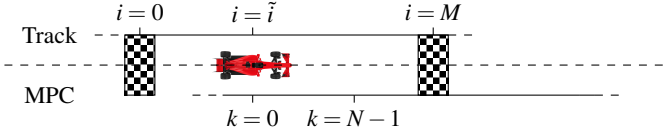}
	\caption{Graphical representation of the discrete reference frames used in this article. Denoted by $i$ we define the frame that is unequivocally linked to the track, while by $k$ we denote the moving \gls{acr:mpc} frame.}
	\label{fig:FrameNotation}
\end{figure}

In the following, we consider two discrete space reference frames shown in Fig. \ref{fig:FrameNotation}.
The track reference frame with the index $i\in\{0,\dots,M\}$, where $i=M$ at the finish line, represents the discretization of the continuous path variable $s$ with a step size of $\Delta s=\SI{2}{m}$.
Further, in the model-predictive control setting, we use the reference frame with index $k\in\{0,\dots,N-1\}$ indicating the step inside the optimization horizon of length $N$.
To link the two frames and be able to compute the position on track, we introduce $i=\tilde{i}$, being the index in the track frame at which the optimization begins (i.e., $k=0$).
This instant is also called the \textit{current step}.
\section{Driver Model}\label{sec:driver}
The real driver is a trained professional always acting to be as fast as possible by exploiting the full grip limits of the tires in corners, while requesting full power on straights.
Since in this paper we focus on the powertrain operation, we assume the steering behavior to be optimal and the ideal racing line to be followed at all times.
\begin{figure}
	\centering
	\tikzstyle{block} = [draw, rectangle, minimum height=2em, minimum width=3em]
\tikzstyle{input} = [coordinate]
\tikzstyle{output} = [coordinate]

\begin{tikzpicture}[scale=1, every node/.append style={outer sep=0pt}, >=stealth, font=\small] %
\def\larghezza{1.6cm}
\def\altezza{4cm}
\def\dist{2cm}
\def\dCirc{0.2cm}
\def\vertdistIN{0.2*\altezza}
\def\vertdistOUT{0.25*\altezza}

\node [block, minimum height=\altezza, minimum width=\larghezza, text centered, text width=\larghezza] at (0,0) (driver) {Driver Model};

\node[input] at ($(driver.west) + (-\dist, {2*\vertdistIN})$) (GL) {};
\node[circle, minimum size=\dCirc, fill = black, inner sep=0cm, label=above right:${b_\mathrm{GL}[i-1]}$] at (GL) {};

\node[input] at ($(driver.west) + (-\dist, \vertdistIN)$) (Ptrac) {};
\node[circle, minimum size=\dCirc, fill = black, inner sep=0cm, label=above right:${P_\mathrm{trac}^\mathrm{NMPC}[i-1]}$] at (Ptrac) {};

\node[input] at ($(driver.west) + (-\dist, 0)$) (vmax) {};
\node[circle, minimum size=\dCirc, fill = black, inner sep=0cm, label=above right:${v_{\mathrm{max}}^{\mathrm{act}}[i+1]}$] at (vmax) {};

\node[input] at ($(driver.west) + (-\dist, -\vertdistIN)$) (v) {};
\node[circle, minimum size=\dCirc, fill = black, inner sep=0cm, label=above right:${v[i]}$] at (v) {};

\node[input] at ($(driver.west) + (-\dist, -{2*\vertdistIN})$) (gamma) {};
\node[circle, minimum size=\dCirc, fill = black, inner sep=0cm, label=above right:${\Gamma^\mathrm{act}[i]}$] at (gamma) {};

\node[output] at ($(driver.east) + (\dist, 0)$) (GLout) {};

\node[output] at ($(driver.east) + (\dist, \vertdistOUT)$) (Preqout) {};

\node[output] at ($(driver.east) + (\dist, -\vertdistOUT)$) (gammaout) {};

\draw[->] (vmax) to (driver);
\draw[->] (Ptrac) to ($(driver.west) + (0,\vertdistIN)$);
\draw[->] (GL) to ($(driver.west) + (0,2*\vertdistIN)$);
\draw[->] (v) to ($(driver.west) + (0,-\vertdistIN)$);
\draw[->] (gamma) to ($(driver.west) + (0,-2*\vertdistIN)$);

\draw[->] (driver.east) to node[above] {$b_\mathrm{GL}[i]$} (GLout);
\draw[->] ($(driver.east) + (0,\vertdistOUT)$) to node[above] {$P_{\mathrm{req}}[i]$} (Preqout);
\draw[->] ($(driver.east) + (0,-\vertdistOUT)$) to node[above] {$\Gamma[i]$} (gammaout);

\end{tikzpicture}
	\caption{Inputs and outputs of the driver subsystem. The index $i$ indicates the discrete online space step of the signal. }
	\label{fig:DriverIO}\vspace{-0.3cm}
\end{figure}
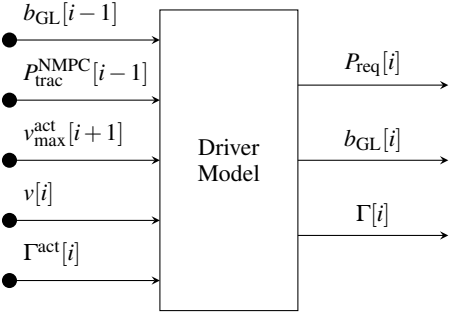
In particular, the relevant human decisions are the power request $P_\mathrm{req}$, set as input for the controller, and the engaged gear $\Gamma$, which is known by the controller and set as input for the plant.
In order to understand the behavior of the driver, we recall the two distinct regions introduced in Section \ref{sec:introduction}.
In power-limited regions, the requested power can be overwritten by the \gls{acr:ecu}, translating to
\begin{equation}\label{eq:PLreq}
		P_{\mathrm{trac}}(s) \leqslant P_\mathrm{req}(s).
\end{equation}
In grip-limited regions, however, the velocity of the car matches the maximum velocity profile, i.e., \eqref{eq:vmaxconstraint} holds with equality. 
The fulfillment of the power requested by the driver reads as
\begin{equation}\label{eq:GLreq}
	P_{\mathrm{trac}}(s) = P_{\mathrm{req}}(s).
\end{equation}
To distinguish between those two different track regions, we introduce the binary variable $b_\mathrm{GL}$, which is set to 0 in power-limited regions and to 1 in grip-limited regions.
\begin{figure}
	\centering
	\tikzstyle{block} = [draw, rectangle, minimum height=2em, minimum width=3em]
\tikzstyle{input} = [coordinate]
\tikzstyle{output} = [coordinate]

\begin{tikzpicture}[scale=1, every node/.append style={outer sep=0pt}, >=stealth, font=\small] %
\def\deltaHorz{1em}
\def\distVert{0.9cm}
\def\deltaLeft{2cm}
\def\heightOneLine{3em}
\def\heightMultiLine{5em}

\node [block, minimum height=\heightOneLine, minimum width=7em] at (0,15) (firstQuestion) {$b_\mathrm{GL}[i-1]= 1$?};

\node [block, minimum height=\heightMultiLine, minimum width=10em, text width=10em] at ($(firstQuestion.south) + (-\deltaLeft, -\distVert-\heightMultiLine/2)$) (vYes) {Compute $v[i+1]$ with the maximum achievable traction power $P_\mathrm{trac}^{\mathrm{max}}$};
\node [block, minimum height=\heightMultiLine, minimum width=10em, text width=10em] at ($(firstQuestion.south) + (\deltaLeft, -\distVert-\heightMultiLine/2)$) (vNo) {Compute $v[i+1]$ with the last delivered traction power $P^\mathrm{NMPC}_{\mathrm{trac}}[i-1]$};

\node [block, minimum height=\heightOneLine, minimum width=9em] at ($(vYes.south) + (\deltaLeft, -\distVert-\heightOneLine/2)$) (secondQuestion) {$v[i+1] \geqslant v_{\mathrm{max}}^{\mathrm{act}}[i+1]$?};

\node [block, minimum height=\heightMultiLine, minimum width=10em, text width = 11em] 
at ($(secondQuestion.south) + (-1.1*\deltaLeft, -\distVert-\heightMultiLine/2)$) (GLyes)
{\vspace{-0.5em} \begin{itemize} \item GL region \\ $\rightarrow$ set $b_\mathrm{GL}[i] = 1$ \item Compute $P_{\mathrm{req}}[i]$ such \\ that $v[i+1] = v_{\mathrm{max}}^{\mathrm{act}}[i+1]$\end{itemize}};
 
\node [block, minimum height=\heightMultiLine, minimum width=10em, text width = 11em]
at ($(secondQuestion.south) + (1.1*\deltaLeft, -\distVert-\heightMultiLine/2)$) (PLno) 
{\vspace{-0.5em}\begin{itemize} \item PL region \\ $\rightarrow$ set $b_\mathrm{GL}[i] = 0$ \item Set $P_{\mathrm{req}}[i] = P_{\mathrm{trac}}^{\mathrm{max}}$\end{itemize}};

\node [output] at ($(firstQuestion.south) + (-\deltaHorz,0)$) (1outyes) {};
\node [output] at ($(firstQuestion.south) + (\deltaHorz,0)$) (1outno) {};

\node [input] at ($(vYes.north)$) (yes1in) {};
\node [output] at ($(vYes.south)$) (yes1out) {};

\node [input] at ($(vNo.north)$) (no1in) {};
\node [output] at ($(vNo.south)$) (no1out) {};

\node [input] at ($(secondQuestion.north) + (-\deltaHorz,0)$) (2inyes) {};
\node [input] at ($(secondQuestion.north) + (\deltaHorz,0)$) (2inno) {};
\node [output] at ($(secondQuestion.south) + (-\deltaHorz,0)$) (2outyes) {};
\node [output] at ($(secondQuestion.south) + (\deltaHorz,0)$) (2outno) {};

\node [input] at ($(GLyes.north)$) (yes2in) {};

\node [input] at ($(PLno.north)$) (no2in) {};

\draw [->] (1outyes) to node[left] {Yes} ($(yes1in) + (2*\deltaHorz, 0)$);
\draw [->] (1outno) to node[right] {No} ($(no1in) + (-2*\deltaHorz, 0)$);
\draw [->] (yes1out) -- ($(2inyes) + (-2*\deltaHorz, 0)$);
\draw [->] (no1out) -- ($(2inno) + (2*\deltaHorz, 0)$);
\draw [->] (2outyes) to node[left] {Yes} ($(yes2in) + (2*\deltaHorz, 0)$);
\draw [->] (2outno) to node[right] {No} ($(no2in) + (-2*\deltaHorz, 0)$);

\end{tikzpicture}
	\caption{Driver heuristics shown as flowchart. The index $i$ indicates the discrete online space step of the signal.}
	\label{fig:Driver_Flowchart}\vspace{-0.3cm}
\end{figure}
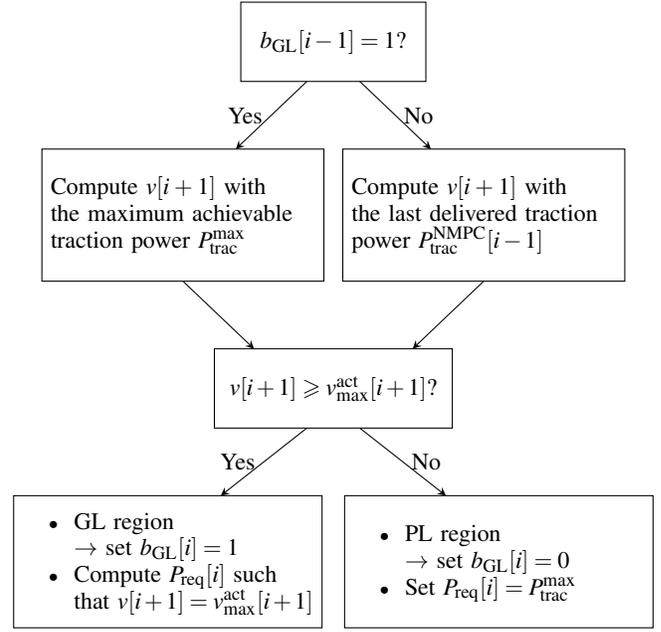

In the present paper, we model the driver with a decision making component with input/output relationship shown in Fig. \ref{fig:DriverIO}.
We need it in our framework to simulate a causal behavior of the system under given disturbances.
In fact, we provide the driver model with actual reference trajectories $\vec{r}^\mathrm{act}$ that are unknown to the controller, i.e.,
\begin{equation}\label{eq:ract}
	\vec{r}^\mathrm{act}(s) = \begin{bmatrix} v^\mathrm{act}_\mathrm{max}(s) & \Gamma^\mathrm{act}(s) \end{bmatrix}^\top.
\end{equation}
Additionally, we provide the driver model with the velocity $v$ of the car, and two internal variables of the controller lumped into the vector $\vec{u}_\mathrm{dr}$ defined as
\begin{equation} \label{eq:udr}
	\vec{u}_\mathrm{dr}(s) = \begin{bmatrix}
		b_\mathrm{GL}(s) & P_\mathrm{trac}^{\mathrm{NMPC}}(s) \end{bmatrix}^\top,
\end{equation}
with $P_\mathrm{trac}^{\mathrm{NMPC}}$ being the traction power optimized by the \gls{acr:nmpc}, representing the driver perception.
Fig. \ref{fig:Driver_Flowchart} illustrates the decision-making inside the driver model, adapted from \cite{salazar2017time}.
In a first step, we need to check whether we were in a grip-limited region at the previous space index $i-1$.
Depending on the outcome of this assessment, we determine the traction power with which we compute the velocity at space index $i+1$ by means of \eqref{eq:eqdiffvelocity}.
If we come from a grip-limited region, we use the maximum traction power achievable by the power unit $P_\mathrm{trac}^{\mathrm{max}}$.
By setting such a high value we account for the eventuality that we are exiting the grip-limited region.
If we come from a power-limited region, however, we use the previously delivered traction power $P_{\mathrm{trac}}^{\mathrm{NMPC}}[i-1]$.
Since in those regions the \gls{acr:ecu} can overwrite the power request of the driver, its current decisions need to be based on that effectively perceived traction power and not prior requests.
The resulting velocity $v[i+1]$ is compared to the actual maximum velocity achievable on the track $v_\mathrm{max}^{\mathrm{act}}[i+1]$, in order to determine whether at the current step $i$ the vehicle is in grip- or power-limited region.
In the former case, we need to compute the power request such that \eqref{eq:vmaxconstraint} holds with equality at step $i+1$.
Otherwise, we set the requested power to the maximum traction power achievable.
As previously mentioned, the \gls{acr:ecu} can overwrite the request, and the power output of the driver model does not influence the optimized traction power determined by the controller.

The last output of the driver is the engaged gear $\Gamma[i]$.
We determine it by considering the vehicle's velocity and an actual reference gear $\Gamma^\mathrm{act}[i]$ that we assume to be followed as indicated below:
\begin{algorithm}
	\begin{algorithmic}
		\STATE $\omega_{\mathrm{e}}[i] = f(v[i], \Gamma^\mathrm{act}[i])$ according to \eqref{eq:vtoomega}
		\IF{\eqref{eq:omegaenginelimits} satisfied} \STATE $\Gamma[i] = \Gamma^\mathrm{act}[i]$ \ELSE \STATE Driver will engage  lower/higher gear if the minimum/maximum bound is violated \ENDIF
	\end{algorithmic}
	\caption{Driver's Gear Choice}
	\label{alg:driver}
\end{algorithm}

\noindent In a disturbance-free scenario, the actual reference gear is equal to the nominal reference gear $\Gamma^\mathrm{nom}[i]$ known by the controller, whereas otherwise (as analyzed in Section \ref{subsec:differingGS}) they differ.
\section{Controller}\label{sec:controller}
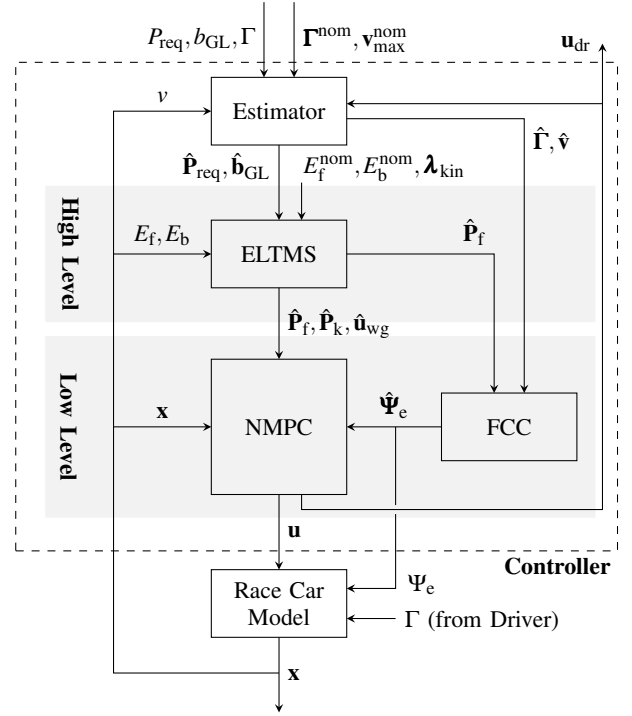
\begin{figure}
	\centering
	\tikzstyle{block} = [draw, rectangle, minimum height=2em, minimum width=3em]
\tikzstyle{input} = [coordinate]
\tikzstyle{output} = [coordinate]

\usetikzlibrary{intersections}

\begin{tikzpicture}[scale=1, every node/.append style={outer sep=0pt}, >=stealth, font=\small] %

\def\altezza{0.9cm}
\def\larghezza{1.8cm}
\def\horzDist{1.9cm}
\def\vertDist{0.8cm}
\def\radius{0.6mm}

\node [block, minimum height=\altezza, minimum width=\larghezza] at (0,0) (estimator) {Estimator};

\node [block, dashed, minimum height=6.5cm, minimum width=8cm, anchor=north west] at ($(estimator.north) + (-3.5cm,0.2cm)$) (dashedhigh) {};
\node [block, fill=black!5, black!5, minimum height=1.8cm, minimum width=7.3cm, anchor=south west] at ($(estimator) + (-3.1cm,-2.8cm)$) (highlevel) {};
\node [block, fill=black!5, black!5, minimum height=2.4cm, minimum width=7.3cm, anchor=south west] at ($(estimator) + (-3.1cm,-5.4cm)$) (lowlevel) {};
\node[rotate=-90] at ($(highlevel.west) + (0.3,0)$) {\textbf{High Level}};
\node[rotate=-90] at ($(lowlevel.west) + (0.3,0)$) {\textbf{Low Level}};
\node[] at ($(dashedhigh.south) + (3.2,-0.2)$) {\textbf{Controller}};

\node [block, minimum height=\altezza, minimum width=\larghezza, anchor=south] at ($(estimator.south) + (0,-\horzDist)$) (eltms) {ELTMS};
\node [block, minimum height=2*\altezza, minimum width=\larghezza, anchor=north] at ($(eltms.south) + (0,-0.5*\horzDist)$) (nmpc) {NMPC};
\node [block, minimum height=\altezza, minimum width=\larghezza] at ($(nmpc.east) + (2.7*\vertDist,0)$) (fcc) {FCC};
\node [block, minimum height=\altezza, minimum width=\larghezza, text centered, text width=1.4cm, anchor=south] at ($(nmpc.south) + (0,-\horzDist)$) (plant) {Race Car Model};

\node[coordinate] at ($(estimator.west) + (0,0)$) (estimatorSinistraSopra) {};
\node[coordinate] at ($(estimator.north) + (-0.2,0)$) (estimatorSopraSinistra) {};
\node[coordinate] at ($(estimator.north) + (0.2,0)$) (estimatorSopraDestra) {};
\node[coordinate] at ($(fcc.north) + (0.2,0)$) (fccSopraDestra) {};
\node[coordinate] at ($(fcc.north) + (-0.2,0)$) (fccSopraSinistra) {};
\node[coordinate] at ($(nmpc.south) + (0.3cm,0)$) (nmpcSottoDestra) {};

\draw [->] ($(estimatorSopraSinistra) + (0,1cm)$) to node[left] {$P_\mathrm{req},b_\mathrm{GL},\Gamma$} (estimatorSopraSinistra);
\draw [->] ($(estimatorSopraDestra) + (0,1cm)$) to node[right] {$\bm{\Gamma}^\mathrm{nom},\vec{v}_\mathrm{max}^{\mathrm{nom}}$} (estimatorSopraDestra);
\draw [->] (plant.south) to node[right] {$\vec{x}$} ($(plant.south) + (0,-1cm)$);
\draw [->] (estimator) -- (eltms);
\draw [->] ($(eltms.north)+(0.3cm,0.5cm)$) -- ($(eltms.north)+(0.3cm,0cm)$);
\draw [->] (eltms) to node[right] {$\hat{\vec{P}}_\mathrm{f},\hat{\vec{P}}_\mathrm{k},\hat{\vec{u}}_\mathrm{wg}$} (nmpc);
\draw [->] (nmpc) to node[right] {$\vec{u}$} (plant);
\draw [->] (fcc) to node[above] {$\hat{\bm{\Psi}}_\mathrm{e}$} (nmpc);
\path [name path=line0] ($(nmpc.east) + (0.65cm,0)$) |- ($(plant.east) + (0,0.2cm)$);
\node[] at ($(nmpc.east) + (1cm,-2.1cm)$) {${\Psi}_\mathrm{e}$};
\draw [->] ($(plant.east) + (0.65cm,-0.2cm)$) -- ($(plant.east) + (0,-0.2cm)$);
\node[] at ($(plant.east) + (1.8cm,-0.23cm)$) {$\Gamma$ (from Driver)};
\draw [->] ($(estimator.east) + (0,-0.1)$) -| node[below right] {$\hat{\bm{\Gamma}},\hat{\vec{v}}$} (fccSopraDestra);
\draw [->] (eltms.east) -| node[above left] {$\hat{\vec{P}}_\mathrm{f}$} (fccSopraSinistra);
\draw [->, name path=baseline2] ($(plant.south) + (0,-0.25*\horzDist)$) |- ($(plant.south) + (-0.5*\larghezza-1.3cm,-0.25*\horzDist)$) -- ($(estimatorSinistraSopra) + (-1.3cm,0)$) to node[above] {$v$} (estimatorSinistraSopra);
\draw [->] ($(eltms.west) + (-1.3cm,0)$) to node[above] {$E_\mathrm{f},E_\mathrm{b}$} (eltms.west);
\draw [->] ($(nmpc.west) + (-1.3cm,0)$) to node[above] {$\vec{x}$} (nmpc.west);
\draw [->, name path=baseline0] (nmpcSottoDestra) |- ($(nmpcSottoDestra) + (4cm,-0.2cm)$) -- ($(nmpcSottoDestra) + (4cm,6cm)$);
\draw [->] ($(estimator.east) + (3.4cm,0.1)$) -- ($(estimator.east) + (0,0.1)$);
\node[] at ($(nmpcSottoDestra) + (3.65cm,6cm)$) {$\vec{u}_\mathrm{dr}$};
\node[] at ($(estimator.south) + (1.4,-0.3)$) {$E_{\mathrm{f}}^{\mathrm{nom}},E_{\mathrm{b}}^{\mathrm{nom}},\bm{\lambda}_{\mathrm{kin}}$};
\node[] at ($(estimator.south) + (-0.7,-0.3)$) {$\hat{\vec{P}}_\mathrm{req},\hat{\vec{b}}_\mathrm{GL}$};

\path [name intersections={of = line0 and baseline0}];
\coordinate (S3)  at (intersection-1);
\path[name path=circle3] (S3) circle(\radius);
\path [name intersections={of = circle3 and line0}];
\coordinate (I31)  at (intersection-1);
\coordinate (I32)  at (intersection-2);
\draw ($(nmpc.east) + (0.65cm,0)$) -- (I31);
\draw[->] (I32) |- ($(plant.east) + (0,0.2cm)$);

\end{tikzpicture}
	\caption{Detailed representation of the internal blocks of the controller shown in Fig. \ref{fig:Overview}. The signals entering the controller at the top represent the driver model outputs (left arrow), the nominal gear trajectory obtained offline by solving Problem \ref{prob:ocpBalerna} and the nominal maximal velocity trajectory (right arrow). Exiting the controller there are the control inputs as defined in \eqref{eq:inputs}, the signals that are fed back to the driver as defined in \eqref{eq:udr}, and the number of active cylinders ${\Psi}_\mathrm{e}$, stored in the first entry of its prediction $\hat{\bm{\Psi}}_\mathrm{e}$. The state vector is fed back completely to the NMPC, while only the energies and the velocity are fed back to the ELTMS and the estimator, respectively.}
	\label{fig:ControllerInside}\vspace{-0.5cm}
\end{figure}

In this section we present the online control framework.
Fig. \ref{fig:ControllerInside} shows an overview of the components and their interaction.
Given that we employ a nonlinear model predictive controller for the low-level actuation, we need an estimator to provide predictive information such as track properties and the driver's possible future intentions. 
Also, we include a high-level supervisory controller, in the form of \gls{acr:eltms}, to take care of slowly changing energy budget dynamics, i.e., the battery state of charge and fuel consumption. 
Furthermore, since the internal combustion engine allows for individual cylinder deactivation, we need to handle integer values in our control problem.
Since solving mixed-integer nonlinear problems in a receding horizon fashion is particularly computationally expensive, we determine the number of active cylinders in a feedforward manner with the \gls{acr:fcc}. 
In the upcoming sections we dive into the details of each of those subsystems. 
\subsection{Estimator}\label{subsec:estimator}
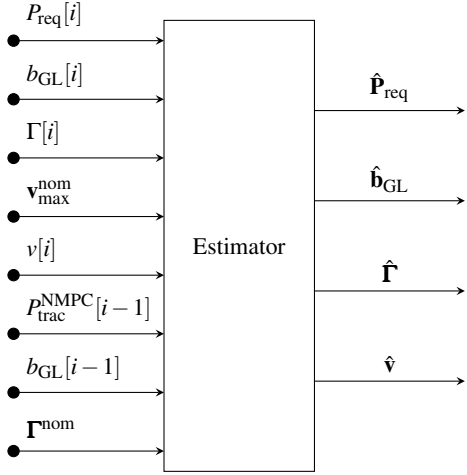
\begin{figure}
	\centering
	\tikzset{x=0cm,y=0cm}

\def\larghezza{2.5cm}
\def\altezza{8cm}
\def\dist{2.2cm}
\def\dCirc{0.15cm}
\def\vertdistIN{\altezza/7.8}
\def\vertdistOUT{1.8cm}

\tikzstyle{block} = [draw, rectangle, minimum height=2em, minimum width=3em]
\tikzstyle{input} = [coordinate]
\tikzstyle{output} = [coordinate]
\tikzstyle{dot} = [draw, circle, minimum size=\dCirc, fill = black, inner sep=0cm]

\begin{tikzpicture}[scale=1, every node/.append style={outer sep=0pt}, >=stealth, font=\small] %

\def\larghezza{2cm}
\def\altezza{6cm}
\def\dist{2cm}
\def\dCirc{0.15cm}
\def\vertdistIN{0.13*\altezza}
\def\vertdistOUT{0.2*\altezza}

\node [block, minimum height=\altezza, minimum width=\larghezza] at (0,0) (estimator) {Estimator};

\node[input] at ($(estimator.west) + (0, \vertdistIN/2)$) (vmax) {};
\node[input] at ($(vmax) + (0, {3*\vertdistIN})$) (Preq) {};
\node[input] at ($(vmax) + (0, 2*\vertdistIN)$) (GL) {};
\node[input] at ($(vmax)+ (0, \vertdistIN)$) (gamma) {};
\node[input] at ($(vmax) + (0, -\vertdistIN)$) (v) {};
\node[input] at ($(vmax) + (0, -2*\vertdistIN)$) (Ptrac) {};
\node[input] at ($(vmax) + (0, -3*\vertdistIN)$) (GLprev) {};
\node[input] at ($(vmax) + (0, -4*\vertdistIN)$) (gammanom) {};

\node[dot, label=above right:$\vec{v}_\mathrm{max}^{\mathrm{nom}}$] at ($(vmax) + (-\dist, 0)$) (circlevmax) {};
\node[dot, label=above right:${P_{\mathrm{req}}[i]}$] at ($(Preq) + (-\dist, 0)$) (circlePreq){};
\node[dot, label=above right:${b_\mathrm{GL}[i]}$] at ($(GL) + (-\dist, 0)$) (circleGL) {};
\node[dot, label=above right:${\Gamma[i]}$] at ($(gamma) + (-\dist, 0)$) (circlegamma) {};
\node[dot, label=above right:${v[i]}$] at ($(v) + (-\dist, 0)$) (circlev) {};
\node[dot, label=above right:${P_{\mathrm{trac}}^{\mathrm{NMPC}}[i-1]}$] at ($(Ptrac) + (-\dist, 0)$) (circlePtrac) {};
\node[dot, label=above right:${b_\mathrm{GL}[i-1]}$] at ($(GLprev) + (-\dist, 0)$) (circleGLprev) {};
\node[dot, label=above right:$\bm{\Gamma}^\mathrm{nom}$] at ($(gammanom) + (-\dist, 0)$) (circlegammanom) {};

\node[output] at ($(estimator.east) + (0, -\vertdistOUT/2)$) (gammaout) {};
\node[output] at ($(gammaout) + (0, \vertdistOUT)$) (GLout) {};
\node[output] at ($(gammaout) + (0, 2*\vertdistOUT)$) (Preqout) {};
\node[output] at ($(gammaout) + (0, -\vertdistOUT)$) (vout) {};

\draw[->] (circlevmax) to (vmax);
\draw[->] (circlePreq) to (Preq);
\draw[->] (circleGL) to (GL);
\draw[->] (circlev) to (v);
\draw[->] (circlegamma) to (gamma);
\draw[->] (circlePtrac) to (Ptrac);
\draw[->] (circleGLprev) to (GLprev);
\draw[->] (circlegammanom) to (gammanom);

\draw[->] (Preqout) to node[above] {$\hat{\vec{P}}_\mathrm{req}$} ($(Preqout) + (\dist,0)$);
\draw[->] (GLout) to node[above] {$\hat{\vec{b}}_\mathrm{GL}$} ($(GLout) + (\dist,0)$);
\draw[->] (gammaout) to node[above] {$\hat{\bm{\Gamma}}$} ($(gammaout) + (\dist,0)$);
\draw[->] (vout) to node[above] {$\hat{\vec{v}}$} ($(vout) + (\dist,0)$);
\end{tikzpicture}
	\caption{Inputs and outputs of the estimator, needed for the predictive nature of the control scheme. The index $i$ indicates the discrete online space step of the signal. In case of missing index, a complete future trajectory is assumed. All trajectories denoted with a hat contain predictive information.}
	\label{fig:EstimatorIO}\vspace{-0.3cm}
\end{figure}

\begin{figure}
	\centering
	\tikzstyle{dot} = [draw, circle, minimum size=\dCirc, fill = black, inner sep=0cm]

\begin{tikzpicture}[trim axis left, trim axis right, >=stealth, font=\small]
	
\def\plotwidth{0.97\columnwidth}%
\def\plotheight{0.46\columnwidth}%
\def\xmin{1300}
\def\xmax{1500}
\def\dCirc{0.2cm}

\begin{axis}[%
width=\plotwidth,
height=\plotheight,
xmin=\xmin,
xmax=\xmax,
xtick={1300,1325,1350,1375,1400,1425,1450,1475},
xticklabels={{},{},{$i$},{},{},{},{$i+N-1$},{}},
xlabel style={font=\color{white!15!black}},
ymin=190,
ymax=320,
ytick={100,200,300,400},
yticklabels={{},{200},{300},{}},
ylabel style={font=\color{white!15!black},at={(0.06,0.5)}},
ylabel={$v$ [kph]},
axis background/.style={fill=white},
xmajorgrids,
ymajorgrids,
legend style={at={(0.01,0.03)}, anchor=south west, legend cell align=left, align=left, draw=white!15!black}
]

\addplot [color=black, line width=1pt, forget plot]
  table[]{./pictures/Chapter2/picData/Estimator_v-2.tsv};

\addplot[mark=*,only marks,mark size=\dCirc/2, fill = black, inner sep=0cm] coordinates{(1350,278)};
\addlegendentry{Feasible velocity}
\node[circle, minimum size=\dCirc, fill = black, inner sep=0cm] at (axis cs:1375,284) (point2) {};
\node[circle, minimum size=\dCirc, fill = black, inner sep=0cm] at (axis cs:1400,289) (point3) {};  
\node[circle, minimum size=\dCirc, fill = black, inner sep=0cm] at (axis cs:1425,261.091) (point4) {};
\addplot[mark=*,only marks,mark size=\dCirc/2, fill = white, inner sep=0.5cm] coordinates{(1425,294)};
\addlegendentry{Infeasible velocity}
\node[circle, minimum size=\dCirc, fill = black, inner sep=0cm] at (axis cs:1450,214.813) (point5) {};
\node[circle, minimum size=\dCirc, inner sep=0cm,] at (axis cs:1450,265) (point5a) {};
\addplot[mark=*,only marks,mark size=\dCirc/2, fill = white, inner sep=0.5cm, forget plot] coordinates{(1450,265)};
\draw[dashed, line width=1pt] (point3) -- (axis cs:1423,294);
\draw[dashed, line width=0.7pt] (axis cs:1425,291) -- (point4);
\draw[dashed, line width=1pt] (point4) -- (axis cs:1448,265);
\draw[dashed, line width=0.7pt] (axis cs:1450,262) -- (point5);

\addplot[area legend, draw=none, fill=black, fill opacity=0.4, forget plot]
table[] {./pictures/Chapter2/picData/Estimator_v-1.tsv}--cycle;

\addplot [color=black, dotted, line width=0.5pt]
table[]{./pictures/Chapter2/picData/Estimator_v-3.tsv};
\addlegendentry{Maximum velocity}

\end{axis}
\end{tikzpicture}%
	\caption{Evolution of the velocity prediction. At index $i$ the estimator output is still deterministic (measured or given by the driver). Therefore, the estimation starts with the computation of the velocity at index $i+1$ according to \eqref{eq:eqdiffvelocity}. From there, the algorithm depicted in Fig. \ref{fig:Driver_Flowchart} is run and the estimation of the GL index and the requested power at that index is computed. Additionally, the estimated velocity of the following step is obtained and the algorithm is repeated until the end of the horizon. In the figure we showcase how in the first two algorithm iterations the velocity profile can evolve freely, while in the last two the $v_{\mathrm{max}}^{\mathrm{act}}$ profile is exceeded and the power needs to be recomputed to respect the maximum velocity profile.}
	\label{fig:estimatorV}\vspace{-0.3cm}
\end{figure}
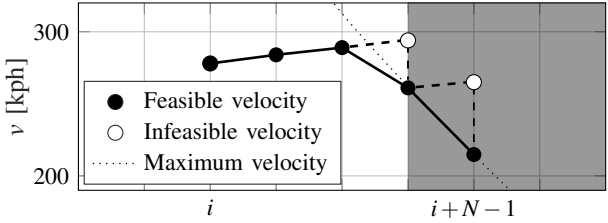

Given the predictive nature of the controller, we need the driver's future intentions and the velocity evolution for the complete \gls{acr:nmpc} horizon of length $N$.
This results in the inputs and outputs shown in Fig. \ref{fig:EstimatorIO}.
Accordingly, the output signals consist of trajectories where deterministic information at step $i$ is augmented with predictions computed by the estimator subsystem.
In particular, the first entries of the output vectors, i.e., at the \textit{current} step, are the driver's decisions (outputs of Fig. \ref{fig:DriverIO}) and the measured velocity, while the remaining entries are predictions denoted with a hat $\hat{(\cdot)}$, i.e.,
\begin{equation}
	\begin{aligned}
		\hat{\vec{P}}_\mathrm{req} &= \big[ P_{\mathrm{req}}[i]  & \hat{P}_{\mathrm{req}}[i+1] && \dots && \hat{P}_{\mathrm{req}}[i+N-1] \big] ,\\
		\hat{\vec{b}}_{\mathrm{GL}} &= \big[ b_\mathrm{GL}[i] & \hat{b}_{\mathrm{GL}}[i+1] && \dots && \hat{b}_{\mathrm{GL}}[i+N-1] \big] ,\\
		\hat{\bm{\Gamma}} &= \big[ \Gamma[i] & \hat{\Gamma}[i+1] && \dots && \hat{\Gamma}[i+N-1] \big] ,\\
		\hat{\vec{v}} &= \big[ v[i] & \hat{v}[i+1] && \dots && \hat{v}[i+N-1] \big] .
	\end{aligned}
\end{equation}
The predictions of the requested power and the GL index are determined by looping the algorithm depicted in Fig. \ref{fig:Driver_Flowchart} over the \gls{acr:nmpc} horizon.
The velocity prediction arises from its evolution according to \eqref{eq:eqdiffvelocity} and the predicted requested power.
In Fig. \ref{fig:estimatorV} we illustrate quantitatively how such a velocity prediction evolves when encountering the $v_{\mathrm{max}}^{\mathrm{act}}$ profile within the horizon.
In the corners, the predictions can be performed without loss of generality because the car is traveling at the maximum velocity of the car.
On the straights, however, we have no knowledge of the \gls{acr:nmpc}'s optimized traction power in the future. 
Therefore, we assume the delivered traction power $P_{\mathrm{trac}}^{\mathrm{NMPC}}[i-1]$ at the previous step to be constant over the entire horizon.
Given that in reality the power rather decreases towards the corner, this assumption is sensible and robust.
In fact, this possibly leads to an overestimation of the future velocity and thus might result in a deviating corner entry point at the end of the straight. 
However, this mismatch disappears within the proposed framework due to the receding horizon property of the controller.

\begin{figure}
	\centering
	\begin{tikzpicture}[trim axis left, trim axis right, >=stealth, font=\small]
	
\def\plotwidth{0.8\columnwidth}%
\def\plotheight{0.17\columnwidth}%
\def\yshift{-0.25cm}%

\begin{axis}[%
name=one,
width=\plotwidth,
height=\plotheight,
scale only axis,
xmin=3,
xmax=15,
xtick={1,3,5,7,9,11,13,15,17,19},
xticklabels={{},{},{},{},{},{},{},{},{},{}},
ymin=2.5,
ymax=4.5,
ylabel style={font=\color{white!15!black}},
ylabel style={at={(0.08,0.5)}},
ylabel={$\Gamma$ [-]},
ytick={3, 4},
yticklabels={{3},{4}},
axis background/.style={fill=white},
title style={at={(0.5,0.85)}},
title={Scenario 1},
xmajorgrids,
ymajorgrids
]
\addplot [color=white!65!black, line width=1.7pt, forget plot]
  table[]{pictures/Chapter2/picData/GearshiftingScenarios-1.tsv};
\addplot [color=black, only marks, mark=o, mark options={solid, black}, forget plot, line width=1pt]
  table[]{pictures/Chapter2/picData/GearshiftingScenarios-2.tsv};
\addplot [color=black, dashed, forget plot, line width=1pt]
  table[]{pictures/Chapter2/picData/GearshiftingScenarios-3.tsv};
\end{axis}

\begin{axis}[%
name=two,
width=\plotwidth,
height=\plotheight,
at=(one.below south west),
yshift=\yshift,
anchor=north west,
scale only axis,
xmin=3,
xmax=15,
xtick={1,3,5,7,9,11,13,15,17,19},
xticklabels={{},{},{},{},{},{},{},{},{},{}},
ymin=3.5,
ymax=5.5,
ylabel style={font=\color{white!15!black}},
ylabel style={at={(0.08,0.5)}},
ylabel={$\Gamma$ [-]},
ytick={4, 5},
yticklabels={{4},{5}},
axis background/.style={fill=white},
title style={at={(0.5,0.85)}},
title={Scenario 2},
xmajorgrids,
ymajorgrids,
legend style={at={(0.99,0.03)}, anchor=south east, legend cell align=left, align=left, draw=white!15!black}
]
\addplot [color=white!65!black, line width=1.7pt]
  table[]{pictures/Chapter2/picData/GearshiftingScenarios-4.tsv};
  \addlegendentry{Nominal}
\addplot [color=black, only marks, mark=o, mark options={solid, black}, line width=1pt]
  table[]{pictures/Chapter2/picData/GearshiftingScenarios-5.tsv};
  \addlegendentry{Estimator Output}
  
\addplot [color=black, dashed, forget plot, line width=1pt]
  table[]{pictures/Chapter2/picData/GearshiftingScenarios-6.tsv};
\end{axis}

\begin{axis}[%
name=three,
width=\plotwidth,
height=\plotheight,
at=(two.below south west),
yshift=\yshift,
anchor=north west,
scale only axis,
xmin=3,
xmax=15,
xtick={1,3,5,7,9,11,13,15,17,19},
xticklabels={{},{},{$i$},{},{},{},{$i+N-1$},{},{},{}},
xlabel style={font=\color{white!15!black}},
ymin=2.5,
ymax=4.5,
ylabel style={font=\color{white!15!black}},
ylabel style={at={(0.08,0.5)}},
ylabel={$\Gamma$ [-]},
ytick={3, 4},
yticklabels={{3},{4}},
axis background/.style={fill=white},
title style={at={(0.5,0.85)}},
title={Scenario 3},
xmajorgrids,
ymajorgrids
]
\addplot [color=white!65!black, line width=1.7pt, forget plot]
  table[]{pictures/Chapter2/picData/GearshiftingScenarios-7.tsv};

\addplot [color=black, only marks, mark=o, mark options={solid, black}, line width=1.2pt, forget plot]
  table[]{pictures/Chapter2/picData/GearshiftingScenarios-8.tsv};

\addplot [color=black, dashed, forget plot, line width=1pt]
  table[]{pictures/Chapter2/picData/GearshiftingScenarios-9.tsv};
\end{axis}

\end{tikzpicture}%
	\caption{Schematic of the three different scenarios that can occur in the estimator at space step $i$ when the nominal gear and the requested gear at that index do not match. The estimator output is a trajectory of length $N=5$. The first entry of that vector is equivalent to the driver's request. In scenario 2 we assume the increase of \SI{200}{rpm} to occur after 2 NMPC iterations (for illustration purposes only).}
	\label{fig:GSScenarios}\vspace{-0.3cm}
\end{figure}
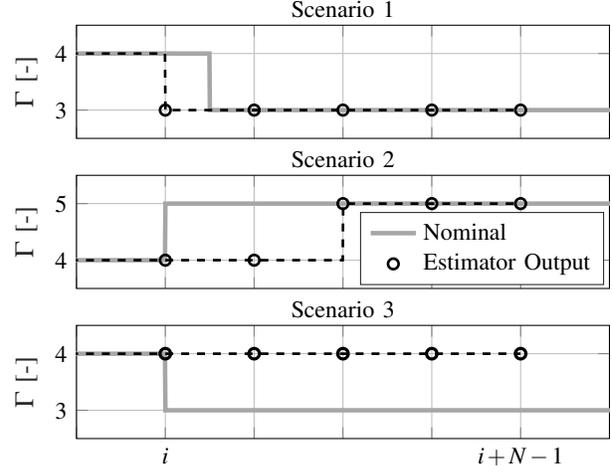

The last component to be predicted is the gearshift trajectory, which depends on the difference between current nominal reference gear $\Gamma^\mathrm{nom}[i]$ and the driver's decision $\Gamma[i]$.
If both are equal, the complete sequence for the predicted trajectory is set equivalent to the nominal one.
If they differ, this could lead to the three possible scenarios shown in Fig. \ref{fig:GSScenarios} and described below:
\begin{itemize}
	\item[1.] The driver performed a gearshift although the nominal trajectory does not show one: The future gear trajectory is set equal to the driver's unexpected new gear, as we assume that in the short \gls{acr:nmpc} horizon of \SI{8}{m} (see Section \ref{subsec:nmpc}) not more than one shift occurs.
	\item[2.] The nominal trajectory shows an upshift although the driver has not shifted: We assume the driver to have missed the ideal shifting point and set the predicted upshift to occur once the engine speed has increased by a further \SI{200}{rpm}.
	\item[3.] The nominal trajectory shows a downshift although the driver has not shifted: As before, we assume the driver to have missed the ideal shifting point. Since this scenario usually only occurs at the entry of a corner while braking, we keep the engaged gear for the entire horizon.
\end{itemize}
As a final step, we verify that the engine speed limits in \eqref{eq:omegaenginelimits} are satisfied: If at some point the estimated gear trajectory leads to a constraint violation, the appropriate up- or downshift is introduced and kept for the subsequent steps.

\subsection{Equivalent Lap Time Minimization Strategies (ELTMS)}\label{subsec:eltms}
\begin{figure}
	\centering
	\definecolor{mycolor1}{rgb}{1.00000,0.76000,0.94000}%
\definecolor{mycolor2}{rgb}{1.00000,0.48000,0.87000}%
\definecolor{mycolor3}{rgb}{1.00000,0.20000,0.80000}%
\definecolor{mycolor4}{rgb}{0.70000,0.91600,0.94000}%
\definecolor{mycolor5}{rgb}{0.35000,0.81800,0.87000}%
\definecolor{mycolor6}{rgb}{0.00000,0.72000,0.80000}%

\def\dCirc{0.0cm}
\tikzstyle{dot} = [draw, circle, minimum size=\dCirc, fill = black, inner sep=0cm]

\tikzstyle{coord} = [coordinate]

\begin{tikzpicture}[>=stealth, font=\small]

\def\plotwidth{0.6\columnwidth}%
\def\plotheight{0.3\columnwidth}%
\def\yshift{-0.1cm}%
\def\xshift{0.2cm}

\begin{axis}[%
name=one,
width=\plotwidth,
height=\plotheight,
xmin=-5.5,
xmax=0.5,
xtick={-4,-2,0},
xticklabels={{}},
ymin=-0.2,
ymax=1.2,
ytick={0,0.5,1},
ylabel style={font=\color{white!15!black},at={(0.09,0.5)}},
ylabel={$P_\mathrm{f}$[-]},
axis background/.style={fill=white},
xmajorgrids,
ymajorgrids,
legend style={legend cell align=left, align=left, draw=white!15!black}
]
\addplot [color=white!70!black, line width=1.2pt, mark size=1.5pt, mark=*, mark options={solid, white!70!black}]
  table[]{./pictures/Chapter2/picData/lookupOCPtikz-1.tsv};

\addplot [color=white!35!black, line width=1.2pt, mark size=1.5pt, mark=*, mark options={solid, white!35!black}]
  table[]{./pictures/Chapter2/picData/lookupOCPtikz-2.tsv};

\addplot [color=black, line width=1.2pt, mark size=1.5pt, mark=*, mark options={solid, black}]
  table[]{./pictures/Chapter2/picData/lookupOCPtikz-3.tsv};

\end{axis}

\begin{axis}[%
name=two,
width=\plotwidth,
height=\plotheight,
at=(one.below south west),
yshift=\yshift,
anchor=north west,
xmin=-5.5,
xmax=0.5,
xtick={-4,-2,0},
xticklabels={{}},
ymin=-1.4018691588785,
ymax=1.4018691588785,
ytick={-1,0,1},
ylabel style={font=\color{white!15!black},at={(0.09,0.5)}},
ylabel={$P_\mathrm{k}$[-]},
axis background/.style={fill=white},
xmajorgrids,
ymajorgrids,
legend style={legend cell align=left, align=left, draw=white!15!black}
]
\addplot [color=white!70!black, line width=1.2pt, mark size=1.5pt, mark=*, mark options={solid, white!70!black}]
  table[]{./pictures/Chapter2/picData/lookupOCPtikz-4.tsv};

\addplot [color=white!35!black, line width=1.2pt, mark size=1.5pt, mark=*, mark options={solid, white!35!black}]
  table[]{./pictures/Chapter2/picData/lookupOCPtikz-5.tsv};

\addplot [color=black, line width=1.2pt, mark size=1.5pt, mark=*, mark options={solid, black}]
  table[]{./pictures/Chapter2/picData/lookupOCPtikz-6.tsv};

\end{axis}

\begin{axis}[%
name=three,
width=\plotwidth,
height=\plotheight,
at=(two.below south west),
yshift=\yshift,
anchor=north west,
xmin=-5.5,
xmax=0.5,
xtick={-4,-2,0},
xlabel style={font=\color{white!15!black}},
xlabel={$\lambda_\mathrm{kin}$[-]},
ymin=-0.2,
ymax=1.2,
ytick={0,0.5,1},
ylabel style={font=\color{white!15!black},at={(0.09,0.5)}},
ylabel={$u_\mathrm{wg}$[-]},
axis background/.style={fill=white},
xmajorgrids,
ymajorgrids,
legend style={legend cell align=left, align=left, draw=white!15!black}
]
\addplot [color=white!70!black, line width=1.2pt, mark size=1.5pt, mark=*, mark options={solid, white!70!black}]
  table[]{./pictures/Chapter2/picData/lookupOCPtikz-7.tsv};

\addplot [color=white!35!black, line width=1.2pt, mark size=1.5pt, mark=*, mark options={solid, white!35!black}]
  table[]{./pictures/Chapter2/picData/lookupOCPtikz-8.tsv};

\addplot [color=black, line width=1.2pt, mark size=1.5pt, mark=*, mark options={solid, black}]
  table[]{./pictures/Chapter2/picData/lookupOCPtikz-9.tsv};

\node[dot] at (axis cs:-5,0.2) (source1) {};
\node[dot, label=right:$\lambda_{\mathrm{b}}$] at (axis cs:-1.5,0.6) (destination1) {};

\draw[->, line width=0.8pt] (source1) to (destination1);

\end{axis}

\begin{axis}[%
name=four,
width=\plotwidth,
height=\plotheight,
at=(one.below south east),
xshift=\xshift,
anchor=below south west,
xmin=-0.437636761487965,
xmax=1.31291028446389,
xtick={0,0.5,1},
xticklabels={{}},
ymin=-0.2,
ymax=1.2,
ytick={0,0.5,1},
yticklabels={{}},
axis background/.style={fill=white},
xmajorgrids,
ymajorgrids,
legend style={legend cell align=left, align=left, draw=white!15!black}
]
\addplot [color=white!70!black, line width=1.2pt, mark size=1.5pt, mark=*, mark options={solid, white!70!black}]
  table[]{./pictures/Chapter2/picData/lookupOCPtikz-10.tsv};

\addplot [color=white!35!black, line width=1.2pt, mark size=1.5pt, mark=*, mark options={solid, white!35!black}]
  table[]{./pictures/Chapter2/picData/lookupOCPtikz-11.tsv};

\addplot [color=black, line width=1.2pt, mark size=1.5pt, mark=*, mark options={solid, black}]
  table[]{./pictures/Chapter2/picData/lookupOCPtikz-12.tsv};

\end{axis}

\begin{axis}[%
name=five,
width=\plotwidth,
height=\plotheight,
at=(two.below south east),
xshift=\xshift,
anchor=below south west,
xmin=-0.437636761487965,
xmax=1.31291028446389,
xtick={0,0.5,1},
xticklabels={{}},
ymin=-1.4018691588785,
ymax=1.4018691588785,
ytick={-1,0,1},
yticklabels={{}},
axis background/.style={fill=white},
xmajorgrids,
ymajorgrids,
legend style={legend cell align=left, align=left, draw=white!15!black}
]
\addplot [color=white!70!black, line width=1.2pt, mark size=1.5pt, mark=*, mark options={solid, white!70!black}]
  table[]{./pictures/Chapter2/picData/lookupOCPtikz-13.tsv};

\addplot [color=white!35!black, line width=1.2pt, mark size=1.5pt, mark=*, mark options={solid, white!35!black}]
  table[]{./pictures/Chapter2/picData/lookupOCPtikz-14.tsv};

\addplot [color=black, line width=1.2pt, mark size=1.5pt, mark=*, mark options={solid, black}]
  table[]{./pictures/Chapter2/picData/lookupOCPtikz-15.tsv};

\end{axis}

\begin{axis}[%
name=six,
width=\plotwidth,
height=\plotheight,
at=(three.below south east),
xshift=\xshift,
anchor=below south west,
xmin=-0.437636761487965,
xmax=1.31291028446389,
xtick={0,0.5,1},
xlabel style={font=\color{white!15!black}},
xlabel={$P_\mathrm{req}$[-]},
ymin=-0.2,
ymax=1.2,
ytick={0,0.5,1},
yticklabels={{}},
axis background/.style={fill=white},
xmajorgrids,
ymajorgrids,
legend style={legend cell align=left, align=left, draw=white!15!black}
]
\addplot [color=white!70!black, line width=1.2pt, mark size=1.5pt, mark=*, mark options={solid, white!70!black}]
  table[]{./pictures/Chapter2/picData/lookupOCPtikz-16.tsv};

\addplot [color=white!35!black, line width=1.2pt, mark size=1.5pt, mark=*, mark options={solid, white!35!black}]
  table[]{./pictures/Chapter2/picData/lookupOCPtikz-17.tsv};

\addplot [color=black, line width=1.2pt, mark size=1.5pt, mark=*, mark options={solid, black}]
  table[]{./pictures/Chapter2/picData/lookupOCPtikz-18.tsv};

\end{axis}

\node[coord, label=above:\textbf{PL}] at (one.north) (PL) {};
\node[coord, label=above:\textbf{GL}] at (four.north) (GL) {};

\end{tikzpicture}%
	\caption{Look-up tables with changing battery costate $\lambda_\mathrm{b}$. On the left: $\lambda_{\mathrm{kin}}$-dependent look-up tables for the power-limited regions. On the right: $P_\mathrm{req}$-dependent look-up tables for the grip-limited regions.}
	\label{fig:OCP}\vspace{-0.3cm}
\end{figure}
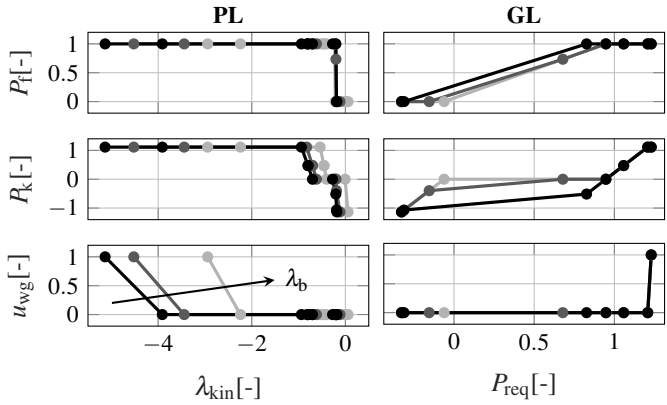

From an energetic point of view, the goal of our online framework is optimizing the low-level powertrain actuators while minimizing lap time and meeting the energy consumption targets introduced in Problem \ref{prob:ocpBalerna}, i.e., $\Delta E_\mathrm{f,target}$ and $\Delta E_\mathrm{b,target}$. 
\begin{figure}
	\centering
	\tikzstyle{block} = [draw, rectangle, minimum height=2em, minimum width=3em]
\tikzstyle{input} = [coordinate]
\tikzstyle{output} = [coordinate]

\begin{tikzpicture}[scale=1, every node/.append style={outer sep=0pt}, >=stealth, font=\small] %
\def\larghezza{1.6cm}
\def\altezza{5.5cm}
\def\dist{2cm}
\def\dCirc{0.2cm}
\def\vertdistIN{0.14*\altezza}
\def\vertdistOUT{0.25*\altezza}

\node [block, minimum height=\altezza, minimum width=\larghezza] at (0,0) (eltms) {ELTMS};

\node[input] at ($(eltms.west) + (0, 3*\vertdistIN)$) (Efnom) {};
\node[input] at ($(eltms.west) + (0, 2*\vertdistIN)$) (Efi) {};
\node[input] at ($(eltms.west) + (0, \vertdistIN)$) (Ebnom) {};
\node[input] at ($(eltms.west) + (0, 0)$) (Ebi) {};
\node[input] at ($(eltms.west) + (0, -{\vertdistIN})$) (Preqest) {};
\node[input] at ($(eltms.west) + (0, -{2*\vertdistIN})$) (GLest) {};
\node[input] at ($(eltms.west) + (0, -{3*\vertdistIN})$) (lambdakin) {};

\node[output] at ($(eltms.east) + (0, \vertdistOUT)$) (Pfest) {};
\node[output] at ($(eltms.east) + (0, 0)$) (Pkest) {};
\node[output] at ($(eltms.east) + (0, -\vertdistOUT)$) (WGest) {};

\node[circle, minimum size=\dCirc, fill = black, inner sep=0cm, label=above right:${E_{\mathrm{f}}^{\mathrm{nom}}[i]}$] at ($(Efnom) + (-\dist,0)$) (circEfnom) {};
\node[circle, minimum size=\dCirc, fill = black, inner sep=0cm, label=above right:${E_{\mathrm{f}}[i]}$] at ($(Efi) + (-\dist,0)$) (circEfi) {};
\node[circle, minimum size=\dCirc, fill = black, inner sep=0cm, label=above right:${E_{\mathrm{b}}^{\mathrm{nom}}[i]}$] at ($(Ebnom) + (-\dist,0)$) (circEbnom) {};
\node[circle, minimum size=\dCirc, fill = black, inner sep=0cm, label=above right:${E_{\mathrm{b}}[i]}$] at ($(Ebi) + (-\dist,0)$) (circEbi) {};
\node[circle, minimum size=\dCirc, fill = black, inner sep=0cm, label=above right:$\hat{\vec{P}}_{\mathrm{req}}$] at ($(Preqest) + (-\dist,0)$) (circPreqest) {};
\node[circle, minimum size=\dCirc, fill = black, inner sep=0cm, label=above right:$\hat{\vec{b}}_\mathrm{GL}$] at ($(GLest) + (-\dist,0)$) (circGLest) {};
\node[circle, minimum size=\dCirc, fill = black, inner sep=0cm, label=above right:$\bm{\lambda}_{\mathrm{kin}}$] at ($(lambdakin) + (-\dist,0)$) (circlambdakin) {};

\draw[->] (circEfnom) to (Efnom);
\draw[->] (circEfi) to (Efi);
\draw[->] (circEbnom) to (Ebnom);
\draw[->] (circEbi) to (Ebi);
\draw[->] (circPreqest) to (Preqest);
\draw[->] (circGLest) to (GLest);
\draw[->] (circlambdakin) to (lambdakin);

\draw[->] (Pfest) to node[above] {$\hat{\vec{P}}_\mathrm{f}$} ($(Pfest) + (\dist,0)$);
\draw[->] (Pkest) to node[above] {$\hat{\vec{P}}_\mathrm{k}$} ($(Pkest) + (\dist,0)$);
\draw[->] (WGest) to node[above] {$\hat{\vec{u}}_\mathrm{wg}$} ($(WGest) + (\dist,0)$);

\end{tikzpicture}
	\caption{Inputs and outputs of the \gls{acr:eltms}, needed to predict low-level actuators to fulfill the high-level energy budgets. The index $i$ indicates the discrete online space step of the signal. In case of missing index, a complete future trajectory is assumed. All trajectories denoted with a hat contain predictive information.}
	\label{fig:ELTMSIO}\vspace{-0.3cm}
\end{figure}
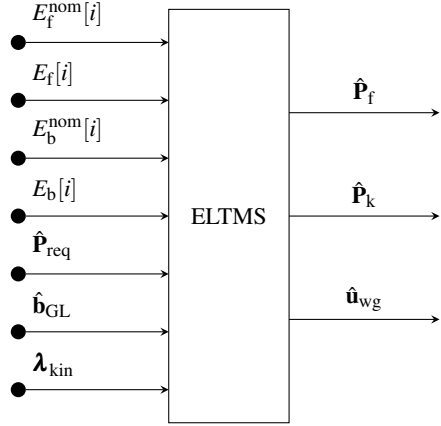
Those energy state variables display much slower dynamics compared to the ones of the internal combustion engine \cite{Balerna}.
Therefore, while the intake manifold pressure and the turbocharger's kinetic energy are controlled by a short-sighted \gls{acr:nmpc}, for the slowly changing states, we employ a dedicated feedback controller structure, which can handle them in a computationally inexpensive and accurate manner.
We opted for the \gls{acr:eltms} controller, which was previously developed in our research group.
The main advantage of this supervisory controller lies in its simplicity.
It relies on computationally inexpensive one-dimensional look-up tables derived with \gls{acr:pmp} \cite{salazar2018minimum}.
Its major disadvantage is that this derivation is based on a simplified high-level model, where the engine is static, hence differs from the plant.
Therefore, to use the high-level \gls{acr:eltms} outputs shown in Fig. \ref{fig:ELTMSIO} as reference trajectories for the \gls{acr:nmpc}, further assumptions need to be made (see Section \ref{subsec:nmpc}).
The outputs of the look-up tables are the fuel power $P_\mathrm{f}$, the MGU-K power $P_\mathrm{k}$ and the waste-gate position $u_\mathrm{wg}$.
In the grip-limited regions, where the total amount of delivered power needs to match the driver's power request, we can directly read the power split shown on the right side of Fig. \ref{fig:OCP}.
In the power-limited regions instead, where we can control also the total traction power, we need to quantify where an increase in power leads to the highest time saving capabilities.
Such a quantification can be defined by the kinetic costate $\lambda_{\mathrm{kin}}$ shown in the lower plot of Fig. \ref{fig:v_max}, which is the position-dependent dual variable of the kinetic energy of the car.
This costate reaches its negative peak at the beginning of a straight, where the lap time sensitivity with respect to propulsive power is high, and increases towards the corner where the car is already at a very high speed.
As a result, on the left side of Fig. \ref{fig:OCP} we show the $\lambda_{\mathrm{kin}}$-dependent look-up table used in power-limited regions.

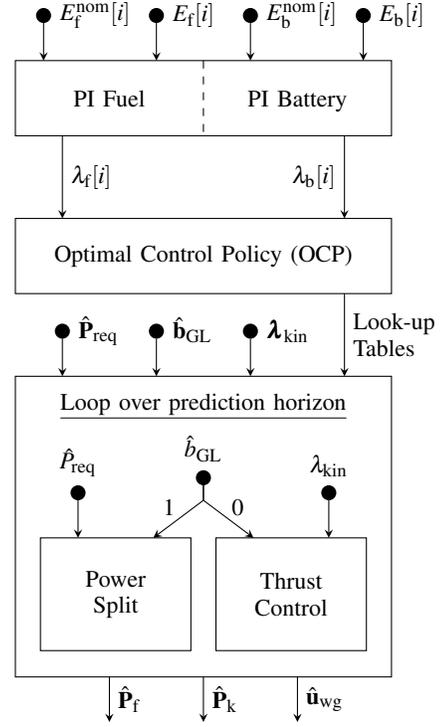
\begin{figure}
	\centering
	\tikzstyle{block} = [draw, rectangle, minimum height=2em, minimum width=3em]
\tikzstyle{input} = [coordinate]
\tikzstyle{output} = [coordinate]

\begin{tikzpicture}[scale=1, every node/.append style={outer sep=0pt}, >=stealth, font=\small] %

\def\larghezza{5cm}
\def\altezza{1cm}
\def\larghezzaPiccolo{2cm}
\def\altezzaPiccolo{1.5cm}
\def\spazioPiccolo{(\larghezza - 2*\larghezzaPiccolo)/6 + \larghezzaPiccolo/2}
\def\altezzaLoop{4*\altezza}
\def\deltaVert{0.4cm}
\def\dist{2cm}
\def\distCircle{0.6cm}
\def\dCirc{0.2cm}
\def\latOffquattro{\larghezza/4}
\def\latOffset{\latOffquattro*3/2}
\def\distInOut{{(\dist - \altezza)/2}}

\node [block, minimum height=\altezza, minimum width=\larghezza] at (0,0)  (PID) {};
\draw [dashed] (PID.north) -- (PID.south);
\node[] at ($(PID)+ (-{\larghezza/4},0)$) {PI Fuel};
\node[] at ($(PID)+ ({\larghezza/4},-1pt)$) {PI Battery};

\node [block, minimum height=\altezza, minimum width=\larghezza, anchor=north] at ($(PID) + (0,-4*\dist/5)$) (OCP) {Optimal Control Policy (OCP)};

\node [block, minimum height=\altezzaLoop, minimum width=\larghezza, anchor=north] at ($(OCP) + (0,-4*\dist/5)$) (Loop) {};
\node at ($(Loop.north) - (0, {\deltaVert})$)  {\underline{Loop over prediction horizon}};

\node[input] at ($(PID.north) + (-1.7*\latOffquattro, 0)$) (Efnom) {};
\node[input] at ($(PID.north) + (-0.5*\latOffquattro, 0)$) (Ef) {};
\node[input] at ($(PID.north) + (0.5*\latOffquattro, 0)$) (Ebnom) {};
\node[input] at ($(PID.north) + (1.7*\latOffquattro, 0)$) (Eb) {};
\node[output] at ($(PID.south) + (-1.5*\latOffquattro, 0)$) (LambdafOut) {};
\node[output] at ($(PID.south) + (1.5*\latOffquattro, 0)$) (LambdabOut) {};

\node[input] at ($(OCP.north) + (-1.5*\latOffquattro, 0)$) (LambdafIn) {};
\node[input] at ($(OCP.north) + (1.5*\latOffquattro, 0)$) (LambdabIn) {};
\node[output] at ($(OCP.south) + (1.5*\latOffquattro, 0)$) (OCPMapsOut) {};

\node[input] at ($(Loop.north) + (-1.5*\latOffquattro, 0)$) (PreqIn) {};
\node[input] at ($(Loop.north) + (-0.5*\latOffquattro, 0)$) (GLIn) {};
\node[input] at ($(Loop.north) + (0.5*\latOffquattro, 0)$) (LambdakinIn) {};
\node[input] at ($(Loop.north) + (1.5*\latOffquattro, 0)$) (OCPMapsIn) {};
\node[output] at ($(Loop.south) + (-\latOffquattro, 0)$) (PfOut) {};
\node[output] at ($(Loop.south) + (0, 0)$) (PkOut) {};
\node[output] at ($(Loop.south) + (\latOffquattro, 0)$) (WGOut) {};

\node[circle, minimum size=\dCirc, fill = black, inner sep=0cm, label=right:${E_{\mathrm{f}}^{\mathrm{nom}}[i]}$] at ($(Efnom) + (0, \distCircle)$) (circEfnom) {};
\node[circle, minimum size=\dCirc, fill = black, inner sep=0cm, label=right:${E_{\mathrm{f}}[i]}$] at ($(Ef) + (0, \distCircle)$) (circEf) {};
\node[circle, minimum size=\dCirc, fill = black, inner sep=0cm, label=right:${E_{\mathrm{b}}^{\mathrm{nom}}[i]}$] at ($(Ebnom) + (0, \distCircle)$) (circEbnom) {};
\node[circle, minimum size=\dCirc, fill = black, inner sep=0cm, label=right:${E_{\mathrm{b}}[i]}$] at ($(Eb) + (0, \distCircle)$) (circEb) {};

\node[circle, minimum size=\dCirc, fill = black, inner sep=0cm, label=right:$\bm{\lambda}_{\mathrm{kin}}$] at ($(LambdakinIn) + (0, \distCircle)$) (circLambdakin) {};
\node[circle, minimum size=\dCirc, fill = black, inner sep=0cm, label=right:$\hat{\vec{b}}_{\mathrm{GL}_{}}$] at ($(GLIn) + (0, \distCircle)$) (circGL) {};
\node[circle, minimum size=\dCirc, fill = black, inner sep=0cm, label=right:$\hat{\vec{P}}_{\mathrm{req}}$] at ($(PreqIn) + (0, \distCircle)$) (circPreq) {};

\draw[->] (circEfnom) to (Efnom);
\draw[->] (circEf) to (Ef);
\draw[->] (circEbnom) to (Ebnom);
\draw[->] (circEb) to (Eb);

\draw[->] (LambdafOut) to node[right] {$\lambda_{\mathrm{f}}[i]$} (LambdafIn);
\draw[->] (LambdabOut) to node[left] {$\lambda_{\mathrm{b}}[i]$} (LambdabIn);

\draw[->] (circLambdakin) to (LambdakinIn);
\draw[->] (circGL) to (GLIn);
\draw[->] (circPreq) to (PreqIn);
\draw[->] (OCPMapsOut) to node[right, text width = 4em] {Look-up Tables} (OCPMapsIn);

\draw[->] (PfOut) to node[right] {$\hat{\vec{P}}_\mathrm{f}$} ($(PfOut) + (0, -\distCircle)$);
\draw[->] (PkOut) to node[right] {$\hat{\vec{P}}_\mathrm{k}$} ($(PkOut) + (0, -\distCircle)$);
\draw[->] (WGOut) to node[right] {$\hat{\vec{u}}_\mathrm{wg}$} ($(WGOut) + (0, -\distCircle)$);

\node [block, minimum height=\altezzaPiccolo, minimum width=\larghezzaPiccolo, text width = 3em, text centered] at ($(Loop) + (-{(\spazioPiccolo)},{-0.6*\altezzaPiccolo})$) (PowerSplit) {Power \\ Split};
\node [block, minimum height=\altezzaPiccolo, minimum width=\larghezzaPiccolo, text width = 3em, text centered] at ($(Loop) + ({\spazioPiccolo},{-0.6*\altezzaPiccolo})$) (ThrustControl) {Thrust Control};

\node[input] at ($(PowerSplit.north) + (-{\larghezzaPiccolo/4},{\distCircle})$) (preq_mini) {};
\node[circle, minimum size=\dCirc, fill = black, inner sep=0cm, label=above:${\hat{P}}_{\mathrm{req}}$] at (preq_mini) {};
\draw[->] (preq_mini) to ($(PowerSplit.north) + (-{\larghezzaPiccolo/4},0)$);

\node[input] at ($(ThrustControl.north) + ({\larghezzaPiccolo/4},{\distCircle})$) (thrust_mini) {};
\node[circle, minimum size=\dCirc, fill = black, inner sep=0cm, label=above:${\lambda}_{\mathrm{kin}}$] at (thrust_mini) {};
\draw[->] (thrust_mini) to ($(ThrustControl.north) + (\larghezzaPiccolo/4,0)$);

\node[coordinate] at ($(PowerSplit.north) + ({\spazioPiccolo},{\dist/2.5})$) (GLmini) {};
\node[circle, minimum size=\dCirc, fill = black, inner sep=0cm, label=above:$\hat{b}_\mathrm{GL}$] at (GLmini) {};
\node[] at ($(GLmini)+ (-1.3em,-1.1em)$) {1};
\node[] at ($(GLmini)+ (1.3em,-1.1em)$) {0};

\draw[->] (GLmini) -- ($(PowerSplit.north) + ({\spazioPiccolo},\dist/4)$) -- ($(PowerSplit.north) + ({\larghezzaPiccolo/4},0)$);
\draw[->] (GLmini) -- ($(PowerSplit.north) + ({\spazioPiccolo},\dist/4)$) -- ($(ThrustControl.north) + (-{\larghezzaPiccolo/4},0)$);

\end{tikzpicture}
	\caption{Architecture of the \gls{acr:eltms}. The upper two blocks are determined once per \gls{acr:nmpc} cycle to determine the look-up tables. The lower block is looped from $k=0,\dots,N-1$ in order to compute the high-level predictions from the estimated driver requests.}
	\label{fig:ELTMS}\vspace{-0.3cm}
\end{figure}

In \cite{salazar2018minimum}, it was demonstrated that the power- and grip-limited look-up tables, characterized by their nodes, are fully defined by the value of the costate variables $\lambda_\mathrm{f}$ and $\lambda_\mathrm{b}$ associated with the states $E_\mathrm{f}$ and $E_\mathrm{b}$, respectively.
Therefore, with changing costates, the resulting look-up tables vary accordingly, as illustrated in Fig. \ref{fig:OCP} for different values of $\lambda_{\mathrm{b}}$. 
Via linear interpolation between nodes it is possible to obtain a unique representation of the \gls{acr:eltms} outputs according to the costate evolution along the optimization.
As a consequence, to track the trajectories of the energy budgets and counteract possible disturbances, we adjust the costate values by means of \gls{acr:pi} control loops, as shown in the upper part of Fig. \ref{fig:ELTMS}.
For example, if the battery trajectory differs from the nominal one, the \gls{acr:pi} controller adapts the battery costate such that the look-up tables change to compensate for the drift.
Finally, to compute the predictions of the fuel power, the MGU-K power, and the waste-gate position, we assume the look-up tables to stay constant over the optimization horizon and feed them with the requested power prediction or the kinetic costate, depending on the track region (see lower part of Fig. \ref{fig:ELTMS}).

\subsection{Feedforward Cylinder Controller (FCC)}\label{subsec:PCC}
\begin{figure}
	\centering
	\tikzstyle{block} = [draw, rectangle, minimum height=2em, minimum width=3em]
\tikzstyle{input} = [coordinate]
\tikzstyle{output} = [coordinate]

\begin{tikzpicture}[scale=1, every node/.append style={outer sep=0pt}, >=stealth, font=\small] %
\def\larghezza{1.6cm}
\def\altezza{3cm}
\def\dist{1.5cm}
\def\dCirc{0.2cm}
\def\vertdistIN{0.3*\altezza}

\node [block, minimum height=\altezza, minimum width=\larghezza] at (0,0) (pcc) {FCC};

\node[input] at ($(pcc.west) + (0, {\vertdistIN})$) (Pf) {};
\node[input] at ($(pcc.west) + (0, 0)$) (Gamma) {};
\node[input] at ($(pcc.west) + (0, -\vertdistIN)$) (v) {};

\node[output] at ($(pcc.east) + (0, 0)$) (Psie) {};

\node[circle, minimum size=\dCirc, fill = black, inner sep=0cm, label=above right:$\hat{\vec{P}}_\mathrm{f}$] at ($(Pf) + (-\dist,0)$) (circPf) {};
\node[circle, minimum size=\dCirc, fill = black, inner sep=0cm, label=above right:$\hat{\bm{\Gamma}}$] at ($(Gamma) + (-\dist,0)$) (circGamma) {};
\node[circle, minimum size=\dCirc, fill = black, inner sep=0cm, label=above right:$\hat{\vec{v}}$] at ($(v) + (-\dist,0)$) (circv) {};

\draw[->] (circGamma) to (Gamma);
\draw[->] (circPf) to (Pf);
\draw[->] (circv) to (v);

\draw[->] (Psie) to node[above] {$\hat{\bm{\Psi}}_\mathrm{e}$} ($(Psie) + (\dist,0)$);

\end{tikzpicture}
	\caption{Inputs and outputs of the feedforward cylinder controller, needed for the predictive nature of the control scheme. By construction, all inputs and outputs are vectors of size $N$.}
	\label{fig:PCCIO}
\end{figure}
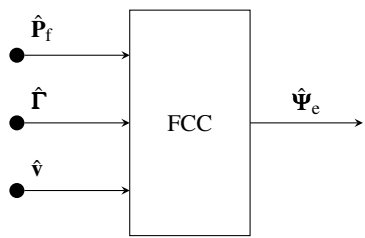

\begin{figure}
	\centering
	\definecolor{mycolor1}{rgb}{0.00000,0.44700,0.74100}%
\definecolor{mycolor2}{rgb}{0.85000,0.32500,0.09800}%
\definecolor{mycolor3}{rgb}{0.92900,0.69400,0.12500}%
\definecolor{mycolor4}{rgb}{0.49400,0.18400,0.55600}%
\definecolor{mycolor5}{rgb}{0.46600,0.67400,0.18800}%
\definecolor{mycolor6}{rgb}{0.30100,0.74500,0.93300}%
\begin{tikzpicture}[trim axis left, trim axis right, >=stealth, font=\small]
\tikzset{>=stealth}
\def\plotwidth{\columnwidth}%
\def\plotheight{0.75\columnwidth}%

\begin{axis}[%
width=\plotwidth,
height=\plotheight,
xmin=0,
xmax=1.22222222222222,
xlabel style={font=\color{white!15!black}},
xlabel={$\dot m_\mathrm{f}$ [-]},
ymin=0,
ymax=0.138888888888889,
ylabel style={font=\color{white!15!black}}, 
ylabel style={at={(0.1,0.5)}},
ylabel={$\dot m_\mathrm{f,cyl}$ [-]},
yticklabels={,,}
axis background/.style={fill=white},
xmajorgrids,
ymajorgrids,
legend style={at={(1.11,0.1)},anchor=south east, legend cell align=center, align=left, draw=white!15!black}
]
\addplot[only marks, mark=*, mark options={fill=mycolor1}, mark size=1pt, draw=mycolor1] table[]{./pictures/Chapter2/picData/mfcyl_Pf-1.tsv};
\addlegendentry{$\Psi_\mathrm{e} = 1$}

\addplot[only marks, mark=*, mark options={fill=mycolor2}, mark size=1pt, draw=mycolor2] table[]{./pictures/Chapter2/picData/mfcyl_Pf-3.tsv};
\addlegendentry{$\Psi_\mathrm{e} = 2$}

\addplot[only marks, mark=*, mark options={fill=mycolor3}, mark size=1pt, draw=mycolor3] table[]{./pictures/Chapter2/picData/mfcyl_Pf-6.tsv};
\addlegendentry{$\Psi_\mathrm{e} = 3$}

\addplot[only marks, mark=*, mark options={fill=mycolor4}, mark size=1pt, draw=mycolor4] table[]{./pictures/Chapter2/picData/mfcyl_Pf-10.tsv};
\addlegendentry{$\Psi_\mathrm{e} = 4$}

\addplot[only marks, mark=*, mark options={fill=mycolor5}, mark size=1pt, draw=mycolor5] table[]{./pictures/Chapter2/picData/mfcyl_Pf-15.tsv};
\addlegendentry{$\Psi_\mathrm{e} = 5$}

\addplot[only marks, mark=*, mark options={fill=mycolor6}, mark size=1pt, draw=mycolor6] table[]{./pictures/Chapter2/picData/mfcyl_Pf-21.tsv};
\addlegendentry{$\Psi_\mathrm{e} = 6$}

\addplot [color=lightgray, forget plot, line width = 0.8]
  table[]{./pictures/Chapter2/picData/mfcyl_Pf-22.tsv};

\addplot [color=lightgray, forget plot, line width = 0.8]
  table[]{./pictures/Chapter2/picData/mfcyl_Pf-23.tsv};

\addplot [color=lightgray, forget plot, line width = 0.8]
  table[]{./pictures/Chapter2/picData/mfcyl_Pf-24.tsv};

\addplot [color=lightgray, forget plot, line width = 0.8]
  table[]{./pictures/Chapter2/picData/mfcyl_Pf-25.tsv};

\addplot [color=lightgray, forget plot, line width = 0.8]
  table[]{./pictures/Chapter2/picData/mfcyl_Pf-26.tsv};

\addplot [color=lightgray, forget plot, line width = 0.8]
  table[]{./pictures/Chapter2/picData/mfcyl_Pf-27.tsv};
  
\addplot[dashed, red, samples=2, line width=1.5] {0.09};
\addlegendentry{$\dot{m}_\mathrm{f,cyl}^\mathrm{max}(\omega_{\mathrm{e}})$}

\draw [mycolor6, fill=mycolor6, text=white] (axis cs:0.6,0) rectangle (axis cs:1.22222222222222,0.005);
\draw [mycolor5, fill=mycolor5, text=white] (axis cs:0.48,0.005) rectangle (axis cs:1,0.01);
\draw [mycolor4, fill=mycolor4, text=white] (axis cs:0.36,0.01) rectangle (axis cs:0.81,0.015);
\draw [mycolor3, fill=mycolor3, text=white] (axis cs:0.25,0.015) rectangle (axis cs:0.61,0.020);
\draw [mycolor2, fill=mycolor2, text=white] (axis cs:0.15,0.020) rectangle (axis cs:0.4,0.025);
\draw [mycolor1, fill=mycolor1, text=white] (axis cs:0.08,0.025) rectangle (axis cs:0.2,0.03);

\draw [->, dashed, black, line width = 1] (axis cs: 0.62,0) -- (axis cs: 0.62,0.09);

\end{axis}

\end{tikzpicture}%
	\caption{Cylinder fuel mass flow as a function of the total fuel mass flow for different numbers of active cylinders. The points are many solutions of Problem \ref{prob:ocpBalerna} with various energy targets. The solid lines represent \eqref{eq:mf}. All variables are normalized for confidentiality reasons.}
	\label{fig:mfCyl}\vspace{-0.3cm}
\end{figure}
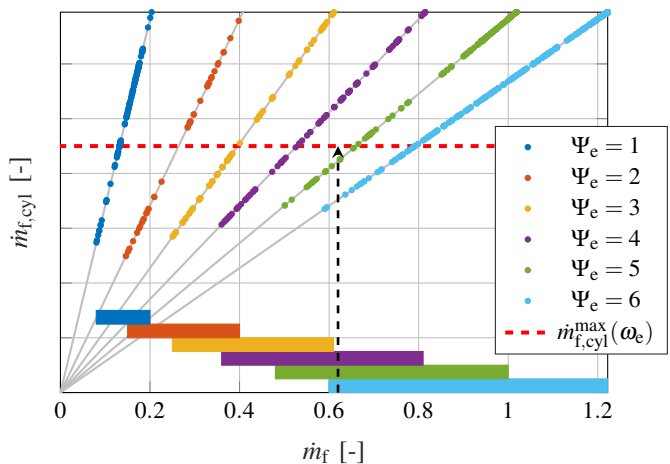

\begin{figure}
	\centering
	\begin{tikzpicture}[trim axis left, trim axis right, >=stealth, font=\small]

\def\plotwidth{\columnwidth}%
\def\plotheight{0.55\columnwidth}%

\begin{axis}[%
width=\plotwidth,
height=\plotheight,
xmin=1495,
xmax=1605,
xlabel style={font=\color{white!15!black}},
xlabel={Position [m]},
xticklabels={, ,1500, , , , , 1600},
ymin=-0.15,
ymax=6.15,
ytick={0,1,2,3,4,5,6},
yticklabels={0, 1, 2, 3, 4, 5, 6},
ylabel style={font=\color{white!15!black}},
ylabel style={at={(0.08,0.5)}},
ylabel={$\Psi_\mathrm{e}$ [-]},
axis background/.style={fill=white},
xmajorgrids,
ymajorgrids,
legend style={at={(0.96,0.2)},anchor=south east, legend cell align=left, align=left, draw=white!15!black}
]
\addplot [color=black, dashed, line width=1.0pt]
  table[]{./pictures/Chapter2/picData/CylinderTunnel-1.tsv};
  \addlegendentry{$\Psi^\mathrm{min}_\mathrm{e}$}
  
\addplot [color=black, line width=1.0pt]
  table[]{./pictures/Chapter2/picData/CylinderTunnel-2.tsv};
  \addlegendentry{$\Psi^\mathrm{max}_\mathrm{e}$}
    
\addplot [color=black, dotted, line width=1.0pt, forget plot]
  table[]{./pictures/Chapter2/picData/CylinderTunnel-3.tsv};
  
\addplot [color=black, dotted, line width=1.0pt, forget plot]
  table[]{./pictures/Chapter2/picData/CylinderTunnel-4.tsv};
  
\addplot[area legend, draw=none, fill=white!65!black, fill opacity=0.4, forget plot]
table[] {./pictures/Chapter2/picData/CylinderTunnel-5.tsv}--cycle;

\end{axis}

\end{tikzpicture}%
	\caption{Maximum and minimum number of active cylinders for a portion of track. The instant where both trajectories leave 0 is the apex of the corner. The gray-shaded area represents the grip-limited region.}
	\label{fig:PCCTunnel}\vspace{-0.3cm}
\end{figure}
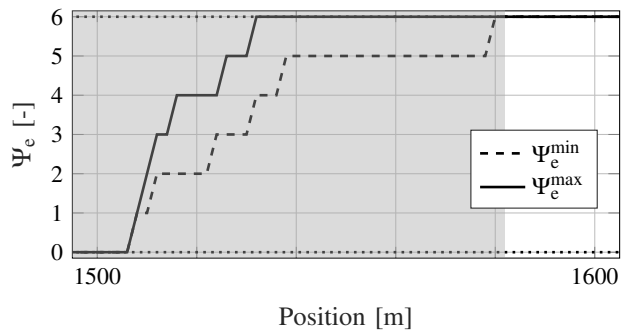

The V6 internal combustion engine of the F1 car features the ability of deactivating single cylinders individually. 
As introduced in \eqref{eq:psidef} we include this degree of freedom by defining the number of active cylinders with the integer variable $\Psi_{\mathrm{e}}$.
However, since the online computation of a mixed-integer \gls{acr:nmpc} is not tractable in a sensible time frame, we employ an algebraic subsystem that computes the number of active cylinders from a range of feasible solutions in a feedforward fashion.
The inputs and outputs of the \gls{acr:fcc} are shown in Fig. \ref{fig:PCCIO}.
In order to define a range of feasible $\Psi_{\mathrm{e}}$, we recall the linear relationship between cylinder fuel mass flow and total fuel mass flow shown in \eqref{eq:mf}.
The total fuel mass flow $\dot{m}_\mathrm{f}$ indicates how much fuel power $P_\mathrm{f}$ can be supplied to the \gls{acr:ICE}, as modeled in \eqref{eq:fuelpower}.
In Fig. \ref{fig:mfCyl} we plot this linear dependency augmenting it with optimization results stemming from Problem \ref{prob:ocpBalerna}, solved with many different energy budget targets and over various track intervals.
We recognize that there exists a clearly defined range of $\dot{m}_\mathrm{f}$, and therefore of $P_\mathrm{f}$, for each number of active cylinders (indicated with the colored bars).
This implies that a set of active cylinders can be inferred from a given $P_\mathrm{f}$ requested by the \gls{acr:eltms}.
As an example, $\dot{m}_\mathrm{f}=0.62$ leads to three possible active cylinders, i.e., ${\Psi_\mathrm{e} = \{4, 5, 6\}}$.
However, from Section \ref{subsec:ice} we know that the control input $\dot{m}_\mathrm{f,cyl}$ is limited by \eqref{eq:fialimit}.
This additional constraint decreases the possible number of active cylinders for $\dot{m}_\mathrm{f}=0.62$ at a given velocity and a certain gear ratio to ${\Psi_\mathrm{e} = \{5, 6\}}$.
By repeating this procedure over the whole prediction we obtain various feasible trajectories.
In Fig. \ref{fig:PCCTunnel} we show two possibilities for a portion of lap: the trajectory of the maximum $\Psi^\mathrm{max}_\mathrm{e}$ and minimum $\Psi^\mathrm{min}_\mathrm{e}$ number of feasible active cylinders. 
We see how at some indices in the acceleration phase up to three possible values arise.
For the prediction we opt for the maximum strategy, i.e.,
\begin{equation} \label{eq:Psimax}
	\hat{\Psi}_{\mathrm{e}}(s) = \hat{\Psi}^\mathrm{max}_{\mathrm{e}}(s),
\end{equation}
because it provides robustness advantages whilst keeping the computational effort very low.  
In particular, choosing more active cylinders than optimal results in a higher fuel power according to \eqref{eq:mf} and \eqref{eq:fuelpower}. 
However, this can be compensated by decreasing the spark advance efficiency $u_\mathrm{sa}$ in the \gls{acr:nmpc}, allowing to regulate the amount of traction power according to \eqref{eq:Pecomb}.
The resulting wasted power can be partially recuperated by the MGU-H.
On the other hand, less than optimal active cylinders might lead to feasibility issues.
In fact, given that $\Psi_{\mathrm{e}}$ directly influences the achievable traction power, less active cylinders might prevent the power matching in grip-limited regions described by \eqref{eq:GLreq} in cases where the MGU-K is already operating at its maximum bound.

\subsection{Nonlinear Model Predictive Control (NMPC)}\label{subsec:nmpc}
With the number of active cylinders and the engaged gear having already been determined, in this section we formulate the \gls{acr:nmpc} that optimizes the remaining inputs to the race car model gathered in the input vector $\vec{u}$ defined in \eqref{eq:inputs}.
To correctly set up the objective of the optimization, we need to address the two track regions introduced in Section \ref{sec:driver} differently.
Therefore, in the following we present two distinct structures and point out their main assumptions. 
The power-limited \gls{acr:nmpc} is used as long as there is at least one power-limited point (i.e., at least one GL index is 0) in the \gls{acr:nmpc} horizon.
This means that on a straight, at a corner entrance, and at a corner exit, the power-limited \gls{acr:nmpc} is used, while the grip-limited \gls{acr:nmpc} is employed only during a corner, i.e., when all GL indices are equal to 1.
Recall Section \ref{subsec:notation} for the index convention throughout the online control.

\subsubsection{Power-limited Region -- Time-optimal Control}\label{subsubsec:PLController}
The \gls{acr:nmpc} for the power-limited regions reads as follows:
\begin{problem}\label{prob:PLNMPC}
	In power-limited regions, the lap time optimal low-level control sequence $\vec{u}^\star[\tilde{i}]$ defined by the entries of $\vec{u}$ at the current step $i=\tilde{i}$ is the solution of
	\begin{equation*}
		\min_{\vec{u},\bm{\varepsilon}} \quad J_\mathrm{time} + \vec{R}^\top \cdot \bm{\varepsilon} + w_{E_\mathrm{f}}\cdot \sum_{k=0}^{N-1}(\hat{P}_{\mathrm{f}}[k] - P_{\mathrm{f}}[\tilde{i}+k])^2
	\end{equation*}
	subject to the following constraints:
	\begin{alignat}{2}
		& \text{Term. Constr.:} \quad &&{E}_{\mathrm{b}}[N-1] \geqslant E_{\mathrm{b}}^\mathrm{nom}[\tilde{i}+N-1] - {\varepsilon}_{E_\mathrm{b}}, \label{eq:PLNLP1}\\
		& \quad &&p_{\mathrm{im}}[N-1] \geqslant p_{\mathrm{im}}^\mathrm{nom}[\tilde{i}+N-1]- {\varepsilon}_{p_\mathrm{im}}, \label{eq:PLNLP2}\\
		& \quad &&E_{\mathrm{tc}}[N-1] \geqslant E_{\mathrm{tc}}^\mathrm{nom}[\tilde{i}+N-1]- {\varepsilon}_{E_\mathrm{tc}}, \label{eq:PLNLP3}\\
		& \text{Slack Variables:} \quad &&\bm{\varepsilon} \geqslant 0, \label{eq:PLNLP4}\\
		& \text{ELTMS Eq.:} \quad &&u_{\mathrm{wg}}[k] =  \hat{u}_{\mathrm{wg}}[k], \label{eq:PLNLP6}\\
		& &&P_{\mathrm{k}}[k] = \hat{P}_{\mathrm{k}}[k], \label{eq:PLNLP7}\\
		& \text{Grip-limited:} \quad &&P_{\mathrm{trac}}[k] = {\hat P}_{\mathrm{req}}[k] \quad \text{if }\hat{b}_\mathrm{GL}[k]=1, \label{eq:PLNLP5}\\
        & \text{model equations a} && \text{ccording to Problem \ref{prob:ocpBalerna}.}\nonumber
    \end{alignat}
\end{problem}
In power-limited regions the car is solely limited by the maximum power output.
These sections represent the only intervals of the track where it is possible to directly gain lap time through the operation of the power unit because we are not hitting the maximum velocity profile.
From a pure lap time point of view, it is therefore optimal to maximize the velocity at each step $k$, translating into
\begin{equation}\label{eq:timeopt}
	J_\mathrm{time} = \sum_{k=0}^{N-1} \frac{\Delta s}{v[k]},
\end{equation}
where $\Delta s$ is the step length.
However, given the importance of meeting the energy targets over the complete lap, we augment the \gls{acr:nmpc} with further components.
In addition to the time minimization, the objective also includes the slack variables
\begin{equation}\label{eq:slack}
	\bm{\varepsilon} = \begin{bmatrix}
		\varepsilon_{E_\mathrm{b}} & \varepsilon_{p_\mathrm{im}} & \varepsilon_{E_\mathrm{tc}}
	\end{bmatrix}^\top,
\end{equation}
and their weights 
\begin{equation}\label{eq:slackweight}
	\vec{R} = \begin{bmatrix}
		w_{E_\mathrm{b}} & w_{p_\mathrm{im}} & w_{E_\mathrm{tc}}
	\end{bmatrix}^\top.
\end{equation}
The slack variables, which can take on only positive values, are needed to introduce the unilateral soft constraints on the system's energy reservoirs shown in \eqref{eq:PLNLP1} $-$ \eqref{eq:PLNLP4}.
For example, if we have consumed more battery energy than the nominal trajectory at the end of the horizon, we penalize the deviation $\varepsilon_{E_\mathrm{b}} = E_{\mathrm{b}}^\mathrm{nom}[\tilde{i}+N-1] - {E}_{\mathrm{b}}[N-1]$ with $w_{E_\mathrm{b}}$.
If instead we saved some battery energy, we do not influence the objective as $\varepsilon_{E_\mathrm{b}}$ takes on the value 0.

Given the model mismatch between \gls{acr:nmpc} and \gls{acr:eltms}, it is not possible to impose an equality constraint on the fuel power.
Therefore, we introduce a fuel power reference tracking component in the last term of the objective, weighted by $w_{E_\mathrm{f}}$.
Next, in \eqref{eq:PLNLP6} and \eqref{eq:PLNLP7} we see that the MGU-K power and the waste-gate position are taken from the \gls{acr:eltms}.
This is done to avoid aggressive action by the short-sighted low-level \gls{acr:nmpc}.
As an example, at the beginning of a straight, changing the MGU-K power from positive (boosting) to negative (recuperating) could fulfill the desired battery target at the end of the horizon, although being highly suboptimal from a lap time point of view.
Finally, we need to account for the possibility that some points inside the horizon are already in the grip-limited region.
To correctly model this effect we include \eqref{eq:PLNLP5}.

\subsubsection{Grip-limited Region -- Power Request Realization}\label{subsubsec:GLController}
The nonlinear model predictive controller for the grip-limited regions reads as follows:
\begin{problem}\label{prob:GLNMPC}
	In grip-limited regions, the lap time optimal low-level control sequence $\vec{u}^\star[\tilde{i}]$ defined by the entries of $\vec{u}$ at the current step $i=\tilde{i}$ is the solution of
	\begin{equation*}
		\min_{\vec{u},\bm{\varepsilon}} \quad \vec{R}^\top \cdot \bm{\varepsilon} + w_{E_\mathrm{f}}\cdot  \sum_{k=0}^{N-1}(\hat{P}_{\mathrm{f}}[k] - P_{\mathrm{f}}[\tilde{i}+k])^2
	\end{equation*}
	subject to the following constraints:
	\begin{alignat}{2}
		& \text{Term. Constr.:} \quad && {E}_{\mathrm{b}}[N-1] \geqslant E_{\mathrm{b}}^\mathrm{nom}[\tilde{i}+N-1] - {\varepsilon}_{E_\mathrm{b}}, \label{eq:GLNLP1}\\
		& \quad && p_{\mathrm{im}}[N-1] \geqslant p_{\mathrm{im}}^\mathrm{nom}[\tilde{i}+N-1]- {\varepsilon}_{p_\mathrm{im}}, \label{eq:GLNLP2}\\
		& \quad && E_{\mathrm{tc}}[N-1] \geqslant E_{\mathrm{tc}}^\mathrm{nom}[\tilde{i}+N-1]- {\varepsilon}_{E_\mathrm{tc}}, \label{eq:GLNLP3}\\
		& \text{Slack Variables:} \quad && \bm{\varepsilon} \geqslant 0, \label{eq:GLNLP4}\\
		& \text{ELTMS Eq.:} \quad && u_{\mathrm{wg}}[k] =  \hat{u}_{\mathrm{wg}}[k], \label{eq:GLNLP5}\\
		& \text{Grip-limited:} \quad && P_{\mathrm{trac}}[k] = {\hat P}_{\mathrm{req}}[k], \label{eq:GLNLP6}\\
		& \text{model equations a} && \text{ccording to Problem \ref{prob:ocpBalerna}.}\nonumber
	\end{alignat}
\end{problem}
While cornering we cannot directly improve the lap time since we are following the maximum velocity constraint, i.e., \eqref{eq:vmaxconstraint} holds with equality.
Accordingly, the time component \eqref{eq:timeopt} is not included in the objective of the grip-limited \gls{acr:nmpc} anymore.
The soft constraints in \eqref{eq:GLNLP1} $-$ \eqref{eq:GLNLP4} stay the same, as well as the waste-gate operation.
In contrast to the power-limited \gls{acr:nmpc}, the MGU-K operation is now freely optimized by the low-level controller to avoid infeasibilities.
This is done for the following reason: Since in grip-limited regions we have to fulfill the power requested by the driver (see \eqref{eq:GLNLP6}) and the fuel power command $\hat{P}_\mathrm{f}$ from the \gls{acr:eltms} might not respect the \gls{acr:FIA} engine-speed dependent fuel limit or the \gls{acr:ICE} dynamics, the MGU-K power can differ from the desired ELTMS value $\hat{P}_\mathrm{k}$.
The MGU-K works as a buffer that can be adapted quickly to respond to situations where the internal combustion engine power is limited (e.g., by the fuel flow limit, the maximum achievable efficiency or the number of active cylinders).

Finally, we discuss the choice of $\Delta s$ and the \gls{acr:nmpc} horizon length.
To properly capture the relevant low-level dynamics, a minimum update frequency of \SI{10}{Hz} is necessary.
In space domain this translates to a maximum step length determined at the critical point of lowest corner velocity.
With $v^{\mathrm{min}}\approx\SI{70}{kph}$ we derive a fixed step length of $\Delta s = \SI{2}{m}$.
Moreover, to keep the average computational time within these \SI{0.1}{s}, the horizon length is set to 5 steps (resulting in an \SI{8}{m} horizon).
\section{Results}\label{sec:results}
In this section we present the simulation results and assess the performance of the control architecture presented in Section \ref{sec:controller} in the presence of disturbances.
First, in Section \ref{subsec:differingGS}, we consider a modified gear trajectory representing a driver that shifts differently than expected.
Such an investigation has high practical relevance due to the fact that the driver can never exactly reproduce a precomputed gearshift strategy.
Second, in Section \ref{subsec:trajectorydeviation}, we increase the maximum velocity profile in the corner by \SI{5}{\percent} to emulate new tires with better grip.
Both disturbances were introduced only over a portion of lap on the Bahrain International Circuit.
To facilitate the analysis, also the evaluation of the low-level trajectories is done only over the same portion of lap.
However, the optimization and the online simulation have been performed over the complete lap.
The online control results are obtained with the simulation environment presented in Fig. \ref{fig:Overview} and compared to the benchmark introduced in Section \ref{sec:offlineoptimization}.
These trajectories are computed a posteriori to exactly match the energy consumption obtained online.
The lap time difference between this non-causal solution and the online simulation represents our performance metric: the lap time suboptimality over a full lap.
For both case studies we illustrate the suboptimalities, where we also included an online scenario with full knowledge of the disturbance and an equivalent case study without cylinder deactivation capability.

\subsection{Differing Gearshift Scenario}\label{subsec:differingGS}
In this case study we want to assess the robustness of the low-level \gls{acr:nmpc} by modifying one of the driver's actuators.
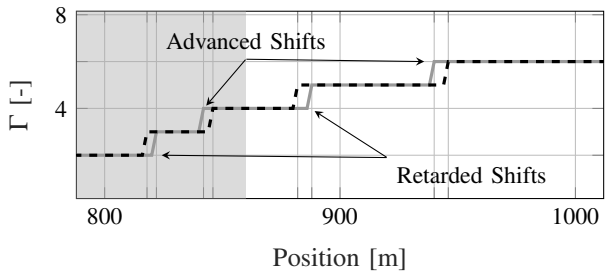
\begin{figure}
	\centering
	\def\dCirc{0.0cm}
\tikzstyle{dot} = [draw, circle, minimum size=\dCirc, fill = black, inner sep=0cm]

\begin{tikzpicture}[trim axis left, trim axis right, >=stealth, font=\small]

\def\plotwidth{0.97\columnwidth}%
\def\plotheight{0.46\columnwidth}%
\def\yshift{-0.1cm}%
\def\xmin{788}
\def\xmax{1012}

\begin{axis}[%
name=one,
width=\plotwidth,
height=\plotheight,
xmin=\xmin,
xmax=\xmax,
xtick={780,800, 818, 822, 842, 846, 882, 888, 900, 940, 946, 1000, 1068, 1078, 1200,1400,1600,1780},
xticklabels={{},{800},{},{},{},{},{},{},{900},{},{},{1000},{},{},{},{},{1600},{}},
xlabel style={font=\color{white!15!black}},
xlabel={Position [m]},
ymin=0.15,
ymax=8.15,
ytick={0,2,4,6,8},
yticklabels={{0},{},{4},{},{8}},
ylabel style={font=\color{white!15!black},at={(0.07,0.5)}},
ylabel={$\Gamma$ [-]},
axis background/.style={fill=white},
xmajorgrids,
ymajorgrids,
legend style={at={(0.3,0.05)}, anchor=south west, legend cell align=left, align=left, draw=white!15!black}
]
\addplot[area legend, draw=none, fill=white!65!black, fill opacity=0.4, forget plot]
table[] {./pictures/Chapter3/picData/CaseStudy1a_final-3.tsv}--cycle;

\addplot [color=white!60!black, line width=1.2pt]
  table[]{./pictures/Chapter3/picData/CaseStudy1a_final-1.tsv};

\addplot [color=black, dashed, line width=1.2pt]
  table[]{./pictures/Chapter3/picData/CaseStudy1a_final-2.tsv};

\node[dot, label=below right:\small{\textcolor{black}{Retarded Shifts}}] at (axis cs:920,1.9) (source1) {};
\node[dot, label=above:\small{\textcolor{black}{Advanced Shifts}}] at (axis cs:860,6.1) (source2) {};
\node[dot] at (axis cs:889,3.9) (destination1) {};
\node[dot] at (axis cs:824,2) (destination3) {};
\node[dot] at (axis cs:843,4.1) (destination4) {};
\node[dot] at (axis cs:936,6) (destination5) {};

\draw[->, color=black] (source1) to (destination1);
\draw[->, color=black] (source1) to (destination3);
\draw[->, color=black] (source2) to (destination4);
\draw[->, color=black] (source2) to (destination5);
\end{axis}
\end{tikzpicture}%
	\caption{Gearshift trajectories that are fed to the control architecture. In gray the disturbed sequence $\bm{\Gamma}^\mathrm{act}$ given by the driver and in dashed black the nominal one $\bm{\Gamma}^\mathrm{nom}$ fed to the estimator and known by the control architecture. The gray-shaded area represents the grip-limited region. To help the reader, vertical lines are shown at the instants of gear shifting.}
	\label{fig:CS1a}\vspace{-0.3cm}
\end{figure}
In particular, the driver's gear choice $\bm{\Gamma}^\mathrm{act}$ in \Cref{alg:driver} is modified without altering the nominal trajectory $\bm{\Gamma}^\mathrm{nom}$ known by the estimator.
As shown in Fig. \ref{fig:CS1a}, we introduce two advanced and two retarded gearshifts.
These disturbances occur at the beginning of the straight, where the impact on lap time is greater.
The advances and delays are set at variable spacing, reaching from \SI{6}{m} too early to \SI{10}{m} too late.
As anticipated in the introduction to the case studies, they capture the imperfection of the human agent in the control loop. Naturally, a person cannot perfectly match a precomputed gearshift strategy and thus the timing of the shifts can differ by some tenths of a second. 

\begin{figure}
	\centering
	\def\dCirc{0.0cm}
\tikzstyle{dot} = [draw, circle, minimum size=\dCirc, fill = black, inner sep=0cm]

\begin{tikzpicture}[trim axis left, trim axis right, >=stealth, font=\small]
	
\def\plotwidth{0.95\columnwidth}%
\def\plotheight{0.36\columnwidth}%
\def\yshift{0.05cm}%
\def\xmin{788}
\def\xmax{1012}

\begin{axis}[%
name=two,
width=\plotwidth,
height=\plotheight,
xmin=\xmin,
xmax=\xmax,
xtick={780,800, 818, 822, 842, 846, 882, 888, 900, 940, 946, 1000, 1068, 1078, 1200,1400,1600,1780},
xticklabels={{}},
ymin=94,
ymax=300,
ytick={100,200,300,400},
yticklabels={{},{},{300},{}},
ylabel style={font=\color{white!15!black},at={(0.06,0.5)}},
ylabel={$v$ [kph]},
axis background/.style={fill=white},
xmajorgrids,
ymajorgrids,
legend style={legend cell align=left, align=left, draw=white!15!black}
]
\addplot[area legend, draw=none, fill=white!65!black, fill opacity=0.4, forget plot]
table[] {./pictures/Chapter3/picData/CaseStudy1b_final-5.tsv}--cycle;

\addplot [color=white!60!black, line width=1.2pt]
  table[]{./pictures/Chapter3/picData/CaseStudy1b_final-1.tsv};

\addplot [color=black, dotted, line width=0.5pt, forget plot]
  table[]{./pictures/Chapter3/picData/CaseStudy1b_final-2.tsv};
\addplot [color=black, line width=0.5pt]
  table[]{./pictures/Chapter3/picData/CaseStudy1b_final-3.tsv};

\end{axis}

\begin{axis}[%
name=three,
width=\plotwidth,
height=\plotheight,
at=(two.below south west),
yshift=\yshift,
anchor=north west,
xmin=\xmin,
xmax=\xmax,
xtick={780,800, 818, 822, 842, 846, 882, 888, 900, 940, 946, 1000, 1068, 1078, 1200,1400,1600,1780},
xticklabels={{}},
ymin=0.7,
ymax=1.15,
ytick={0.9130,1.2},
yticklabels={{1},{}},
ylabel style={font=\color{white!15!black},at={(0.06,0.5)}},
ylabel={$\omega_{\mathrm{e}}$ [-]},
axis background/.style={fill=white},
xmajorgrids,
ymajorgrids,
legend style={legend cell align=left, align=left, draw=white!15!black}
]
\addplot[area legend, draw=none, fill=white!65!black, fill opacity=0.4, forget plot]
table[] {./pictures/Chapter3/picData/CaseStudy1b_final-11.tsv}--cycle;

\addplot [color=white!60!black, line width=1.2pt]
  table[]{./pictures/Chapter3/picData/CaseStudy1b_final-9.tsv};

\addplot [color=black, line width=0.5pt]
  table[]{./pictures/Chapter3/picData/CaseStudy1b_final-10.tsv};

\draw [color=black, dashed, line width=1pt] (axis cs:780,0.9130) -- (axis cs:1100,0.9130);
\end{axis}

\begin{axis}[%
name=four,
width=\plotwidth,
height=\plotheight,
at=(three.below south west),
yshift=\yshift,
anchor=north west,
xmin=\xmin,
xmax=\xmax,
xtick={780,800, 818, 822, 842, 846, 882, 888, 900, 940, 946, 1000, 1068, 1078, 1200,1400,1600,1780},
xticklabels={{}},
ymin=0.69,
ymax=0.91,
ytick={0.2,0.4,0.6,0.8,1},
yticklabels={{},{},{},{1},{}},
ylabel style={font=\color{white!15!black},at={(0.06,0.5)}},
ylabel={$p_{\mathrm{im}}$ [-]},
axis background/.style={fill=white},
xmajorgrids,
ymajorgrids,
legend style={at={(0.751,0.7)}, anchor=south west, legend cell align=left, align=left, draw=white!15!black}
]
\addplot[area legend, draw=none, fill=white!65!black, fill opacity=0.4, forget plot]
table[] {./pictures/Chapter3/picData/CaseStudy1b_final-18.tsv}--cycle;

\addplot [color=white!60!black, line width=1.2pt, forget plot]
  table[]{./pictures/Chapter3/picData/CaseStudy1b_final-15.tsv};

\addplot [color=black, line width=0.5pt, forget plot]
  table[]{./pictures/Chapter3/picData/CaseStudy1b_final-16.tsv};
  
\addplot [color=white!65!black, dashed]
table[]{./pictures/Chapter3/picData/CaseStudy1b_final-17.tsv};
\addlegendentry{$p_\mathrm{surge}$}
\end{axis}

\begin{axis}[%
name=six,
width=\plotwidth,
height=\plotheight,
at=(four.below south west),
yshift=\yshift,
anchor=north west,
xmin=\xmin,
xmax=\xmax,
xtick={780,800, 818, 822, 842, 846, 882, 888, 900, 940, 946, 1000, 1068, 1078, 1200,1400,1600,1780},
xticklabels={{}},
ymin=0.75,
ymax=0.86,
ytick={0.8},
yticklabels={{1}},
ylabel style={font=\color{white!15!black},at={(0.06,0.5)}},
ylabel={$\omega_{\mathrm{tc}}$ [-]},
axis background/.style={fill=white},
xmajorgrids,
ymajorgrids,
legend style={legend cell align=left, align=left, draw=white!15!black}
]
\addplot[area legend, draw=none, fill=white!65!black, fill opacity=0.4, forget plot]
table[] {./pictures/Chapter3/picData/CaseStudy1b_final-30.tsv}--cycle;

\addplot [color=white!60!black, line width=1.2pt]
  table[]{./pictures/Chapter3/picData/CaseStudy1b_final-28.tsv};

\addplot [color=black, line width=0.5pt]
  table[]{./pictures/Chapter3/picData/CaseStudy1b_final-29.tsv};

\end{axis}

\begin{axis}[%
name=seven,
width=\plotwidth,
height=\plotheight,
at=(six.below south west),
yshift=\yshift,
anchor=north west,
xmin=\xmin,
xmax=\xmax,
xtick={780,800, 818, 822, 842, 846, 882, 888, 900, 940, 946, 1000, 1068, 1078, 1200,1400,1600,1780},
xticklabels={{}},
ymin=-1.1,
ymax=1.1,
ytick={-1,-0.5,0,0.5,1},
yticklabels={{$-1$},{},{0},{},{1}},
ylabel style={font=\color{white!15!black},at={(0.06,0.5)}},
ylabel={$P_{\mathrm{h}}$ [-]},
axis background/.style={fill=white},
xmajorgrids,
ymajorgrids,
legend style={at={(0.4,0.05)}, anchor=south west, legend cell align=left, align=left, draw=white!15!black}
]
\addplot[area legend, draw=none, fill=white!65!black, fill opacity=0.4, forget plot]
table[] {./pictures/Chapter3/picData/CaseStudy1b_final-36.tsv}--cycle;

\addplot [color=white!60!black, line width=1.2pt]
  table[]{./pictures/Chapter3/picData/CaseStudy1b_final-34.tsv};

\addplot [color=black, line width=0.5pt]
  table[]{./pictures/Chapter3/picData/CaseStudy1b_final-35.tsv};

\end{axis}

\begin{axis}[%
name=eight,
width=\plotwidth,
height=\plotheight,
at=(seven.below south west),
yshift=\yshift,
anchor=north west,
xmin=\xmin,
xmax=\xmax,
xtick={780,800, 818, 822, 842, 846, 882, 888, 900, 940, 946, 1000, 1068, 1078, 1200,1400,1600,1780},
xticklabels={{}},
ymin=-1.1,
ymax=1.1,
ytick={-0.928, -0.465, 0, 0.465, 0.928},
yticklabels={{$-1$},{},{0},{},{1}},
ylabel style={font=\color{white!15!black},at={(0.06,0.5)}},
ylabel={$P_{\mathrm{k}}$ [-]},
axis background/.style={fill=white},
xmajorgrids,
ymajorgrids,
legend style={at={(0.616,0.05)}, anchor=south west, legend cell align=left, align=left, draw=white!15!black}
]
\addplot[area legend, draw=none, fill=white!65!black, fill opacity=0.4, forget plot]
table[] {./pictures/Chapter3/picData/CaseStudy1b_final-42.tsv}--cycle;

\addplot [color=white!60!black, line width=1.2pt]
  table[]{./pictures/Chapter3/picData/CaseStudy1b_final-40.tsv};
\addlegendentry{Online}
\addplot [color=black, line width=0.5pt]
  table[]{./pictures/Chapter3/picData/CaseStudy1b_final-41.tsv};
\addlegendentry{Benchmark}

\end{axis}

\begin{axis}[%
name=nine,
width=\plotwidth,
height=\plotheight,
at=(eight.below south west),
yshift=\yshift,
anchor=north west,
xmin=\xmin,
xmax=\xmax,
xtick={780,800, 818, 822, 842, 846, 882, 888, 900, 940, 946, 1000, 1068, 1078, 1200,1400,1600,1780},
xticklabels={{}},
ymin=0.6,
ymax=1.05,
ytick={0.6,0.8,1},
yticklabels={{},{},{1}},
ylabel style={font=\color{white!15!black},at={(0.06,0.5)}},
ylabel={$P_{\mathrm{e}}$ [-]},
axis background/.style={fill=white},
xmajorgrids,
ymajorgrids,
legend style={at={(0.03,0.03)}, anchor=south west, legend cell align=left, align=left, draw=white!15!black}
]
\addplot[area legend, draw=none, fill=white!65!black, fill opacity=0.4, forget plot]
table[] {./pictures/Chapter3/picData/CaseStudy1b_final-48.tsv}--cycle;

\addplot [color=white!60!black, line width=1.2pt]
  table[]{./pictures/Chapter3/picData/CaseStudy1b_final-46.tsv};

\addplot [color=black, line width=0.5pt]
  table[]{./pictures/Chapter3/picData/CaseStudy1b_final-47.tsv};

\end{axis}

\begin{axis}[%
name=ten,
width=\plotwidth,
height=\plotheight,
at=(nine.below south west),
yshift=\yshift,
anchor=north west,
xmin=\xmin,
xmax=\xmax,
xtick={780,800, 818, 822, 842, 846, 882, 888, 900, 940, 946, 1000, 1068, 1078, 1200,1400,1600,1780},
xticklabels={{},{800},{},{},{},{},{},{},{900},{},{},{1000},{},{},{},{},{1600},{}},
xlabel style={font=\color{white!15!black}},
xlabel={Position [m]},
ymin=0.8,
ymax=1.05,
ytick={1},
yticklabels={{1}},
ylabel style={font=\color{white!15!black},at={(0.06,0.5)}},
ylabel={$\dot{m}_{\mathrm{f}}$ [-]},
axis background/.style={fill=white},
xmajorgrids,
ymajorgrids,
legend style={legend cell align=left, align=left, draw=white!15!black}
]
\addplot[area legend, draw=none, fill=white!65!black, fill opacity=0.4, forget plot]
table[] {./pictures/Chapter3/picData/CaseStudy1b_final-54.tsv}--cycle;

\addplot [color=white!60!black, line width=1.2pt]
  table[]{./pictures/Chapter3/picData/CaseStudy1b_final-52.tsv};

\addplot [color=black, line width=0.5pt]
  table[]{./pictures/Chapter3/picData/CaseStudy1b_final-53.tsv};

\end{axis}

\end{tikzpicture}%
	\caption{State and input trajectories over roughly \SI{200}{m} at the Bahrain International Circuit for the online control and the benchmark. The gray area represents the grip-limited region, the dotted signal in the velocity is the maximum velocity trajectory, the dashed line in the engine speed is set to \SI{10500}{rpm}, and the dashed line in the pressure plot represents the compressor surge pressure. To facilitate the comparison, vertical lines have been added at each gearshift. All variables have been normalized for confidentiality reasons.}
	\label{fig:CS1b}\vspace{-0.3cm}
\end{figure}
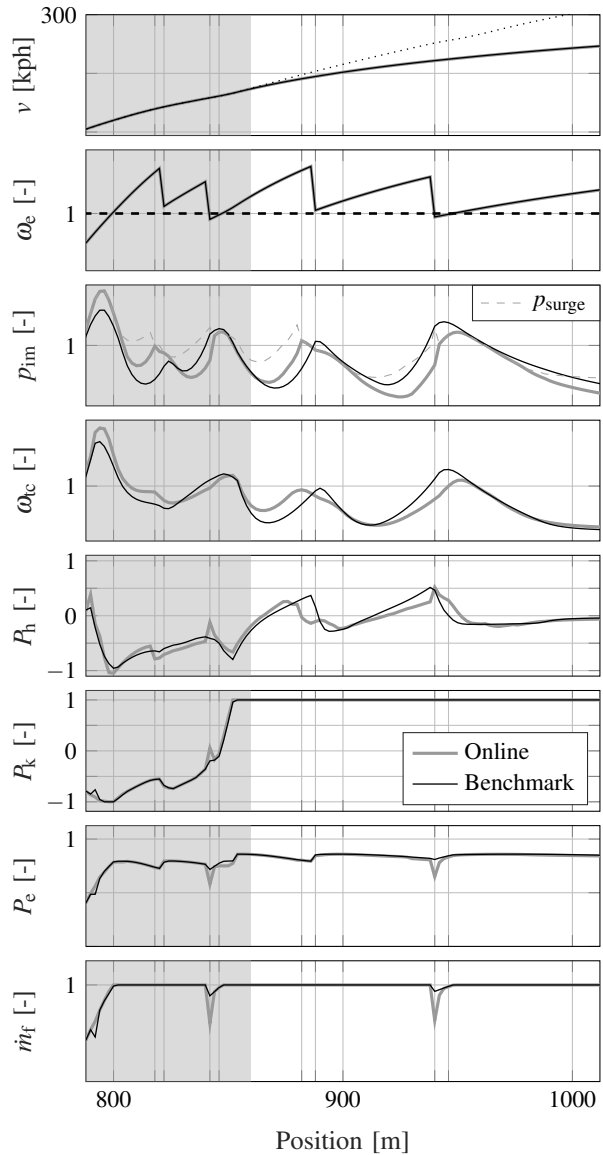
Fig. \ref{fig:CS1b} shows the state and input trajectories for the online control and the benchmark.
The first analysis concerns the \textit{advanced} upshifts, resulting in a sudden and unexpected change of scenario.
As explained in Section \ref{subsec:estimator}, the estimator handles this kind of disturbance by simply adapting future shifts. 
A quantitative illustration of this scenario is shown in the first plot of Fig. \ref{fig:GSScenarios} (advanced up- and downshifts are treated equally).
In the case of an advanced gearshift, it keeps the newly engaged gear constant for the rest of the horizon of the controller. 
From a low-level perspective, the \gls{acr:nmpc} needs to handle a more complex chain of events.
The disturbance introduces an unexpected drop in engine speed $\omega_{\mathrm{e}}$, leading to a sudden decrease of air mass flow entering the cylinders according to \eqref{eq:mbeta}.
To compensate for this deficit, the low-level controller tries to quickly increase the intake manifold pressure $p_\mathrm{im}$ by boosting with the MGU-H connected to the turbocharger, resulting in the visible peaks.
Unfortunately, due to the inertia of the intake manifold this pressure increase does not entirely compensate for the lack of air mass flow instantly.
Additionally, the sudden changing operating conditions of the compressor lead to potential surge issues, meaning that a further intake manifold pressure increase would cause unwanted flow instabilities.
Because in a gasoline engine we cannot differ from the optimal air-to-fuel ratio significantly, the lack of air means that less fuel can be injected, resulting in less engine power $P_\mathrm{e}$.
In the grip-limited region the lack of power is compensated by the MGU-K power $P_\mathrm{k}$.
In the power-limited region where the MGU-K is already at its maximum power, the deficit directly affects the lap time suboptimality $\Delta T$.
In addition to the previously explained pressure deficit, there is another effect decreasing $\dot{m}_\mathrm{f}$.
As can be seen in the benchmark solution, at the instants of advanced upshift even the non-causal solution experiences a small lack of fuel mass flow.
In fact, the engine speed drops below the crucial value of \SI{10500}{rpm}, allowing for even less fuel to be injected according to \eqref{eq:fialimit}.
It is important to point out that this repercussion on $\dot{m}_\mathrm{f}$ cannot be avoided by means of control action in this scenario.
The opposite scenario of \textit{retarded} upshifts poses a different challenge for the controller.
As soon as the estimator realizes that the shift does not occur in the nominal instant but rather later, we enter the retarded upshift scenario.
This implies that the prediction delays the upshift by a fixed amount of engine revolutions, as quantitatively shown in the second plot of Fig. \ref{fig:GSScenarios}.
From a low-level perspective, these disturbances are easier to handle, since the \gls{acr:FIA} fuel mass flow limit is not an issue and all energy adaptations needed at the gearshifts (e.g., increase of the kinetic energy of the turbocharger) are already performed.
The challenge arising is the optimal usage of the energy stored in the reservoirs at the instants where a nominal gearshift is missed.
For example, in the second retarded upshift interval, the turbocharger speed $\omega_\mathrm{tc}$ was raised to increase the intake manifold pressure and react to the gearshift.
However, since the nominal shift does not occur, the rotational kinetic energy in that instant is too high for that unexpected operating condition, resulting in a suboptimal energy surplus.
The \gls{acr:nmpc} stores this excess in the battery by recuperating with the MGU-H.
A similar scenario can be observed in the intake manifold, where the pressure level was raised too early.
The benchmark solution avoids this suboptimal scenario, by retarding the energy adaptations accordingly, representing the optimal solution (without global optimality guarantees since it is an \gls{acr:nlp}).
In Fig. \ref{fig:CS_Sub} we summarize the suboptimalities in terms of lap time over an entire lap lasting about \SI{94}{s}.
The presented control architecture is \SI{49}{ms} suboptimal compared to the benchmark solution.
However, this value also includes the suboptimality due to the tuning of the \gls{acr:pi} controllers inside the \gls{acr:eltms}, responsible for the MGU-K and \gls{acr:ICE} power cuts (see \ref{subsec:eltms}).
By tuning them very aggressively, we would perform power cuts at highly suboptimal track positions to quickly compensate for energy drifts.
Likewise, when relaxing the supervisory controller we could diminish the suboptimality at the cost of losing the ability of drift compensation in a reasonable amount of time.
Therefore, in the presented results, the \gls{acr:eltms} tuning is set such that energy drifts are almost completely compensated after one lap at the latest ($\Delta E_\mathrm{b}\approx\SI{0.002}{MJ}$ and $\Delta E_\mathrm{f}\approx\SI{0.5}{\percent}$), given that we assume each lap to repeat itself.
To assess the effective suboptimality due to the gearshift disturbance, we consider the online simulation with full knowledge, meaning that the first two entries of $\vec{r}^\mathrm{nom}$ in \eqref{eq:rnom} are set equal to $\vec{r}^\mathrm{act}$ from \eqref{eq:ract}.
The difference in suboptimality between the two online simulations is \SI{2}{ms}.
When removing the single cylinder deactivation capability, i.e., either all 6 cylinders are active or the entire engine is switched off (as shown in Fig. \ref{fig:CS2c} for the second case study), the suboptimality rises above \SI{55}{ms}.
Since the suboptimality difference does not increase, this suggests that the disturbance rejection in this scenario is not affected by the cylinder deactivation.
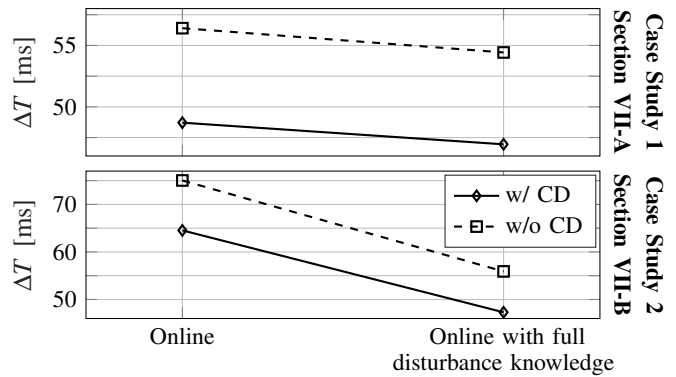
\begin{figure}
	\centering
	\begin{tikzpicture}[trim axis left, trim axis right, >=stealth, font=\small]
\def\plotwidth{0.95\columnwidth}%
\def\plotheight{0.4\columnwidth}%
\def\yshift{0.04cm}%
\def\xmin{0.7}%
\def\xmax{2.3}%

\begin{axis}[%
name=one,
width=\plotwidth,
height=\plotheight,
xmin=\xmin,
xmax=\xmax,
xtick={1,2},
xticklabels={{},{}},
ymin=46,
ymax=58,
ytick={45,47.5,50,52.5,55,57.5,60},
yticklabels={{},{},{50},{},{55},{},{}},
ylabel style={font=\color{white!15!black},at={(0.06,0.5)}},
ylabel={$\Delta T$ [ms]},
axis background/.style={fill=white},
xmajorgrids,
ymajorgrids
]
\addplot [color=black, mark=diamond, mark options={solid, black}, forget plot, line width=0.8pt]
  table[]{./pictures/Chapter3/picData/CaseStudy_Suboptimality_new-1.tsv};
\addplot [color=black, dashed, mark=square, mark options={solid, black}, forget plot, line width=0.8pt]
  table[]{./pictures/Chapter3/picData/CaseStudy_Suboptimality_new-2.tsv};
\end{axis}

\node[rotate=-90] at ($(one.east) + (0.7cm,0)$) {\textbf{Case Study 1}};
\node[rotate=-90] at ($(one.east) + (0.3cm,0)$) {\textbf{Section \ref{subsec:differingGS}}};
 
\begin{axis}[%
name=two,
width=\plotwidth,
height=\plotheight,
at=(one.below south west),
yshift=\yshift,
anchor=north west,
xmin=\xmin,
xmax=\xmax,
xtick={1,2},
xticklabels={{Online},{ Online with full disturbance knowledge}},
xticklabel style={align=center,text width=30mm},
ymin=46,
ymax=77,
ytick={50,55,60,65,70,75},
yticklabels={{50},{},{60},{},{70},{}},
ylabel style={font=\color{white!15!black},at={(0.06,0.5)}},
ylabel={$\Delta T$ [ms]},
axis background/.style={fill=white},
xmajorgrids,
ymajorgrids,
legend style={at={(0.993,0.48)}, anchor=south east, legend cell align=left, align=left, draw=white!15!black}
]
\addplot [color=black, mark=diamond, mark options={solid, black}, line width=0.8pt]
  table[]{./pictures/Chapter3/picData/CaseStudy_Suboptimality_new-3.tsv};
\addlegendentry{w/ CD}

\addplot [color=black, dashed, mark=square, mark options={solid, black}, line width=0.8pt]
  table[]{./pictures/Chapter3/picData/CaseStudy_Suboptimality_new-4.tsv};
\addlegendentry{w/o CD}

\end{axis}
\node[rotate=-90] at ($(two.east) + (0.7cm,0)$) {\textbf{Case Study 2}};
\node[rotate=-90] at ($(two.east) + (0.3cm,0)$) {\textbf{Section \ref{subsec:trajectorydeviation}}};
\end{tikzpicture}%
	\caption{Lap time suboptimalities $\Delta T$ of the online frameworks of both case studies with respect to their benchmark solution over an entire lap. The benchmark solution was computed a posteriori with the same energy budgets obtained in the respective online scenario. The online simulation with full knowledge of the disturbance was obtained by setting the first two entries of $\vec{r}^\mathrm{nom}$ in \eqref{eq:rnom} equal to $\vec{r}^\mathrm{act}$ from \eqref{eq:ract}. For all scenarios also an online simulation without cylinder deactivation (only engine completely on or off possible) was included.}
	\label{fig:CS_Sub}\vspace{-0.1cm}
\end{figure}

\subsection{Race Trajectory Deviation}\label{subsec:trajectorydeviation}
\begin{figure}
	\centering
	\def\dCirc{0.0cm}
\tikzstyle{dot} = [draw, circle, minimum size=\dCirc, fill = black, inner sep=0cm]

\begin{tikzpicture}[trim axis left, trim axis right, >=stealth, font=\small]
	
\def\plotwidth{0.97\columnwidth}%
\def\plotheight{0.45\columnwidth}%
\def\xmin{1100}
\def\xmax{1700}

\begin{axis}[%
width=\plotwidth,
height=\plotheight,
xmin=\xmin,
xmax=\xmax,
xtick={780,800,1000,1200,1400,1600,1780},
xticklabels={{},{800},{},{1200},{1400},{1600},{}},
xlabel style={font=\color{white!15!black}},
xlabel={Position [m]},
ymin=100,
ymax=349,
ytick={100,200,300,400},
yticklabels={{},{200},{300},{}},
ylabel style={font=\color{white!15!black},at={(0.06,0.5)}},
ylabel={$v$ [kph]},
axis background/.style={fill=white},
xmajorgrids,
ymajorgrids,
legend style={legend cell align=left, align=left, draw=white!15!black}
]

\addplot[area legend, draw=none, fill=black, fill opacity=0.4, forget plot]
table[] {./pictures/Chapter3/picData/CaseStudy2a_final-8.tsv}--cycle;

\addplot[area legend, draw=none, fill=white!90!black, fill opacity=0.4, forget plot]
table[] {./pictures/Chapter3/picData/CaseStudy2a_final-6.tsv}--cycle;

\addplot [color=white!60!black, line width=1.2pt]
  table[]{./pictures/Chapter3/picData/CaseStudy2a_final-1.tsv};

\addplot [color=black, line width=0.5pt]
  table[]{./pictures/Chapter3/picData/CaseStudy2a_final-2.tsv};

\addplot [color=black, dotted, line width=0.5pt, forget plot]
  table[]{./pictures/Chapter3/picData/CaseStudy2a_final-3.tsv};
\addplot [color=black, dashed, line width=0.5pt, forget plot]
  table[]{./pictures/Chapter3/picData/CaseStudy2a_final-4.tsv};

\node[dot, label=left:\small{\textcolor{black}{Increased $v_\mathrm{max}^{\mathrm{act}}(s)$}}] at (axis cs:1625,320) (source1) {};
\node[dot, label=left:\small{\textcolor{black}{Nominal $v_\mathrm{max}^{\mathrm{nom}}(s)$}}] at (axis cs:1408,150) (source2) {};
\node[dot] at (axis cs:1675,238) (destination1) {};
\node[dot] at (axis cs:1450,200) (destination2) {};

\draw[->, color=black] (source1) to (destination1);
\draw[->, color=black] (source2) to (destination2);

\end{axis}
\end{tikzpicture}%
	\caption{Velocity trajectories of the online control (gray) and the benchmark solution (black). The dashed signal represents the nominal maximum velocity trajectory $v_\mathrm{max}^\mathrm{nom}$, while the dotted signal is its increased counterpart $v_\mathrm{max}^\mathrm{act}$. The gray areas indicate the grip-limited regions, which begin in two slightly shifted instants: in light gray the online scenario and in darker gray the benchmark.}
	\label{fig:CS2a}\vspace{-0.3cm}
\end{figure}
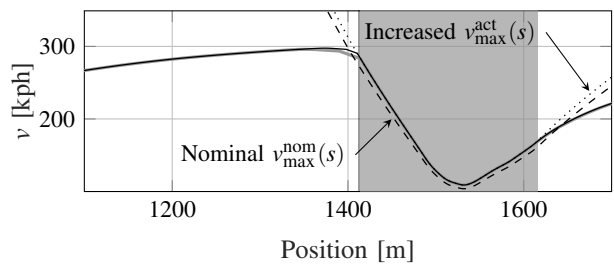

In the second case study we increase the $v_\mathrm{max}$ profile of a corner by \SI{5}{\percent} to emulate higher achievable lateral accelerations, e.g., due to newer tires or more favorable track conditions.
In Fig. \ref{fig:CS2a} we show the velocity trajectories over another interval, including a cornering maneuver.
In gray we depict the online control and in black the benchmark.
Additionally, we include the nominal maximum velocity profile $v_\mathrm{max}^{\mathrm{nom}}$ fed to the estimator and its increased counterpart $v_\mathrm{max}^{\mathrm{act}}$, known and followed by the driver.
In the analysis we focus on the handling of corrupt predictive information given by the estimator and its effects on the supervisory control action.
Moreover, we consider the repercussions of the cylinder deactivation on the low-level \gls{acr:nmpc}.

We can observe how the grip-limited region of the online scenario (in light gray) begins slightly later compared to the benchmark (in darker gray).
This can be affiliated to the estimator, which is only aware of the lower maximum velocity trajectory $v_\mathrm{max}^{\mathrm{nom}}$.
Given that its velocity prediction fulfills \eqref{eq:vmaxconstraint} with equality at an earlier index, the grip-limited region is predicted to start sooner.
This leads to a lower requested power and therefore a lower velocity.
As a result, the effective $v_\mathrm{max}^{\mathrm{act}}$ trajectory is reached later.
\begin{figure}
	\centering
	\input{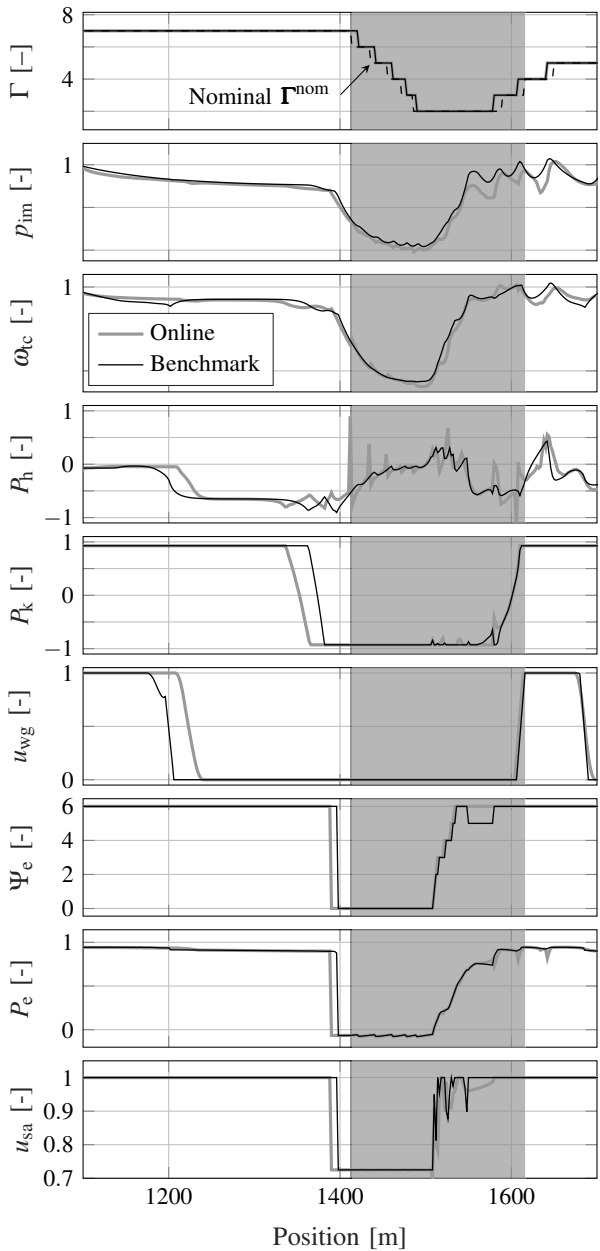}
	\caption{State and input trajectories over roughly \SI{600}{m} at the Bahrain International Circuit for the online control and the benchmark. The gray area represents the grip-limited region. The dotted signal in the velocity is the maximum velocity trajectory. For confidentiality reasons some variables are normalized.}
	\label{fig:CS2b}\vspace{-0.4cm}
\end{figure}
Nevertheless, once the grip-limited region is reached, the corrupted prediction made by the estimator is rejected in a receding horizon fashion, since the requested power $P_\mathrm{req}[i]$ is defined by the driver.
Therefore, the increased maximum velocity profile $v_\mathrm{max}^\mathrm{act}$ is followed even though $v_\mathrm{max}^\mathrm{nom}$ is off.
In Fig. \ref{fig:CS2b} we show additional state and input trajectories.
To explain the MGU-H power peaks optimized by the low-level controller, we need to consider the driver model.
In fact, as it can be observed in the gear trajectory $\Gamma$, from a low-level perspective the disturbance is twofold: Not only the maximum velocity profile, but also the engaged gear differ from nominal.
Accordingly, there is an unexpected decrease in the engine speed, influencing the air path as stated in \eqref{eq:pIM}.
As shown in the previous case study, the electric motor on the turbocharger shaft is able to swiftly compensate such a lack of air in a close-to-optimal fashion.
In the online cylinder activation phase we can observe the influence of the maximum cylinder strategy introduced in \eqref{eq:Psimax} and the robustness of this approach.
When more cylinders are active (between \SI{1500}{m} and \SI{1580}{m}), the MGU-K and the spark advance efficiency are used to match the requested power.
Whenever one of these two actuators is at its saturation, the other one takes over and either increases or decreases the overall power unit power.
It can be noticed that in the regions where the ignition is retarded, the waste-heat recuperation capability by the \gls{acr:ers} is increased.
In Fig. \ref{fig:CS_Sub} we expose the suboptimalities in terms of lap time of this case study.
With respect to the benchmark, the online simulation is \SI{64}{ms} slower, although only \SI{17}{ms} are incurred due to the introduced disturbance.
This aligns with the fact that the disturbance introduced in this case study has a higher impact on the energy budget than the one introduced in Section \ref{subsec:differingGS} owing to its repercussions on MGU-K and MGU-H recuperation.
In fact, even after the disturbance has occurred, the lap time that is lost compared to the benchmark keeps increasing until the end of the lap.
As expected, removing the cylinder deactivation as shown in Fig. \ref{fig:CS2c} results in a suboptimality increase of 7-\SI{8}{ms} in both online scenarios.
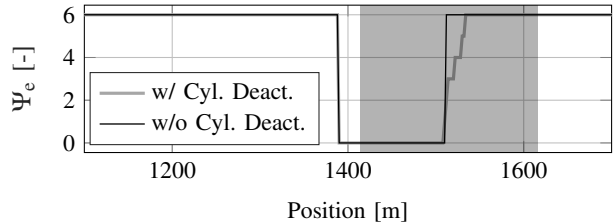
\begin{figure}
	\centering
	\begin{tikzpicture}[trim axis left, trim axis right, >=stealth, font=\small]

\def\plotwidth{0.97\columnwidth}%
\def\plotheight{0.4\columnwidth}%
\def\xmin{1100}
\def\xmax{1700}

\begin{axis}[%
width=\plotwidth,
height=\plotheight,
xmin=\xmin,
xmax=\xmax,
xtick={780,800,1000,1200,1400,1600,1780},
xticklabels={{},{800},{},{1200},{1400},{1600},{}},
xlabel={Position [m]},
ymin=-0.45,
ymax=6.45,
ylabel style={font=\color{white!15!black},at={(0.06,0.5)}},
ylabel={$\Psi_{\mathrm{e}}$ [-]},
axis background/.style={fill=white},
xmajorgrids,
ymajorgrids,
legend style={at={(0.01,0.05)}, anchor=south west, legend cell align=left, align=left, draw=white!15!black}
]
\addplot [color=white!65!black, line width=1.2pt]
  table[]{./pictures/Chapter3/picData/CaseStudy2c-1.tsv};
  \addlegendentry{w/ Cyl. Deact.}
\addplot [color=black, line width=0.5pt]
  table[]{./pictures/Chapter3/picData/CaseStudy2c-2.tsv};
  \addlegendentry{w/o Cyl. Deact.}

\addplot[area legend, draw=none, fill=black, fill opacity=0.3, forget plot]
table[] {./pictures/Chapter3/picData/CaseStudy2c-6.tsv}--cycle;
\end{axis}

\end{tikzpicture}%
	\caption{Number of active cylinders for two online simulations. In gray we show the same as the one analysed in Fig. \ref{fig:CS2a} and Fig. \ref{fig:CS2b}. The black trajectory is the online simulation for a scenario where the engine is either fully running, i.e., all 6 cylinders are active, or the engine is switched off. }
	\label{fig:CS2c}\vspace{-0.3cm}
\end{figure}
\section{Conclusion}\label{sec:conclusion}
In this paper, we presented a low-level online control structure for the F1 hybrid electric vehicle. 
First, we proposed a mathematical model of the powertrain and introduced its offline optimization needed for reference trajectory generation and as benchmark.
Second, we devised a control framework in which we split the high-level control of the slow changing energy budgets, i.e., the fuel and the battery, from the fast changing low-level power unit state variables, i.e., the intake manifold pressure and turbocharger dynamics.
The high-level supervisory controller is based on \gls{acr:pi} controllers that translate online energy budget deviations into time-varying one-dimensional look-up tables, while the low-level controller is based on a track region-dependent nonlinear model predictive controller.
Third, to cope with the computationally expensive and highly undesirable online control of integer decision variables, we define the cylinder deactivation strategy of the engine by means of heuristic look-up tables generated through the analysis of a high number of benchmark solutions. 
Finally, we integrated the control framework in a suitable simulation environment to provide a tool to assess online controllers without incurring major testing costs.
In a first case study, our results showed that the online controller is able to handle disturbances introduced by the driver's gear choice.
The lap time lost over an entire lap including four differing gear shifts is \SI{49}{ms} compared to the benchmark solution. 
However, the nature of this suboptimality lies mainly in the tuning of the high-level \gls{acr:eltms} controller.
When comparing the online simulation with the same control architecture, but with full knowledge of the differing gearshifts, the suboptimality is decreased to only \SI{2}{ms} over the full lap.
In a second case study we analyzed a cornering scenario, where the actual maximum velocity achievable is above the predicted one, e.g., due to varying track conditions.
We were able to show that the online control of the low-level actuators and the cylinder deactivation heuristic were robust and compensated for sudden unexpected changes within the system's constraints. 
The suboptimality reached over a full lap is of \SI{64}{ms} compared to the benchmark, and \SI{17}{ms} compared to the online solution with perfect knowledge.
For both case studies we also included the suboptimalities incurred in a scenario without single cylinder deactivation.
We showed that this degree of freedom decreases the suboptimality by between 7 and \SI{8}{ms} in all scenarios.

Since the prediction of future gearshifts is of crucial importance for the control architecture, further research could focus on its improvement.
Machine learning techniques considering track- and driver-dependent quantities could significantly improve the quality of the overall framework.

\section*{Acknowledgment}
We would like to thank Ferrari S.p.A. for supporting this project. 
Moreover, we would like to express our gratitude to Dr. Ilse New for her helpful and valuable comments during the proofreading phase. 

\bibliographystyle{IEEEtran}
\bibliography{bibtex/Literature}

\begin{thebibliography}{10}
\providecommand{\url}[1]{#1}
\csname url@samestyle\endcsname
\providecommand{\newblock}{\relax}
\providecommand{\bibinfo}[2]{#2}
\providecommand{\BIBentrySTDinterwordspacing}{\spaceskip=0pt\relax}
\providecommand{\BIBentryALTinterwordstretchfactor}{4}
\providecommand{\BIBentryALTinterwordspacing}{\spaceskip=\fontdimen2\font plus
\BIBentryALTinterwordstretchfactor\fontdimen3\font minus
  \fontdimen4\font\relax}
\providecommand{\BIBforeignlanguage}[2]{{%
\expandafter\ifx\csname l@#1\endcsname\relax
\typeout{** WARNING: IEEEtran.bst: No hyphenation pattern has been}%
\typeout{** loaded for the language `#1'. Using the pattern for}%
\typeout{** the default language instead.}%
\else
\language=\csname l@#1\endcsname
\fi
#2}}
\providecommand{\BIBdecl}{\relax}
\BIBdecl

\bibitem{FIA2021Sporting}
FIA, ``2021 {F}ormula one sporting regulations,'' Geneva, Switzerland, Tech.
  Rep., 2020.

\bibitem{FIA2021Technical}
------, ``2021 formula one technical regulations,'' Geneva, Switzerland, Tech.
  Rep., 2020.

\bibitem{hooker1988optimal}
J.~Hooker, ``Optimal driving for single-vehicle fuel economy,''
  \emph{Transportation Research Part A: General}, vol.~22, no.~3, pp. 183--201,
  1988.

\bibitem{perez2009optimal}
L.~V. P{\'e}rez and E.~A. Pilotta, ``Optimal power split in a hybrid electric
  vehicle using direct transcription of an optimal control problem,''
  \emph{Math. Comput. Simul.}, vol.~79, no.~6, pp. 1959--1970, 2009.

\bibitem{heppeler2014fuel}
G.~Heppeler, M.~Sonntag, and O.~Sawodny, ``Fuel efficiency analysis for
  simultaneous optimization of the velocity trajectory and the energy
  management in hybrid electric vehicles,'' \emph{IFAC Proceedings Volumes},
  vol.~47, no.~3, pp. 6612--6617, 2014.

\bibitem{sciarretta2004optimal}
A.~Sciarretta, M.~Back, and L.~Guzzella, ``Optimal control of parallel hybrid
  electric vehicles,'' \emph{IEEE Trans. Control Syst. Technol.}, vol.~12,
  no.~3, pp. 352--363, 2004.

\bibitem{nuesch2014convex}
T.~N{\"u}esch, P.~Elbert, M.~Flankl, C.~Onder, and L.~Guzzella, ``Convex
  optimization for the energy management of hybrid electric vehicles
  considering engine start and gearshift costs,'' \emph{Energies}, vol.~7,
  no.~2, pp. 834--856, 2014.

\bibitem{sciarretta2007control}
A.~Sciarretta and L.~Guzzella, ``Control of hybrid electric vehicles,''
  \emph{IEEE Control syst.}, vol.~27, no.~2, pp. 60--70, 2007.

\bibitem{kim2011optimal}
N.~Kim, S.~Cha, and H.~Peng, ``Optimal control of hybrid electric vehicles
  based on pontryagin's minimum principle,'' \emph{IEEE Trans. Control Syst.
  Technol.}, vol.~19, no.~5, pp. 1279--1287, 2011.

\bibitem{sciarretta2015optimal}
A.~Sciarretta, G.~De~Nunzio, and L.~L. Ojeda, ``Optimal ecodriving control:
  Energy-efficient driving of road vehicles as an optimal control problem,''
  \emph{IEEE Control Syst.}, vol.~35, no.~5, pp. 71--90, 2015.

\bibitem{robuschi2018minimum}
N.~Robuschi, M.~Salazar, P.~Duhr, F.~Braghin, and C.~H. Onder, ``Minimum-fuel
  engine on/off control for the energy management of hybrid electric vehicles
  via iterative linear programming,'' in \emph{\uppercase{IFAC} Symposium on
  Advances in Automotive Control (\uppercase{AAC})}, 2019.

\bibitem{BALERNA2020115248}
C.~Balerna, N.~Lanzetti, M.~Salazar, A.~Cerofolini, and C.~Onder, ``Optimal
  low-level control strategies for a high-performance hybrid electric power
  unit,'' \emph{Applied Energy}, vol. 276, 2020, {A}rt. no. 115248.

\bibitem{ritzmann2019}
J.~Ritzmann, A.~Christon, M.~Salazar, and C.~H. Onder, ``Fuel-optimal power
  split and gear selection strategies for a hybrid electric vehicle,'' in
  \emph{\uppercase{SAE} Technical Paper 2019-24-0205}, 2019.

\bibitem{limebeer2014optimal}
D.~Limebeer, G.~Perantoni, and A.~Rao, ``Optimal control of formula one car
  energy recovery systems,'' \emph{Int. J. Control}, vol.~87, no.~10, pp.
  2065--2080, 2014.

\bibitem{perantoni2015optimal}
G.~Perantoni and D.~Limebeer, ``Optimal control of a formula one car on a
  three-dimensional track -- part 1: Track modeling and identification,''
  \emph{ASME J. Dyn. Syst., Meas., Control}, vol. 137, no.~5, 2015, {A}rt. no.
  051018.

\bibitem{limebeer2015optimal}
D.~Limebeer and G.~Perantoni, ``Optimal control of a formula one car on a
  three-dimensional track -- part 2: Optimal control,'' \emph{ASME J. Dyn.
  Syst., Meas., Control}, vol. 137, no.~5, 2015, {A}rt. no. 051019.

\bibitem{perantoni2014optimal}
G.~Perantoni and D.~Limebeer, ``Optimal control for a formula one car with
  variable parameters,'' \emph{Vehicle System Dynamics}, vol.~52, no.~5, pp.
  653--678, 2014.

\bibitem{ebbesen2018time}
S.~Ebbesen, M.~Salazar, P.~Elbert, C.~Bussi, and C.~H. Onder, ``Time-optimal
  control strategies for a hybrid electric race car,'' \emph{IEEE Trans.
  Control Syst. Technol.}, vol.~26, no.~1, pp. 233--247, 2018.

\bibitem{Balerna}
C.~Balerna, M.-P. Neumann, N.~Robuschi, P.~Duhr, A.~Cerofolini, V.~Ravaglioli,
  and C.~Onder, ``Time-optimal low-level control and gearshift strategies for
  the formula 1 hybrid electric powertrain,'' \emph{Energies}, vol.~14, p. 171,
  2021.

\bibitem{serrao2011comparative}
L.~Serrao, S.~Onori, and G.~Rizzoni, ``A comparative analysis of energy
  management strategies for hybrid electric vehicles,'' \emph{ASME J. Dyn.
  Syst., Meas., Control}, vol. 133, no.~3, 2011, {A}rt. no. 031012.

\bibitem{paganelli2002equivalent}
G.~Paganelli, S.~Delprat, T.-M. Guerra, J.~Rimaux, and J.-J. Santin,
  ``Equivalent consumption minimization strategy for parallel hybrid
  powertrains,'' in \emph{Proc. 2002 IEEE VTC}, vol.~4, 2002, pp. 2076--2081.

\bibitem{Pisu2007}
P.~Pisu and G.~Rizzoni, ``A comparative study of supervisory control strategies
  for hybrid electric vehicles,'' \emph{IEEE Trans. Control Syst. Technol.},
  vol.~15, pp. 506--518, 2007.

\bibitem{nuesch2014equivalent}
T.~N{\"u}esch, A.~Cerofolini, G.~Mancini, N.~Cavina, C.~Onder, and L.~Guzzella,
  ``Equivalent consumption minimization strategy for the control of real
  driving {NO}x emissions of a diesel hybrid electric vehicle,''
  \emph{Energies}, vol.~7, no.~5, pp. 3148--3178, 2014.

\bibitem{ebbesen2012battery}
S.~Ebbesen, P.~Elbert, and L.~Guzzella, ``Battery state-of-health perceptive
  energy management for hybrid electric vehicles,'' \emph{IEEE Trans. Veh.
  Technol.}, vol.~61, no.~7, pp. 2893--2900, 2012.

\bibitem{zhao2015real}
D.~Zhao, R.~Stobart, G.~Dong, and E.~Winward, ``Real-time energy management for
  diesel heavy duty hybrid electric vehicles,'' \emph{IEEE Trans. Control Syst.
  Technol.}, vol.~23, no.~3, pp. 829--841, 2015.

\bibitem{Borhan2009}
H.~A. Borhan, A.~Vahidi, A.~M. Phillips, M.~L. Kuang, and I.~V. Kolmanovsky,
  ``Predictive energy management of a power-split hybrid electric vehicle,'' in
  \emph{Proc. 2009 ACC}, 2009, pp. 3970--3976.

\bibitem{borhan2012mpc}
H.~A. Borhan, A.~Vahidi, A.~M. Phillips, M.~L. Kuang, I.~V. Kolmanovsky, and
  S.~{Di Cairano}, ``Mpc-based energy management of a power-split hybrid
  electric vehicle,'' \emph{IEEE Trans. Control Syst. Technol.}, vol.~20,
  no.~3, pp. 593--603, 2012.

\bibitem{zhao2017characterisation}
D.~Zhao, E.~Winward, Z.~Yang, R.~Stobart, and T.~Steffen, ``Characterisation,
  control, and energy management of electrified turbocharged diesel engines,''
  \emph{Energy Conversion and Management}, vol. 135, pp. 416--433, 2017.

\bibitem{Zhou2021}
Q.~Zhou and C.~Du, ``A two-term energy management strategy of hybrid electric
  vehicles for power distribution and gear selection with intelligent
  state-of-charge reference,'' \emph{Journal of Energy Storage}, vol.~42, 2021,
  {A}rt. no. 103054.

\bibitem{lot2013lap}
R.~Lot and S.~Evangelou, ``Lap time optimization of a sports series hybrid
  electric vehicle,'' in \emph{2013 World Congress on Engineering}, 2013, pp.
  1--6.

\bibitem{salazar2017time}
M.~Salazar, P.~Elbert, S.~Ebbesen, C.~Bussi, and C.~H. Onder, ``Time-optimal
  control policy for a hybrid electric race car,'' \emph{IEEE Trans. Control
  Syst. Technol.}, vol.~25, no.~6, pp. 1921--1934, 2017.

\bibitem{salazar2018equivalent}
M.~Salazar, C.~Balerna, E.~Chisari, C.~Bussi, and C.~H. Onder, ``Equivalent lap
  time minimization strategies for a hybrid electric race car,'' in \emph{Proc.
  2018 IEEE CDC}, 2018, pp. 6125--6131.

\bibitem{salazar2018minimum}
M.~Salazar, P.~Duhr, C.~Balerna, L.~Arzilli, and C.~H. Onder, ``Minimum lap
  time control of hybrid electric race cars in qualifying scenarios,''
  \emph{IEEE Trans. Veh. Technol.}, vol.~68, no.~8, pp. 7296--7308, 2019.

\bibitem{salazar2017real}
M.~Salazar, C.~Balerna, P.~Elbert, F.~P. Grando, and C.~H. Onder, ``Real-time
  control algorithms for a hybrid electric race car using a two-level model
  predictive control scheme,'' \emph{IEEE Trans. Veh. Technol.}, vol.~66,
  no.~12, pp. 10\,911--10\,922, 2017.

\bibitem{Borsboom2020}
O.~Borsboom, C.~A. Fahdzyana, and M.~Salazar, ``Time-optimal control strategies
  for electric race cars with different transmission technologies,'' in
  \emph{Proc. 2020 IEEE VPPC}, 2020, pp. 1--5.

\bibitem{Borsboom2021}
O.~Borsboom, C.~A. Fahdzyana, T.~Hofman, and M.~Salazar, ``A convex
  optimization framework for minimum lap time design and control of electric
  race cars,'' \emph{IEEE Trans. Veh. Technol.}, vol.~70, pp. 8478--8489, 2021.

\bibitem{Locatello2021}
A.~Locatello, M.~Konda, O.~Borsboom, T.~Hofman, and M.~Salazar, ``Time-optimal
  control of electric race cars under thermal constraints,'' in \emph{Proc.
  2021 ECC}, 2021, pp. 905--912.

\bibitem{Michelini2003}
J.~Michelini and C.~Glugla, ``Control system design for steady state operation
  and mode switching of an engine with cylinder deactivation,'' in \emph{Proc.
  2003 ACC}, 2003, pp. 3125--3129 vol.4.

\bibitem{elbert2014engine}
P.~Elbert, T.~N{\"u}esch, A.~Ritter, N.~Murgovski, and L.~Guzzella, ``Engine
  on/off control for the energy management of a serial hybrid electric bus via
  convex optimization,'' \emph{IEEE Trans. Veh. Technol.}, vol.~63, no.~8, pp.
  3549--3559, 2014.

\bibitem{6476747}
N.~Murgovski, L.~M. Johannesson, and J.~Sjöberg, ``Engine on/off control for
  dimensioning hybrid electric powertrains via convex optimization,''
  \emph{IEEE Trans. Veh. Technol.}, vol.~62, no.~7, pp. 2949--2962, 2013.

\bibitem{Corno2019}
M.~Corno, L.~D'Avico, S.~Marelli, M.~Galvani, and S.~M. Savaresi, ``Predictive
  cylinder deactivation control for large displacement automotive engines,''
  \emph{IEEE Trans. Veh. Technol.}, vol.~68, pp. 9554--9563, 2019.

\bibitem{Fujiwara2008}
M.~Fujiwara, K.~Kumagai, M.~Segawa, R.~Sato, and Y.~Tamura, ``{Development of a
  6-cylinder Gasoline Engine with New Variable Cylinder Management
  Technology},'' Tech. Rep., 2008.

\bibitem{Sujan}
V.~A. Sujan, T.~R. Frazier, K.~Follen, and S.~M. Moon, ``System and method of
  cylinder deactivation for optimal engine torque-speed map operation,'' Patent
  US 8,886.422 B2, Nov. 11, 2014.

\bibitem{Josevski2016}
M.~Josevski and D.~Abel, ``Gear shifting and engine on/off optimal control in
  hybrid electric vehicles using partial outer convexification,'' in
  \emph{Proc. 2016 CCA}, 2016, pp. 562--568.

\bibitem{Bekker2009}
J.~Bekker and W.~Lotz, ``Planning formula one race strategies using
  discrete-event simulation,'' \emph{Journal of the Operational Research
  Society}, vol.~60, no.~7, pp. 952--961, 2009.

\bibitem{Heilmeier2018}
A.~Heilmeier, M.~Graf, and M.~Lienkamp, ``A race simulation for strategy
  decisions in circuit motorsports,'' in \emph{Proc. 2018 IEEE ITSC}, 2018, pp.
  2986--2993.

\bibitem{RacingLimited15/02/2022}
\BIBentryALTinterwordspacing
McLaren-Racing-Limited, ``Formula one race strategy,'' Royal Academy of
  Engineering, Tech. Rep., 15/02/2022. [Online]. Available:
  \url{https://www.raeng.org.uk/publications/other/14-car-racing}
\BIBentrySTDinterwordspacing

\bibitem{guzzella2004introduction}
L.~Guzzella and C.~H. Onder, \emph{Introduction to modeling and control of
  internal combustion engine systems}, 2nd~ed.\hskip 1em plus 0.5em minus
  0.4em\relax Berlin: Springer, 2010.

\bibitem{9769925}
P.~Duhr, A.~Sandeep, A.~Cerofolini, and C.~H. Onder, ``Convex performance
  envelope for minimum lap time energy management of race cars,'' \emph{IEEE
  Trans. Veh. Technol.}, vol.~71, no.~8, pp. 8280--8295, 2022.

\bibitem{Kirches2011}
\BIBentryALTinterwordspacing
C.~Kirches, \emph{Fast Numerical Methods for Mixed-Integer Nonlinear
  Model-Predictive Control}, 1st~ed.\hskip 1em plus 0.5em minus 0.4em\relax
  Wiesbaden: Vieweg+Teubner Verlag, 2011. [Online]. Available:
  \url{http://dx.doi.org/10.1007/978-3-8348-8202-8}
\BIBentrySTDinterwordspacing

\bibitem{Sager2005}
S.~Sager, \emph{Numerical methods for mixed-integer optimal control
  problems}.\hskip 1em plus 0.5em minus 0.4em\relax Tönning [u.a.]: Der Andere
  Verl., 2005.

\bibitem{sager2007solving}
S.~Sager, H.~G. Bock, and M.~Diehl, ``Solving mixed--integer control problems
  by sum up rounding with guaranteed integer gap,'' \emph{SIAM Journal on
  Control and Optimization}, 2007.

\bibitem{Casadi}
J.~A.~E. Andersson, J.~Gillis, G.~Horn, J.~B. Rawlings, and M.~Diehl,
  ``{CasADi} -- {A} software framework for nonlinear optimization and optimal
  control,'' \emph{Mathematical Programming Computation}, vol.~11, no.~1, pp.
  1--36, 2019.

\end{thebibliography}

\vspace{-0.3cm}
\begin{IEEEbiography}[{\includegraphics[width=1in,height=1.25in,clip,keepaspectratio]{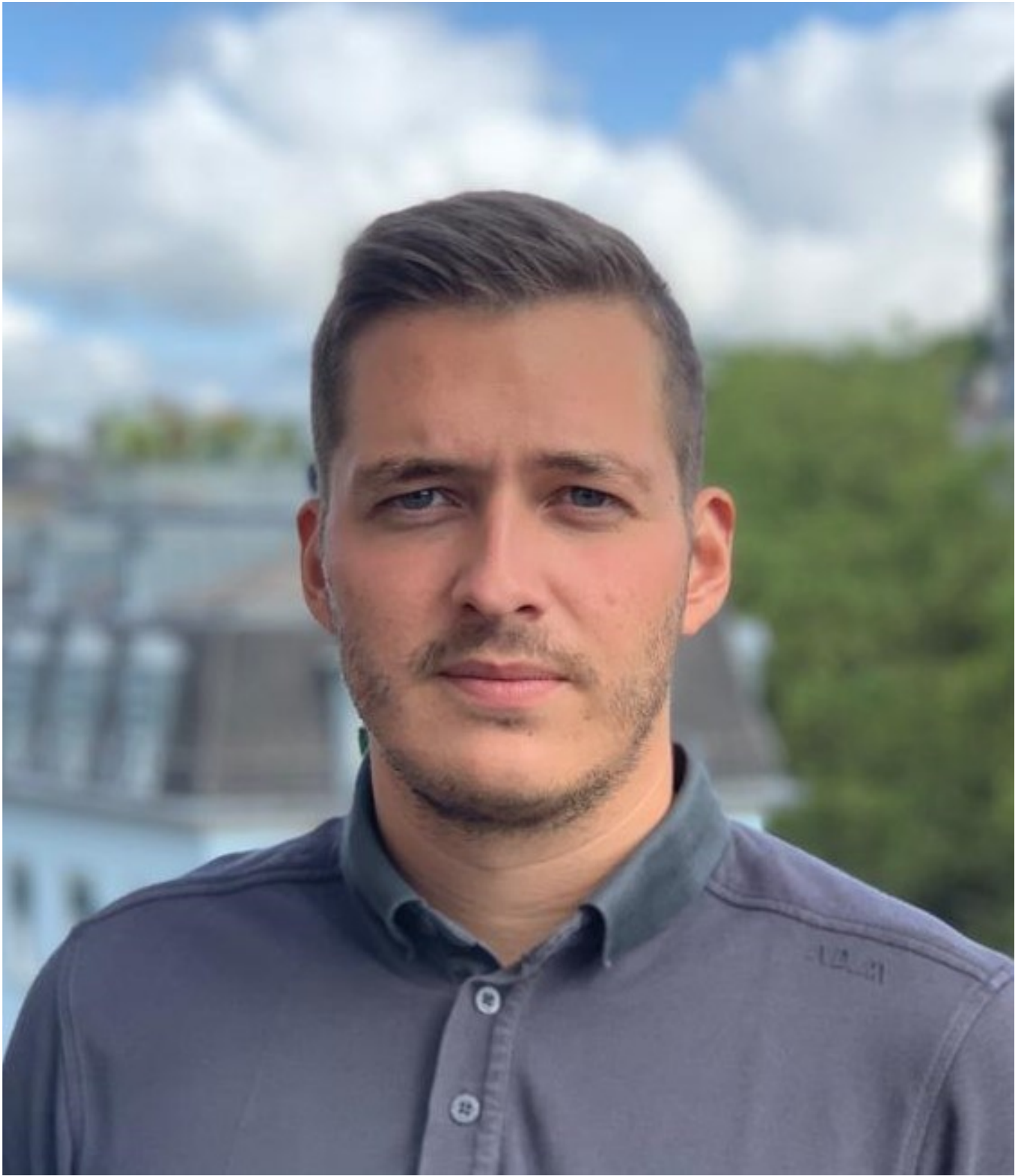}}]{Marc-Philippe Neumann}
	was born in Ludwigshafen am Rhein, Germany, and grew up near Lugano, Switzerland. 
	He received his B.Sc. and M.Sc. degree in mechanical engineering from ETH Zürich in 2016 and 2019, respectively. 
	Since July 2021 he pursues the Ph.D. degree with the Institute for Dynamic Systems and Control at ETH Zürich. 
	His research focuses on hybrid electric race car powertrains, as well as optimal control theory and model predictive control.
\end{IEEEbiography}
\begin{IEEEbiography}[{\includegraphics[width=1in,height=1.25in,clip,keepaspectratio]{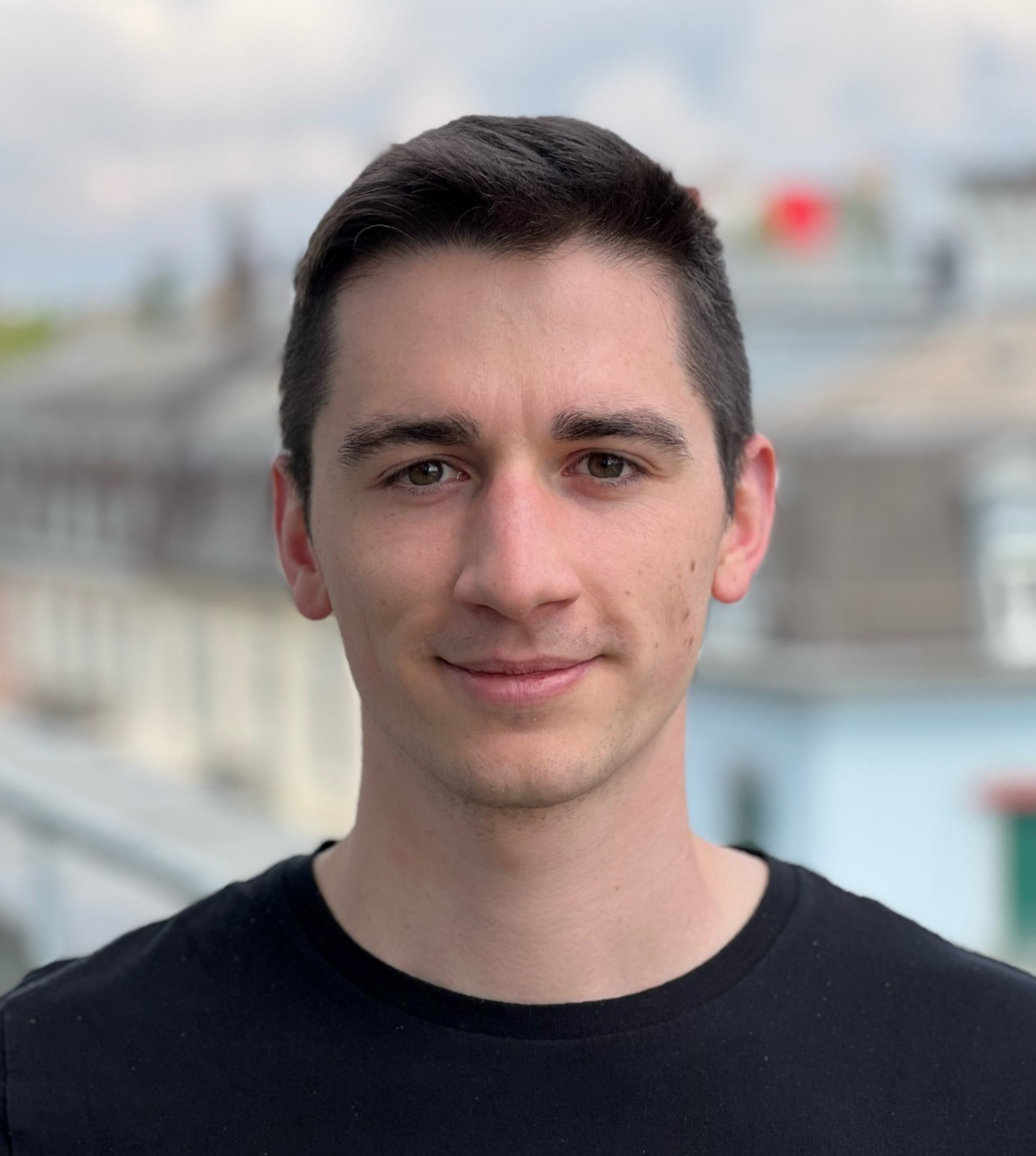}}]{Giona Fieni}
	was born in Mendrisio, Switzerland. He received his B.Sc. and M.Sc. degree in mechanical engineering from ETH Zürich in 2018 and 2021, respectively. 
	Since November 2022 he is enrolled as Ph.D. Student at the Institute for Dynamic Systems and Control at ETH Zürich. 
	His research fields include engine systems, race car power unit optimal control and modeling of large marine two-stroke dual-fuel engines.
	He focuses on optimal control theory and model predictive control.
\end{IEEEbiography}
\begin{IEEEbiography}[{\includegraphics[width=1in,height=1.25in,clip,keepaspectratio]{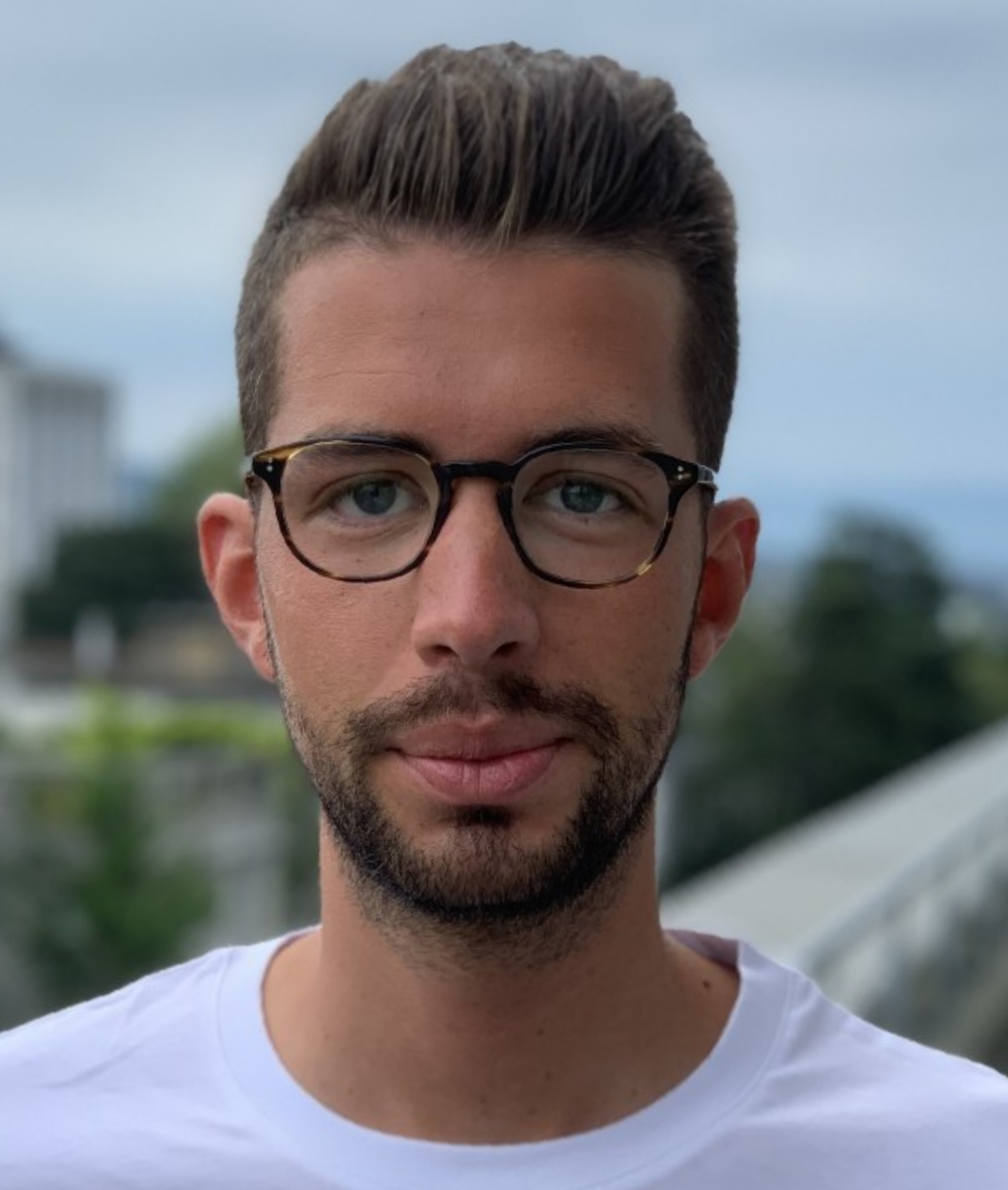}}]{Camillo Balerna}
	was born in Locarno, Switzerland. He received the B.Sc.\ degree in mechanical engineering from ETH Z\"urich in 2015, and the M.Sc.\ degree in mechanical engineering in 2016. He received the Doctor of Science degree in 2021 with the Institute for Dynamic Systems and Control, ETH Z\"urich.
	His research interests include engine systems, hybrid electric vehicles, and model predictive control.
	Mr.\ Balerna received the ABB Turbocharging and Engine Systems Dream internship award and the Johann Puch Automotive Award.
\end{IEEEbiography}
\begin{IEEEbiography}[{\includegraphics[width=1in,height=1.25in,clip,keepaspectratio]{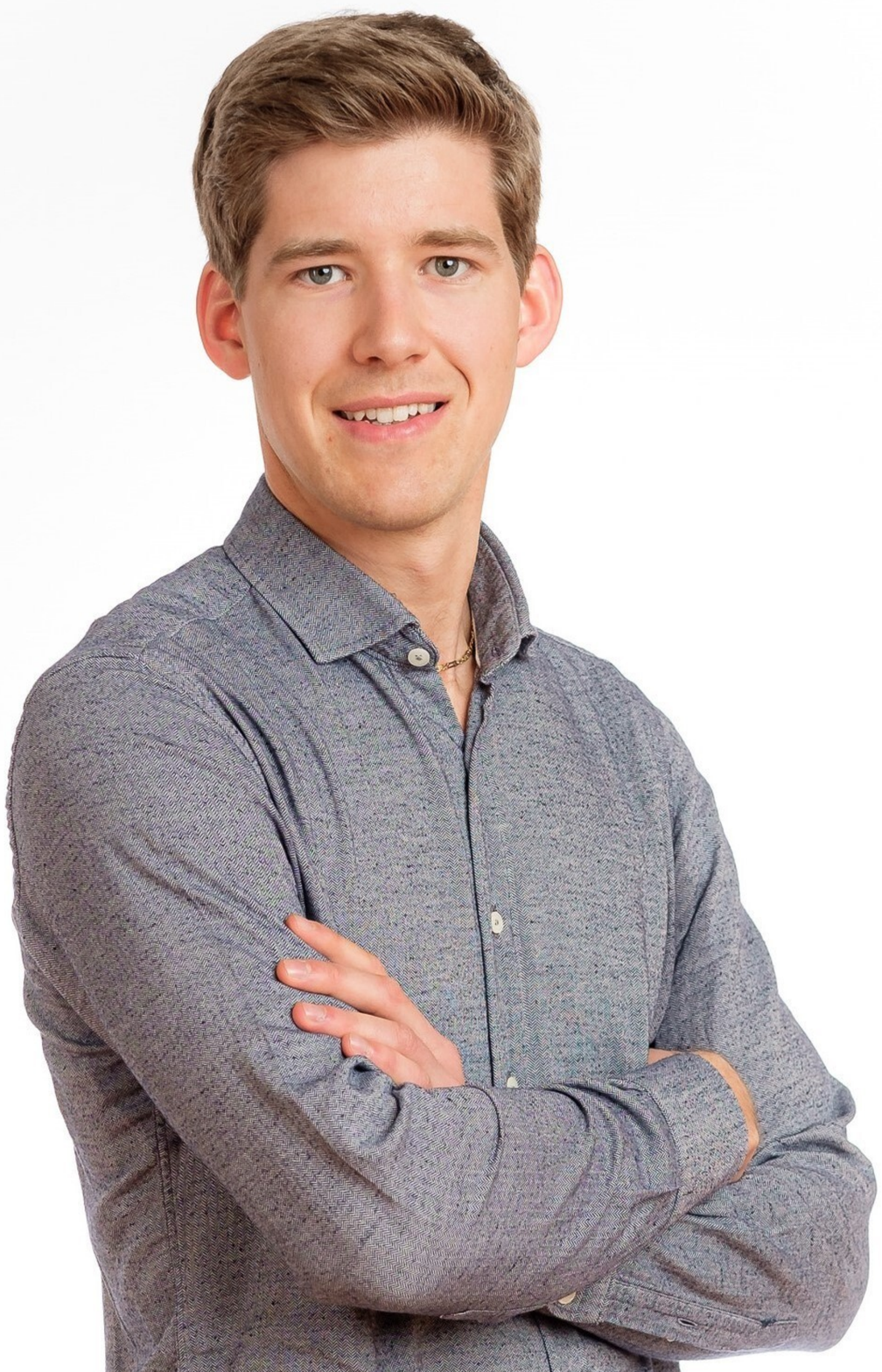}}]{Pol Duhr}
	received the B.Sc.\ and M.Sc.\ degrees in Mechanical Engineering in 2015 and 2018, respectively, both from ETH Z\"urich, Switzerland. He received the Doctor of Science degree in 2022 with the Institute for Dynamic Systems and Control at ETH Z\"urich.
	His research focuses on the time-optimal control of race car powertrains.
	For his Master thesis on the control of the Formula 1 power unit, he was awarded the ETH Medal, and he recently received the Student Award at the 2021 FISITA World Congress.
\end{IEEEbiography}
\begin{IEEEbiography}[{\includegraphics[width=1in,height=1.25in,clip,keepaspectratio]{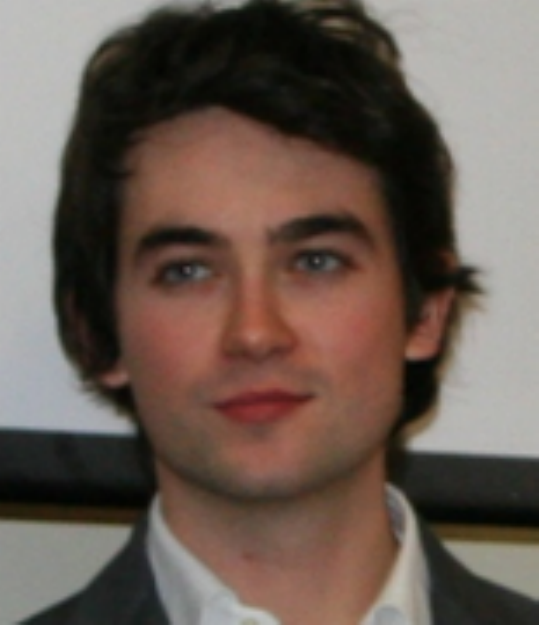}}]{Alberto Cerofolini}
	received the M.Sc. degree in Mechanical Engineering and the Ph.D. degree in Mechanics and Engineering Advanced Science from Universit\`{a} di Bologna, Italy, in 2009 and 2014, respectively. He currently holds a position as Power Unit Performance Engineer with the Power Unit Performance and Control Strategies Group of the Formula 1 team Scuderia Ferrari. His research focuses on lap-time-optimal and robust control strategies for the energy management of the Formula 1 car.
\end{IEEEbiography}
\begin{IEEEbiography}[{\includegraphics[width=1in,height=1.25in,clip,keepaspectratio]{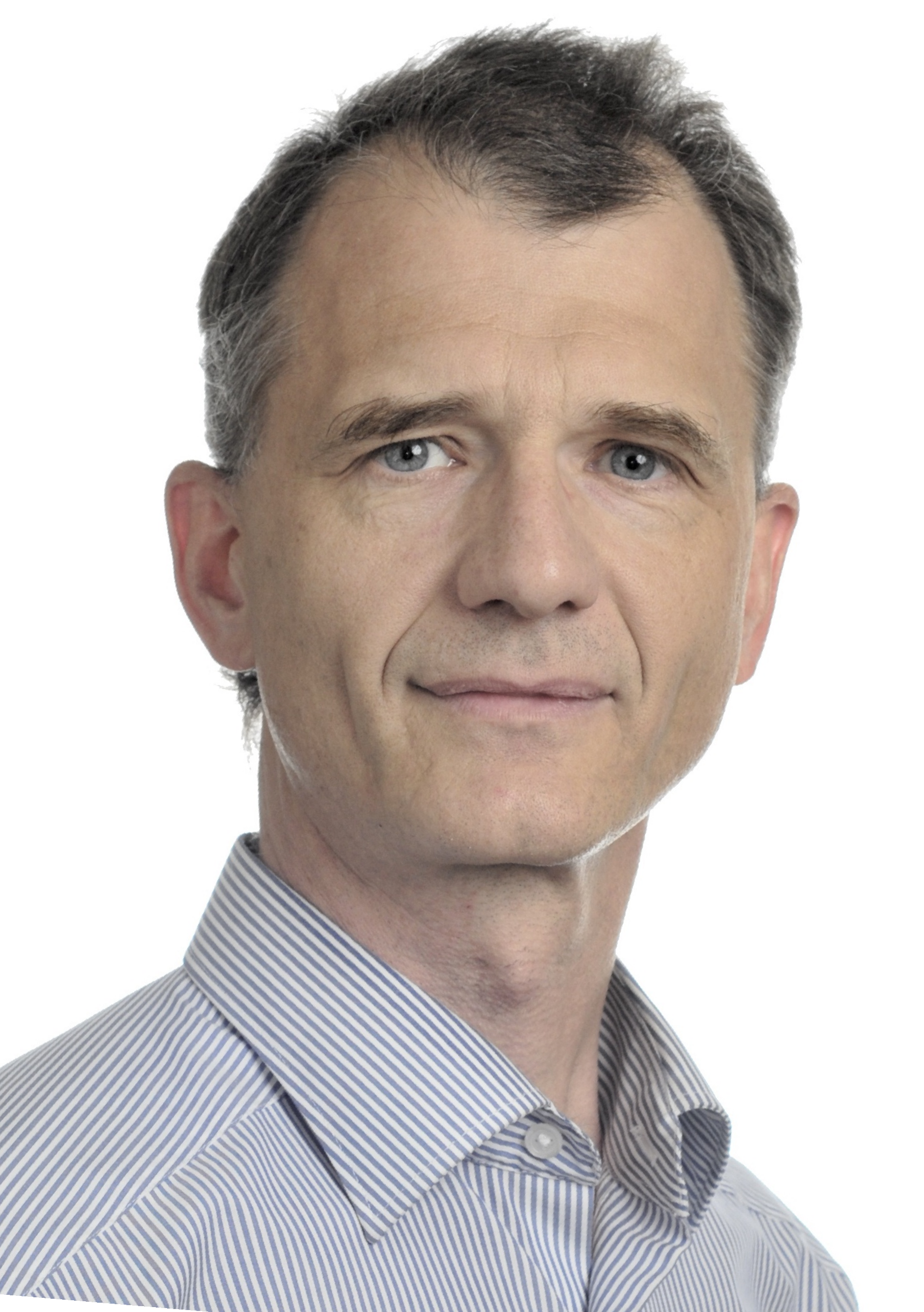}}]{Christopher H. Onder}
	received the Diploma and Ph.D.\ degrees in Mechanical Engineering from ETH Z\"urich, Switzerland. He is currently a Professor with the Institute for Dynamic Systems and Controls, ETH Z\"urich. He has authored or coauthored numerous articles and a book on modeling and control of engine systems. Prof.\ Dr.\ Onder was the recipient of the BMW Scientific Award, ETH Medal, Vincent Bendix Award, and Watt d'Or Energy Prize.
\end{IEEEbiography}
\vfill

\end{document}